\documentclass[showpacs,eqsecnum,preprintnumbers]{revtex4}
\usepackage{amsmath}
\usepackage{amsfonts}
\usepackage{amssymb}

\begin{document}

\title{Baryon wave function: Large-$N_{c}$ QCD and lessons from models}
\author{P.V. Pobylitsa}
\affiliation{Institute for Theoretical Physics II, Ruhr University Bochum, D-44780
Bochum, Germany\\ {\em and} \\
Petersburg Nuclear Physics Institute, Gatchina, St. Petersburg, 188300,
Russia}

\begin{abstract}
The structure of the $1/N_{c}$ expansion for the baryon distribution amplitude
in QCD is tested using quark models. Earlier conjectures about this structure
based on the evolution equation and on the soft-pion theorem are confirmed by
the model analysis. The problem of the calculation of the baryon wave function
at large $N_{c}$ is reduced to the analysis of equations of motion for an
effective classical dynamical system.
\end{abstract}
\maketitle

\section{Introduction}

\setcounter{equation}{0} 

\subsection{Large $N_{c}$: QCD and models}

The limit of the large number of colors $N_{c}$ and the $1/N_{c}$ expansion
\cite{Hooft-74}--\cite{Witten-79} are important nonperturbative methods in
QCD. However, in spite of the long studies of the large-$N_{c}$ limit, QCD\ is
not solved even in the leading order of the $1/N_{c}$ expansion. In the
applications of the $1/N_{c}$ expansion to QCD, one usually deals not with the
order-by-order \emph{calculation} of the $1/N_{c}$ corrections but with the
analysis of the \emph{structure} of the $1/N_{c}$ expansion.

The failure of the dynamical approach to the $1/N_{c}$ expansion in QCD led to
the redistribution of the effort towards various models imitating
large-$N_{c}$ QCD. This approach proved to be rather fruitful. Instead of
being trapped into the study of model artifacts (which often happens in the
phenomenology of strong interactions), the analysis of various large-$N_{c}$
models has revealed a number of their common properties including the
so-called large-$N_{c}$ spin-flavor symmetry \cite{Bardakci-84}--\cite{DJM-95}
which is believed to be a symmetry of large-$N_{c}$ QCD as well.

A classical example of results, historically first derived in the Skyrme model
but actually based on the spin-flavor symmetry of large-$N_{c}$ QCD, is the
ratio of the pion-nucleon ($g_{\pi NN}$) and pion-nucleon-$\Delta$ ($g_{\pi
N\Delta}$) constants \cite{Adkins-83}
\begin{equation}
\frac{g_{\pi N\Delta}}{g_{\pi NN}}=\frac{3}{2}\left[  1+O(N_{c}^{-2})\right]
\quad(\mathrm{exp.}\,1.48)\,.\label{g-pi-N-Delta-g-pi-NN}
\end{equation}
The same ratio $3/2$ can be derived in the quark model \cite{Manohar-84} in
the limit $N_{c}\rightarrow\infty$.

This simple example shows that the study of large-$N_{c}$ models can be
helpful for establishing the properties of large-$N_{c}$ QCD itself.

\subsection{Baryon wave function in the large-$N_{c}$ limit}

This paper is devoted to the $1/N_{c}$ expansion for the baryon wave function.
This problem was raised in Refs. \cite{PP-03,DP-04}. The construction of the
$1/N_{c}$ expansion for the baryon wave function is rather nontrivial, since
one has to deal with the exponential large-$N_{c}$ behavior. In Refs.
\cite{Pobylitsa-04,Pobylitsa-05} it was shown that this exponential behavior
can be consistently described using a specially designed generating functional
for the baryon wave function. Strictly speaking, the results of Refs.
\cite{Pobylitsa-04,Pobylitsa-05} are based on \emph{conjectures}, whose
consistency was checked using various indirect methods like evolution equation
\cite{Pobylitsa-04} and soft-pion theorem \cite{Pobylitsa-05}. In this paper
we provide additional arguments supporting these conjectures, using models of
large-$N_{c}$ QCD.

Before describing the main features of the $1/N_{c}$ expansion for the baryon
wave function we would like to agree about the definition of the baryon wave
function. The concept of the wave function is natural for nonrelativistic
quantum mechanics but its incorporation in QCD requires some extra
specifications. Usually one keeps in mind the transition matrix element
between baryon $|B\rangle$ and vacuum $|0\rangle$
\begin{equation}
\langle0|\prod_{c=1}^{N_{c}}\psi_{c}(z_{c})|B\rangle\,,\label{ME-B-0}
\end{equation}
where $\psi_{c}(z)$ is the quark field with the color index $c$. Strictly
speaking one has to settle several questions:

1) ordering of quark operators $\psi_{c}(z_{c})$,

2) insertion of Wilson lines providing the gauge invariance,

3) choice of the space-time or momentum kinematics (equal-time wave function,
light-cone wave function etc.).

From the point of view of QCD applications to hard exclusive processes the
most interesting object is the \emph{baryon distribution amplitude}
\cite{Lepage-Brodsky-80,CZ-84,BL-89} which corresponds to the choice
\begin{equation}
z_{c}=\lambda_{c}n,\quad n^{2}=0
\end{equation}
in Eq. (\ref{ME-B-0}). Here $n$ is a fixed light-cone vector. The standard
definition of the baryon distribution amplitude also involves the Fourier
transformation from parameters $\lambda_{c}$ to variables $x_{c}$
corresponding to the momentum fractions of quarks in the infinite momentum
frame:
\begin{equation}
\int d\lambda e^{i\lambda_{k}x_{k}}\langle0|\prod_{k=1}^{N_{c}}\psi_{c_{k}
}(\lambda_{k}n)|B\rangle\sim\varepsilon_{c_{1}c_{2}\ldots c_{N_{c}}}\Psi
_{B}(x_{1},x_{2},\ldots,x_{N_{c}})\,.\label{Psi-B-function}
\end{equation}
We have omitted various normalization factors which are not important for the
discussion of the general properties of the $1/N_{c}$ expansion.

Although the light-cone elements of the definition of the baryon distribution
amplitude are important from the \emph{dynamical} point of view, they are
rather insignificant for the \emph{structure of the} $1/N_{c}$
\emph{expansion}. What is really important for the construction of the
$1/N_{c}$ expansion is that the baryon wave function depends on $N_{c}$ quark
variables. At this moment one meets a serious problem:\ how can one construct
the $1/N_{c}$ expansion for the function (\ref{Psi-B-function}) which depends
on $N_{c}$ variables $x_{k}$? This problem is much more general than the
context of the baryon distribution amplitude. Below we will consider the
general case without specifying the precise meaning of the wave function
$\Psi_{B}(x_{1},x_{2},\ldots,x_{N_{c}})$ and of its arguments
$x_{1},x_{2},\ldots,x_{N_{c}}$. For example, the variables $x_{k}$
can be composite
objects carrying information about the spin and flavor of quarks in addition
to their coordinates or momenta. Since the antisymmetric color tensor
$\varepsilon_{c_{1}c_{2}\ldots c_{N_{c}}}$ is factored out in Eq.
(\ref{Psi-B-function}), the function $\Psi_{B}(x_{1},x_{2},\ldots,x_{N_{c}}) $
is symmetric in all its arguments $x_{1},x_{2},\ldots,x_{N_{c}}$.

In Refs. \cite{Pobylitsa-04,Pobylitsa-05} the problem of the construction the
$1/N_{c}$ expansion for function depending on $N_{c}$ variables was solved by
introducing a generating functional for the baryon wave function. Omitting
irrelevant technical details (spin and flavor structure, light-cone
kinematics, $1/N_{c}$ rescaling of the quark momentum, etc.) and concentrating
on the general structure which is important for the large-$N_{c}$ limit, we
can define this generating functional as
\begin{equation}
\Phi_{B}(g)=\int\Psi_{B}(x_{1},x_{2},\ldots,x_{N_{c}})g(x_{1})g(x_{2})\ldots
g(x_{N_{c}})dx_{1}dx_{2}\ldots dx_{N_{c}}\,,\label{Phi-B-def-0}
\end{equation}
where $g(x)$ is an arbitrary ``source'' function. Since function $\Psi
_{B}(x_{1},x_{2},\ldots,x_{N_{c}})$ is symmetric in permutations of
$x_{1},x_{2},\ldots,x_{N_{c}}$, the transition from the wave function
$\Psi_{B}(x_{1},x_{2},\ldots,x_{N_{c}})$ to the functional $\Phi_{B}(g)$ does
not lead to any loss of information at finite $N_{c}$.

As was argued in Refs. \cite{Pobylitsa-04,Pobylitsa-05}, the functional
$\Phi_{B}(g)$ has an exponential behavior at large $N_{c}$:

\begin{equation}
\Phi_{B}(g)\overset{N_{c}\rightarrow\infty}{\sim}\exp\left[  N_{c}W(g)\right]
\,.\label{Phi-B-exponential}
\end{equation}
where $W(g)$ is some ($N_{c}$ independent) functional of $g$. An important
property of the functional $W(g)$ is its universality: all low-lying baryons
[with $O(N_{c}^{-1})$ or $O(N_{c}^{0})$ excitation energy] are described by
the same functional $W(g)$.

The dependence on the type of the baryon $B$ appears in the preexponential
factor $A_{B}(g)$:
\begin{equation}
\Phi_{B}(g)\overset{N_{c}\rightarrow\infty}{=}N_{c}^{\nu_{B}}A_{B}
(g)\exp\left[  N_{c}W(g)\right]  \,\left[  1+O(N_{c}^{-1})\right]
.\label{Phi-A-W}
\end{equation}
Although the functional $A_{B}(g)$ depends on the baryon $B$, this dependence
has simple factorization properties studied in Ref. \cite{Pobylitsa-05}.

Strictly speaking, these properties of functionals $W(g)$ and $A_{B}(g)$ were
not properly derived in Refs. \cite{Pobylitsa-04,Pobylitsa-05}. Instead the
consistency of these properties with a number of basic features of QCD was checked:

-- QCD evolution equation \cite{Pobylitsa-04},

-- asymptotic behavior in terms of anomalous dimensions of leading twist
operators \cite{Pobylitsa-04},

-- spin-flavor symmetry of large-$N_{c}$ QCD \cite{Pobylitsa-05},

-- soft-pion theorem for the baryon distribution amplitude \cite{Pobylitsa-05}.

Although all these consistency checks successfully passed, they cannot be
considered as a final proof. In this paper we turn to the analysis of
large-$N_{c}$ models in order to find additional arguments supporting the
large-$N_{c}$ behavior (\ref{Phi-A-W}) and illustrating the basic properties
of functionals $W(g)$ and $A_{B}(g)$.

\subsection{Consequences of the spin-flavor symmetry}

In the absence of the dynamical solution for large-$N_{c}$ QCD, one cannot
expect to find much more than relations like the $N$-$\Delta$ ratio
(\ref{g-pi-N-Delta-g-pi-NN}) in the pure large-$N_{c}$ approach. In Ref.
\cite{Pobylitsa-05} a relation was derived for the generating functionals
$\Phi_{T_{3}J_{3}}^{(N)}(g)$ and $\Phi_{T_{3}J_{3}}^{(\Delta)}(g)$ describing
the baryon wave functions of nucleon and Delta resonance where $T_{3}$ and
$J_{3}$ are projections of isospin and spin, respectively.

In order to write this relation let us first notice that
$\Phi_{T_{3}J_{3}}^{(N)}(g)$ is a $2\times2$ matrix with indices
$T_{3},J_{3}$. Let us consider the $2\times2$ matrix
\begin{equation}
R=\frac{\Phi^{(N)}(g)}{\sqrt{\det\Phi^{(N)}(g)}}\,.
\end{equation}
Obviously
\begin{equation}
\det R=1
\end{equation}
so that matrix $R$ belong to the group $SL(2,C)$. The irreducible
representations of $SL(2,C)$ are described by Wigner functions 
$D^{(j_{1},j_{2})}(R)$ depending on two ``spins'' $j_{1},j_{2}$.
Let us consider the
Wigner function corresponding to $j_{1}=3/2$, $j_{2}=0$. For brevity we will
denote it simply $D_{T_{3}J_{3}}^{3/2}$. Now we are ready to write the
$N$-$\Delta$ relation derived in Ref. \cite{Pobylitsa-05}:
\begin{equation}
\Phi_{T_{3}J_{3}}^{(\Delta)}(g)=D_{T_{3}J_{3}}^{3/2}\left[  \frac{\Phi
^{(N)}(g)}{\sqrt{\det\Phi^{(N)}(g)}}\right]  \sqrt{2\det\Phi^{(N)}(g)}\left[
1+O(N_{c}^{-1})\right]  \,.\label{Phi-Delta-N-relation}
\end{equation}
Relation (\ref{Phi-Delta-N-relation}) was derived in Ref. \cite{Pobylitsa-05}
using the realization of the large-$N_{c}$ spin-flavor symmetry in the space
of the preexponential functionals $A_{B}(g)$ appearing in Eq. (\ref{Phi-A-W}).

Due to the universality of the functional $W(g)$, relation
(\ref{Phi-Delta-N-relation}) is equivalent to the analogous relation for the
preexponential functionals $A_{T_{3}J_{3}}^{(\Delta)}(g)$ and 
$A_{T_{3}J_{3}}^{(N)}(g)$ appearing in Eq. (\ref{Phi-A-W}):
\begin{equation}
A_{T_{3}J_{3}}^{(\Delta)}(g)=D_{T_{3}J_{3}}^{3/2}\left[  \frac{A^{(N)}
(g)}{\sqrt{\det A^{(N)}(g)}}\right]  \sqrt{2\det A^{(N)}(g)}\left[
1+O(N_{c}^{-1})\right]  \,.
\end{equation}

\subsection{What do we want of models?}

It should be stressed that the construction of viable realistic models of the
baryon distribution amplitude is not the aim of this paper. We are interested
in another problem. We want to trace how the exponential large-$N_{c}$
behavior (\ref{Phi-A-W}) appears within a certain class of models. The main
problem of large-$N_{c}$ QCD is that its dynamics is not solved. Therefore the
analysis of Refs. \cite{Pobylitsa-04,Pobylitsa-05} was based on the
\emph{assumption} about the exponential large-$N_{c}$ behavior (\ref{Phi-A-W})
with the universal functional $W(g)$.

In QCD we cannot compute the functional $W(g)$. On the contrary, in the models
studied in this paper both the exponential behavior (\ref{Phi-A-W}) and the
calculation of $W(g)$ are under good theoretical control. From the point of
view of the aim of this paper (model check of the assumptions made in
large-$N_{c}$ QCD), the result of this work is trivial: all assumptions
involved in the theoretical construction \cite{Pobylitsa-04,Pobylitsa-05} are
confirmed in the models studied in this paper. This puts the results of Refs.
\cite{Pobylitsa-04,Pobylitsa-05} on more solid ground.

Several comments must be made about the status of the spin-flavor symmetry in
our model analysis. Most of this work is based on the Hamiltonian
(Schr\"{o}dinger) analysis of models. As is well known, the mass difference
$M_{\Delta}-M_{N}$ between the $\Delta$ resonance and nucleon is 
$O(N_{c}^{-1})$, whereas the masses $M_{N}$ and $M_{\Delta}$ grow as 
$O(N_{c})$.
Therefore the mass splitting $M_{\Delta}-M_{N}$ is an effect of the
next-to-next-to-leading order of the $1/N_{c}$ expansion. In this paper we do
not go so deeply into the $1/N_{c}$ expansion. Actually the main part is
devoted to the calculation of the functional $W(g)$ corresponding to the
leading order. Therefore the subtleties of the realization of the spin-flavor
symmetry are not important for the most of our analysis. Moreover, in our
analysis we make no special assumptions about the symmetries of the
Hamiltonian [only the $SU(N_{c})$ color symmetry is important for us].

On the other hand, the readers interested in model implementations of the
spin-flavor symmetry can be addressed to Ref. \cite{Pobylitsa-04} where
functionals $W(g)$ and $A_{B}(g)$ are computed in the naive quark model and
the results explicitly agree with all constraints imposed by the spin-flavor symmetry.

\subsection{Mean field approximation and large-$N_{c}$ limit}

In contrast to ``unsolvable'' large-$N_{c}$ QCD, many large-$N_{c}$ models
allow for a straightforward construction of the $1/N_{c}$ expansion. Although
the precise form of the $1/N_{c}$ expansion can be model dependent, many
features are common since they are based on the mean field approximation which
is justified in these models (but not in QCD) in the large-$N_{c}$ limit. The
manifestations of the underlying mean field approximation can be different:
saddle point approximation in the path integral approach, the Hartree equation
in the Hamiltonian approach to the models with explicit quark degrees of
freedom, classical equations of motion in models based on meson fields, etc.

In this paper we work with models explicitly containing quark degrees of
freedom. In the leading order of the $1/N_{c}$ expansion the solution of these
models is described by the Hartree equation.

\subsection{Density matrix or wave function?}

This paper deals with the $1/N_{c}$ expansion for the functional $\Phi_{B}(g)
$ (\ref{Phi-B-def-0}) describing the baryon wave function. One can wonder how
this approach is related to the mean field method. Note that the mean-field
approach to fermion systems is based on equations for the \emph{density
matrix} whereas the functional $\Phi_{B}(g)$ is constructed in terms of the
\emph{wave function} $\Psi_{B}(x_{1},x_{2},\ldots,x_{N_{c}})$. Strictly
speaking the transition from the multiparticle wave function to the
single-particle density matrix
\begin{equation}
\rho(x_{1}^{\prime},x_{1})=\int dx_{2}\ldots dx_{N_{c}}\Psi_{B}(x_{1}^{\prime
},x_{2},\ldots,x_{N_{c}})\Psi_{B}^{\ast}(x_{1},x_{2},\ldots,x_{N_{c}})
\end{equation}
leads to a \emph{loss} of information. Therefore the knowledge of the mean
field solution in terms of the density matrix $\rho(x_{1}^{\prime},x_{1})$ is
not sufficient for the calculation of the functionals $W(g)$ and $A_{B}(g) $
describing the large-$N_{c}$ behavior (\ref{Phi-A-W}) of the baryon wave
function. Nevertheless one can derive a closed equation for the functional
$W(g)$ [See Sec. \ref{Large-Nc-section} and Eq. (\ref{W-WKB})]. An essential
part of this paper is devoted to the analysis of this equation and to the
construction of its solutions.

In the leading order of the $1/N_{c}$ expansion the density matrix
$\rho(x^{\prime},x)$ is described by the Hartree equation which leads to the
representation for $\rho(x^{\prime},x)$ in terms of single particle wave
functions $\psi_{n}(x)$ of occupied states:
\begin{equation}
\rho(x^{\prime},x)\overset{N_{c}\rightarrow\infty}{\longrightarrow}
\sum\limits_{n:\mathrm{occupied}}\psi_{n}(x^{\prime})\psi_{n}^{\ast}(x)\,.
\end{equation}
In the simplest models with only one ``valence'' level $\psi_{0}$ occupied by
$N_{c}$ quarks, the density matrix factorizes into the product
\begin{equation}
\rho(x^{\prime},x)\overset{N_{c}\rightarrow\infty}{\longrightarrow}\psi
_{0}(x^{\prime})\psi_{0}^{\ast}(x)\,.\label{rho-factorized}
\end{equation}
Here we assume that the trivial color term $\delta_{cc^{\prime}}$ is factored out.

In the large-$N_{c}$ limit the factorization of the density matrix
(\ref{rho-factorized}) becomes asymptotically exact. Naively one could think
that the wave function $\Psi(x_{1},x_{2},\ldots,x_{N_{c}})$ has a similar
factorization at large $N_{c}$:
\begin{equation}
\Psi(x_{1},x_{2},\ldots,x_{N_{c}})\overset{N_{c}\rightarrow\infty
}{\longrightarrow}\prod\limits_{k=1}^{N_{c}}\psi_{0}(x_{k})\quad
\mathrm{(wrong!)}\,.\label{Psi-factorization-naive}
\end{equation}
Unfortunately this statement is wrong. In order to see the violation of Eq.
(\ref{Psi-factorization-naive}) let us multiply it by 
$\prod_{k=1}^{N_{c}}g(x_{k})$ and integrate over all $x_{k}$. If Eq.
(\ref{Psi-factorization-naive}) were correct, then we would obtain using Eq.
(\ref{Phi-A-W})
\begin{equation}
W(g)=\ln\left[  \int\psi_{0}(x)g(x)dx\right]  \quad\mathrm{(wrong!)}
\,.\label{W-naive}
\end{equation}
In Sec. \ref{Asymmetric-rotator-section} we describe a model for which the
functional $W(g)$ can be computed analytically [Eq. (\ref{W-g-rotator-res})].
This result explicitly shows the breakdown of the naive expectation
(\ref{W-naive}).

Although the naive statement (\ref{Psi-factorization-naive}) about the
factorization of the wave function can be often met in literature, this
statement can be trusted only in the sense of the factorization of the density
matrix (\ref{rho-factorized}) in the leading order of the $1/N_{c}$ expansion.

To summarize, we cannot use Eq. (\ref{W-naive}) for the calculation of the
functional $W(g)$. Moreover, the knowledge of the solution of the Hartree
equation is not sufficient for the calculation of $W(g)$. In this paper we
derive a differential equation for $W(g)$ and construct its solution.

\subsection{Factorization of the preexponential terms}

The universality of the large-$N_{c}$ behavior (\ref{Phi-B-exponential}) holds
only with the exponential accuracy. The preexponential functional $A_{B}(g)$
in Eq. (\ref{Phi-A-W}) depends on the baryon (baryon-meson) state $B$. In
Refs. \cite{Pobylitsa-04,Pobylitsa-05} several important factorization
properties of these functionals were suggested and checked using the evolution
equation, the spin-flavor symmetry and the soft-pion theorem. In this paper we
verify these factorization properties by explicit calculations in
large-$N_{c}$ models.

One should keep in mind that the factorization properties of functionals
$A_{B}(g)$ are determined by two ingredients:

1) the breakdown of the spin and flavor symmetries by the mean field solution
and the restoration of these symmetries via the standard method of ``the
quantization of collective coordinates'',

2) the spectrum of the harmonic fluctuations above the mean field solution.

If the mean field solution for the baryon had the same symmetries as the
vacuum mean field solution, then we would have a simple oscillator-like
spectrum of baryons
\begin{equation}
\mathcal{E}_{B}\equiv\mathcal{E}_{\{n_{k}\}}=N_{c}E_{0}+\left(  \Delta
E_{0}+\sum\limits_{k}n_{k}\Omega_{k}\right)  +O(N_{c}^{-1})\,,
\end{equation}
where the baryon excitations $B$ are labeled by sets of integer numbers
$\{n_{k}\}$ associated with elementary $\Omega_{k}$ excitations. In this case
the factorization of $A_{B}(g)$ is described by the formula
\begin{equation}
A_{B}\left(  g\right)  \equiv A_{\{n_{k}\}}(g)=A^{(0)}(g)\prod\limits_{k}
\left[  A_{k}(g)\right]  ^{n_{k}}\,.\label{A-factorization-1}
\end{equation}

In the case when the mean field solution breaks some symmetries (which are
nevertheless restored by the collective quantization) the situation is more
subtle and the precise expression for $A_{B}\left(  g\right)  $ depends on the
involved symmetries. As is well known, the effects of broken symmetries in the
mean field approach to large-$N_{c}$ models correspond to the manifestation of
the spin-flavor symmetry in large-$N_{c}$ QCD. The structure of the functional
$A_{B}\left(  g\right)  $ in large-$N_{c}$ QCD and the role of the spin-flavor
symmetry were studied in Ref. \cite{Pobylitsa-05}.

In this paper we concentrate on models where the effects of broken symmetries
and zero modes are absent so that one has the simple factorization formula
(\ref{A-factorization-1}). In Sec. \ref{Asymmetric-rotator-section} we
describe a model where the factorization relation (\ref{A-factorization-1})
can be checked explicitly.

\subsection{Schr\"{o}dinger equation versus path integral}

One can use two methods for the analysis of the models for the baryon wave
function: the stationary Schr\"{o}dinger equation or the path integral
approach. In principle, these two methods should give equivalent results. It
is well known that the large-$N_{c}$ limit has many common features with the
semiclassical limit. In the path integral approach this leads to a
representation for $W(g)$ in terms of classical \emph{time-dependent}
trajectories. In the case of the stationary Schr\"{o}dinger equation one works
with the \emph{time-independent} formalism.

In the simplest models (i.e. models with the trivial vacuum), the connection
between the two representations can be easily seen. However, in the general
case the equivalence of results obtained in the two approaches is less
obvious. Therefore in this paper we use both methods. The first part of the
paper (Secs. \ref{models-section}--\ref{Nontrivial-vacuum-section}) is based
on the Schr\"{o}dinger equation. In Sec. \ref{Path-integral-section} we
describe the path integral approach and derive a representation for $W(g)$ in
terms of trajectories. After that we demonstrate the equivalence of the two
representations for $W(g)$ in Sec. \ref{Equivalence-section}.

\subsection{Large-$N_{c}$ limit and classical dynamics}

An essential part of this paper is devoted to the representation of $W(g)$ in
terms of classical trajectories described by an effective Hamiltonian. In
fact, the functional $W(g)$ can be interpreted as a classical action depending
on the ``coordinate'' $g$. This action obeys the classical Hamilton--Jacobi
equation. The method of trajectories can be used for the construction of the
solution of this Hamilton--Jacobi equation.

The connection between the large-$N$ systems and classical dynamics was
extensively discussed in literature \cite{Berezin-78} (see also review
\cite{Yaffe-82} and references therein). In the path integral approach to the
large-$N$ systems, the $1/N$ expansion can be constructed using the saddle
point method. In a straightforward approach the corresponding saddle point
equations have a nonlocal form. Nevertheless in many cases it is possible to
find new degrees of freedom which reduce the saddle point equations to a local
form described by a local Hamiltonian dynamics.

In this paper we use similar methods and reduce the problem of the calculation
of the functional $W(g)$ to the analysis of trajectories described by an
effective Hamiltonian. However, the problem studied in this paper is different
from the ``traditional'' time-dependent problems in large-$N$ systems
\cite{Berezin-78}. Most of the works on the effective classical dynamics of
large-$N_{c}$ systems concentrate on the time dependence of \emph{expectation
values} of observables so that the classical trajectories correspond to the
time-dependent \emph{diagonal} matrix elements of the corresponding quantum
operators. On the contrary, the calculation of the baryon wave function is a
problem of \emph{nondiagonal transition matrix elements.} In spite of the
difference between the two cases, we will see that the corresponding effective
classical dynamical systems have many common features.

\subsection{Large $N_{c}$ models and traditional many-body physics}

In our analysis of the large-$N_{c}$ models we will meet many equations which
are well known in the ``traditional'' many-body physics (see e.g. Ref.
\cite{RS-80}). The machinery of the old many-body physics includes many
interesting equations: stationary Hartree--Fock, time-dependent Hartree--Fock
equation, random phase approximation (RPA) equations. A common feature of
these equations is that their derivation is usually based on \emph{ad hoc}
approximations so that in a general case the validity of these approximations
is a matter of luck. For example, the standard derivation of the Hartree--Fock
equation is based on the variational principle with a special ansatz for the
wave function. Nevertheless in some cases one can find a justification for
these approximations. In particular, the $1/N$ expansion ($N$ not necessarily
being color) puts these equations on solid ground \cite{GLM-65,HL-82}. In the
framework of the $1/N$ expansion, the Hartree equation corresponds to the
leading order of the $1/N$ expansion whereas the contribution of the Fock term
is $1/N$ suppressed. Similarly, in the next-to-leading order of the $1/N$
expansion one arrives at the RPA equation \cite{GLM-65,Pobylitsa-03} with
slight modifications caused by the large-$N$ limit

As was explained above, the functional $W(g)$ cannot be expressed via the
solution of the \emph{stationary} Hartree equation. However, in Sec.
\ref{Classical-dynamics-section} we will see that $W(g)$ can be expressed via
solutions of the so-called \emph{time-dependent} Hartree equation (TDHE).
Indeed, as was mentioned above, the functional $W(g)$ is just an action for
some effective mechanical system. The equations of motion describing these
trajectories of this mechanical system are similar to the TDHE. The properties
and solutions of the TDHE were extensively studied in the context of the
problems of the many-body physics (typically in its Hartree--Fock version
\cite{RS-80}). Strictly speaking, the analogy between our equations and TDHE
is not complete since our problem the large-$N_{c}$ wave functions differs
from the traditional problems of the many-body physics. Nevertheless many
interesting connections can be found. For example, our analysis of the
effective large-$N_{c}$ dynamics makes use of the Hamiltonian structure of the
classical equations of motion. In the context of the ``traditional'' approach
to TDHE (with Fock term included), the classical Hamiltonian structure of the
TDHE was studied in Ref. \cite{KK-76}.

\subsection{Antiquarks in large-$N_{c}$ models}

In the simplest models the baryon consists of $N_{c}$ quarks without any
quark-antiquark pairs. This class of models is studied in the first part of
the paper (Secs. \ref{Simplest-models-section}
--\ref{Asymmetric-rotator-section}). The second part of this paper (Secs.
\ref{Nontrivial-vacuum-section}--\ref{Equivalence-section}) deals more
complicated models where the quark-antiquark pairs are described using the
well-known ``old-fashioned'' Dirac representation for antiparticles. In these
models we have a nontrivial \emph{physical} vacuum made of $MN_{c}$ quarks put
into the \emph{bare} vacuum (with some integer $M$), whereas the baryon is
considered as a state containing $(M+1)N_{c}$ quarks above the bare vacuum.
The baryon wave function [defined by Eq. (\ref{ME-B-0}) as a transition matrix
element between the baryon and the \emph{physical} vacuum] still can be used
for the construction of the generating functional $\Phi_{B}(g)$
(\ref{Phi-B-def-0}). We show that in this class of models we also have the
exponential behavior (\ref{Phi-B-exponential}) of $\Phi_{B}(g)$ and develop
methods for the calculation of the functional $W(g)$.

\subsection{From quantum mechanics to quantum field theory}

The general structure of the $1/N_{c}$ expansion for the functional $\Phi
_{B}(g)$ is the same in quantum mechanics and in quantum field theory. For
simplicity we use the \emph{discrete} quantum mechanical notation in this
paper. For example, instead of the continuous variables $x_{k}$ of Eq.
(\ref{Phi-B-def-0}) we write discrete indices $i_{k}$:
\begin{equation}
\Phi_{B}(g)=\sum\limits_{i_{1}i_{2}\ldots i_{N_{c}}}\Psi_{i_{1}i_{2}\ldots
i_{N_{c}}}^{B}g_{i_{1}}g_{i_{2}}\ldots g_{i_{k}}\,.
\end{equation}
This compact discrete notation allows us to simplify equations. By inertia we
will often call $\Phi_{B}(g)$ generating \emph{functional,} although the word
\emph{function} could be more appropriate.

One can ask whether our compact discrete notation is only a matter of language
or some serious problems will come in the case of field theoretical models.
The main problem is that our quantum-mechanical models are based on the
four-fermion interaction. Their field theory analog of these models is
represented by model of the Nambu--Jona-Lasinio (NJL) type. Formally our
quantum-mechanical equations can be generalized for case of the field theory.
However, there is a problem with the nonrenormalizability of the four-fermion
interaction. From the physical point of view, models of the NJL\ type can be
considered only as low-energy effective models which should be simply cut at
high momenta. But problem is that our treatment is based on the Hamiltonian
(or path integral) formalism which assumes the locality in time. Therefore
applications of our formalism to NJL and similar models would require
regularizations preserving the locality in time. Unfortunately most of
regularization local in time conflict with the Lorentz invariance.

It should be stressed that the methods developed in this paper can be applied
only to the large-$N_{c}$ models of QCD but not for QCD itself. Nevertheless
some special problems of large-$N_{c}$ QCD these methods still can be used::

1) the problem of heavy-quark baryons \cite{Witten-79-80},

2) the problem of the diagonalization of the matrix of anomalous dimensions
for the leading-twist baryon operators \cite{Pobylitsa-04}.

\subsection{Structure of the paper}

As was already mentioned, the paper consists of two parts. The first part
(Secs. \ref{Simplest-models-section}--\ref{Asymmetric-rotator-section}) is
devoted to models with the trivial vacuum (i.e. without antiquarks) whereas in
Secs. \ref{Nontrivial-vacuum-section}--\ref{Equivalence-section} more powerful
methods are developed for models imitating antiquarks in terms of the old
Dirac picture. The general class of models is discussed in Sec.
\ref{models-section}. In Sec. \ref{Simplest-models-section} we describe the
simplest models of the baryon wave function where the baryon appears as a
bound state of $N_{c}$ quarks (without quark-antiquark pairs). In Sec.
\ref{Large-Nc-section} we derive a nonlinear differential equation for the
functional $W(g)$ and show that this equation agrees with the standard Hartree
equation for large-$N_{c}$ systems. In Sec. \ref{Large-Nc-RPA-section} we turn
to the analysis of the preexponential functional $A(g)$ (\ref{Phi-A-W}). In
Sec. \ref{Classical-dynamics-section} we show how the problem of the
calculation of $W(g)$ can be reduced to the analysis of trajectories in some
effective classical dynamics. In Sec. \ref{Asymmetric-rotator-section} we
study a simple model where the functional $W(g)$ can be computed analytically.
In Sec. \ref{Nontrivial-vacuum-section} we turn to systems with the nontrivial
vacuum, using the Schr\"{o}dinger equation. In Sec.
\ref{Path-integral-section} we describe the path integral approach to the
large-$N_{c}$ models of the baryon wave function. In Sec.
\ref{Equivalence-section} we check the agreement of the results based on the
Schr\"{o}dinger equation and on the path integral approach.

\section{Models}

\label{models-section}

\setcounter{equation}{0} 

\subsection{Choice of models}

\label{Constraints-on-models-section}

As was explained in the introduction, our aim is to check conjectures about
the structure of the $1/N_{c}$ expansion for the baryon wave function which
were put forward in Refs. \cite{Pobylitsa-04}, \cite{Pobylitsa-05}. We want to
test these conjectures in simple large-$N_{c}$ models. Our choice of these
models is determined by the balance between the intention to preserve the main
features of the $1/N_{c}$ expansion in QCD and the possibility to solve the
model. As was explained in the introduction, the aspects of the
phenomenological relevance will play a secondary role (if any) in our choice
of models.

The minimal constraints on the models include two conditions: the model must have

1) quarks degrees of freedom,

2) $SU(N_{c})$ symmetry.

The quark degrees of freedom are understood here in a rather weak sense.
Neither full-fledged quantum field theory nor complete quantum mechanics is
needed for our aims. For example, we can work with models ignoring the space
motion of quarks where the dynamics of quarks is described only by color and
some other discrete quantum numbers (e.g. spin and flavor). For the
construction of the $1/N_{c}$ expansion we need the global $SU(N_{c})$
symmetry but there is no need in the local gauge invariance.

Now let us turn to the dynamics of the models. Most of our work will be done
in the Hamiltonian approach based on the analysis of the stationary
Schr\"{o}dinger equation. In principle, we can include both quarks and
antiquarks in our models using the old Dirac picture. In this case the
nontrivial physical vacuum will contain the ``Dirac'' sea of quarks above the
vacuum and we must solve the Schr\"{o}dinger equation both for the vacuum
$|0\rangle$ and for the baryon $|B\rangle$
\begin{align}
H|0\rangle &  =\mathcal{E}_{\mathrm{vac}}|0\rangle
\,\,,\label{Schroedinger-vacuum}\\
H|B\rangle &  =\mathcal{E}_{B}|B\rangle\,.\label{Schroedinger-baryon}
\end{align}

The dynamics of our models will be formulated in terms of the fermionic
annihilation and creation operators $a_{nc}$, $a_{nc}^{+}$
\begin{equation}
\{a_{nc},a_{n^{\prime}c^{\prime}}^{+}\}=\delta_{nn^{\prime}}\delta
_{cc^{\prime}}\,,\label{a-a-dagger-anticommutator}
\end{equation}
where
\begin{equation}
c=1,2,\ldots,N_{c}
\end{equation}
is the color index and $n=1,\ldots,K$ is some index which may have the meaning
of spin, isospin, etc.

We will be interested in the baryon wave function given by the transition
matrix element
\begin{equation}
\langle0|\prod\limits_{c=1}^{N_{c}}a_{m_{c}c}|B\rangle
\,.\label{bar-vac-transition}
\end{equation}
where $a_{mc}$ are quark annihilation operators. We can define the generating
functional for this wave function by analogy with Eq. (\ref{Phi-B-def-0})
\begin{equation}
\Phi_{B}(g)=\langle0|\prod\limits_{c=1}^{N_{c}}\left(  \sum\limits_{m}
g_{m}a_{mc}\right)  |B\rangle\,.\label{Phi-transition-def}
\end{equation}
At large $N_{c}$ we expect the exponential behavior (\ref{Phi-B-exponential})
\begin{equation}
\Phi_{B}(g)\overset{N_{c}\rightarrow\infty}{\sim}\exp\left[  N_{c}W(g)\right]
\,.\label{Phi-mod-exponential}
\end{equation}

\subsection{Hamiltonian}

We will work with models described by the Hamiltonian
\begin{equation}
H=\frac{1}{2N_{c}}\sum\limits_{n_{1}n_{2}n_{3}n_{4}}V_{n_{1}n_{2}n_{3}n_{4}
}\left(  \sum\limits_{c^{\prime}=1}^{N_{c}}a_{n_{1}c^{\prime}}^{+}
a_{n_{2}c^{\prime}}\right)  \left(  \sum\limits_{c=1}^{N_{c}}a_{n_{3}c}
^{+}a_{n_{4}c}\right)  \,+\sum\limits_{n_{1}n_{2}}L_{n_{1}n_{2}}\left(
\sum\limits_{c=1}^{N_{c}}a_{n_{1}c}^{+}a_{n_{2}c}\right) \label{H-general}
\end{equation}
assuming that
\begin{equation}
V_{n_{1}n_{2}n_{3}n_{4}}=V_{n_{3}n_{4}n_{1}n_{2}}\,.\label{V-sym}
\end{equation}
We we will be interested in color singlet states. The color singlet states can
contain $MN_{c}$ ``quarks'' where $M$ is an integer number. For any fixed $M$
we can study the problem of the lowest color-singlet state containing $MN_{c}$ fermions.

In our models of the baryon function we associate notation $M$ with the vacuum
whereas the baryon will correspond to $M+1$ so that one has a nonzero matrix
element (\ref{bar-vac-transition}) which will be interpreted as the baryon
wave function.

\subsection{Hartree equation}

\label{Hartree-general-section}

Below we will study the limit of large $N_{c}$ at fixed values of the
parameters $V_{n_{1}n_{2}n_{3}n_{4}}$ and $L_{n_{1}n_{2}}$ in the Hamiltonian
(\ref{H-general}). It is well known (see e.g. \cite{GLM-65,Pobylitsa-03}) that
in the leading order of the $1/N_{c}$ expansion the energy of the ground state
can be found by solving the Hartree equation
\begin{equation}
\sum\limits_{n_{2}}h_{n_{1}n_{2}}\phi_{n_{2}}^{s}=\varepsilon_{a}\phi_{n_{1}
}^{s}\,,\label{h-phi-Hartree-general}
\end{equation}
where the single particle Hamiltonian $h_{n_{1}n_{2}}$ is given by
\begin{equation}
h_{n_{1}n_{2}}=\sum\limits_{s=1}^{M}\sum\limits_{n_{3}n_{4}}V_{n_{1}n_{2}
n_{3}n_{4}}\phi_{n_{3}}^{s\ast}\phi_{n_{4}}^{s}+L_{n_{1}n_{2}}
\,\label{h-phi-Hartree-general-2}
\end{equation}
and the single particle eigenstates are normalized by the condition
\begin{equation}
\sum\limits_{n}\phi_{n}^{r}\phi_{n}^{s\ast}=\delta^{rs}\,.
\end{equation}
In the leading order of the $1/N_{c}$ expansion the energy of the
corresponding state is
\begin{equation}
\mathcal{E}=N_{c}E_{0}+O(N_{c}^{0})\,,
\end{equation}
where $E_{0}$ is given by
\begin{equation}
E_{0}=\frac{1}{2}\sum\limits_{r,s=1}^{M}\sum\limits_{n_{1}n_{2}n_{3}n_{4}
}V_{n_{1}n_{2}n_{3}n_{4}}\phi_{n_{1}}^{r\ast}\phi_{n_{2}}^{r}\phi_{n_{3}
}^{s\ast}\phi_{n_{4}}^{s}+\sum\limits_{s=1}^{M}\sum\limits_{n_{1}n_{2}
}L_{n_{1}n_{2}}\phi_{n_{1}}^{s\ast}\phi_{n_{2}}^{s}\,.\label{E-0-general}
\end{equation}

In the case of Hamiltonians with $L_{n_{1}n_{2}}=0$ the expression for $E_{0}
$ can be simplified using Eq. (\ref{h-phi-Hartree-general}):
\begin{equation}
E_{0}=\frac{1}{2}\sum\limits_{s=1}^{M}\varepsilon_{s}\quad(\mathrm{if}
\,L_{mn}=0)\,.
\end{equation}

Below we will see that the knowledge of the solutions of the Hartree equations
for the vacuum $|0\rangle$ and for the baryon $|B\rangle$ is not sufficient
for the calculation of the generating functional $W(g)$ appearing in Eq.
(\ref{Phi-mod-exponential}). One has to derive a special equation for $W(g)$.

\subsection{Time-dependent Hartree equation}

\label{TDHE}

The time-dependent Hartree equation (TDHE) appears in various problems of the
traditional many-body physics (where its Hartree--Fock modification is usually
discussed, see e.g. \cite{RS-80}). In the case of the Hamiltonian
(\ref{H-general}) TDHE has the form
\begin{equation}
i\frac{d}{dt}\phi_{n_{1}}^{s}(t)=\sum\limits_{n_{2}}h_{n_{1}n_{2}}
(t)\phi_{n_{2}}^{s}(t)\,,\label{TDHE-1}
\end{equation}
where
\begin{equation}
h_{n_{1}n_{2}}(t)=\sum\limits_{r=1}^{M}\sum\limits_{n_{3}n_{4}}V_{n_{1}
n_{2}n_{3}n_{4}}\phi_{n_{3}}^{r\ast}(t)\phi_{n_{4}}^{r}(t)+L_{n_{1}n_{2}
}\,.\label{TDHE-2}
\end{equation}
Although the context of the traditional problems of the many-body physics
where TDHE appears differs from our problem of the calculation of the
functional $W(g)$, we will see in Sec. \ref{TDHE-interpretation-section} that
the functional $W(g)$ allows for an interesting representation in terms of the
solutions of the TDHE (\ref{TDHE-1}), (\ref{TDHE-2}). The possibility of the
Hamiltonian interpretation of the Hartree--Fock equations was discussed in
Ref. \cite{KK-76}. This Hamiltonian interpretation plays an important role in
our analysis.

\section{Models with the trivial vacuum}

\label{Simplest-models-section}

\setcounter{equation}{0} 

\subsection{Simplest models}

As was explained in Sec. \ref{Constraints-on-models-section}, we want to study
the baryon wave function in models described by the Hamiltonian
(\ref{H-general}) using the states containing $N_{c}M$ and $N_{c}(M+1)$
fermions for the vacuum and baryon, respectively.

We want to start from the simplest case $M=0$ (the generalization to arbitrary
$M$ will be considered in 
Secs. \ref{Nontrivial-vacuum-section}--\ref{Equivalence-section}).
In these $M=0$ models the physical vacuum
coincides with the bare one, and the baryon consists of $N_{c}$ quarks only.
We will refer to the $M=0$ models as \emph{models with the trivial vacuum.}
The main advantage of these models is that the solution $\Psi_{\mathrm{bar}}$
of the Schr\"{o}dinger equation (\ref{Schroedinger-baryon}) directly gives us
the baryon wave function (\ref{bar-vac-transition}).

Our first aim is to define the generating functional $\Phi_{B}(g)$ for the
baryons in these models and to rewrite the stationary Schr\"{o}dinger equation
in terms of $\Phi_{B}(g)$. This can be easily done using the well known
holomorphic representation (also known as boson representation in the context
of the many-body physics \cite{RS-80}).

In our simple models ``baryons'' are described by states containing $N_{c}$
quarks:
\begin{align}
|\psi^{B}\rangle &  =\sum\limits_{n_{1}n_{2}\ldots n_{N_{c}}}\psi_{n_{1}\ldots
n_{N_{c}}}^{B}a_{n_{N_{c}}N_{c}}^{+}\ldots a_{n_{1}1}^{+}|0\rangle\nonumber\\
&  =\frac{1}{N_{c}!}\sum\limits_{n_{1}n_{2}\ldots n_{N_{c}}}\sum
\limits_{c_{1}c_{2}\ldots c_{N_{c}}}\varepsilon_{c_{1}\ldots c_{N_{c}}}
\psi_{n_{1}\ldots n_{N_{c}}}^{B}a_{n_{N_{c}}c_{N_{c}}}^{+}\ldots a_{n_{1}
c_{1}}^{+}|0\rangle\,.\label{psi-B-decompos}
\end{align}
Here $\psi_{n_{1}\ldots n_{N_{c}}}$ are coefficients. Due to the antisymmetry
of fermionic operators coefficients $\psi_{n_{1}\ldots n_{N_{c}}}$ are
completely symmetric in all indices $n_{1}\ldots n_{N_{c}}$.

\subsection{Generating function for baryon wave function}

In our simple model the baryon wave function (\ref{bar-vac-transition})
coincides with the coefficients $\psi_{n_{1}\ldots n_{N_{c}}}^{B}$ of the
decomposition (\ref{psi-B-decompos}):
\begin{equation}
\psi_{n_{1}\ldots n_{N_{c}}}^{B}=\langle0|a_{n_{1}1}^{+}\ldots a_{n_{N_{c}
}c_{N_{c}}}^{+}|\psi^{B}\rangle\,.
\end{equation}

Now we can construct the model analog of the generating functional
(\ref{Phi-B-def-0}) describing the baryon wave function:
\begin{equation}
\Phi_{B}(g)=\sum\limits_{n_{1}n_{2}\ldots n_{N_{c}}}\psi_{n_{1}\ldots
n_{N_{c}}}^{B}g_{n_{1}}\ldots g_{n_{N_{c}}}=\sum\limits_{n_{1}n_{2}\ldots
n_{N_{c}}}\langle0|a_{n_{1}1}^{+}\ldots a_{n_{N_{c}}c_{N_{c}}}^{+}|\psi
^{B}\rangle g_{n_{1}}\ldots g_{n_{N_{c}}}\,.\label{Phi-g-general-def}
\end{equation}
Note that function $\Phi_{B}(g)$ (\ref{Phi-g-general-def}) is a homogeneous
polynomial of degree $N_{c}$:
\begin{equation}
\Phi_{B}(\lambda g)=\lambda^{N_{c}}\Phi_{B}(g)\,.\label{Phi-B-Homogeneous}
\end{equation}

The correspondence between the states $|\psi^{B}\rangle$ and functions
$\Phi_{B}(g)$ is useful not only for the description of the eigenstates of the
Hamiltonian (\ref{H-general}) but for any color singlet states containing
$N_{c}$ quarks. Let us consider operator
\begin{equation}
T_{mn}=\sum\limits_{c=1}^{N_{c}}a_{mc}^{+}a_{nc}\label{T-via-a-0}
\end{equation}
on the state $|\psi\rangle$. In terms of the $\Phi_{B}(g)$ representation for
states $|\psi^{B}\rangle$
\begin{equation}
|\psi^{B}\rangle\rightarrow\Phi_{B}(g)
\end{equation}
operator $T_{mn}$ takes the form
\begin{equation}
T_{mn}=\sum\limits_{c=1}^{N_{c}}a_{mc}^{+}a_{nc}\quad\rightarrow\quad
g_{m}\frac{\partial}{\partial g_{n}}\,.\label{a-h-correspondence}
\end{equation}

\subsection{Schr\"{o}dinger equation for $\Phi_{B}(g)$}

Using the $g$ representation (\ref{a-h-correspondence}) for the operators
$T_{mn}$, we can rewrite the Hamiltonian (\ref{H-general}) in the form
\begin{equation}
H=\frac{1}{2N_{c}}\sum\limits_{n_{1}n_{2}n_{3}n_{4}}V_{n_{1}n_{2}n_{3}n_{4}
}\left(  g_{n_{1}}\frac{\partial}{\partial g_{n_{2}}}\right)  \left(
g_{n_{3}}\frac{\partial}{\partial g_{n_{4}}}\right)  \,+\sum\limits_{n_{1}
n_{2}}L_{n_{1}n_{2}}\left(  g_{n_{1}}\frac{\partial}{\partial g_{n_{2}}
}\right)  \,.\label{H-g}
\end{equation}
Then the Schr\"{o}dinger equation
\begin{equation}
H|\psi^{B}\rangle=\mathcal{E}_{B}|\psi^{B}\rangle
\end{equation}
becomes
\begin{gather}
\frac{1}{2N_{c}}\sum\limits_{n_{1}n_{2}n_{3}n_{4}}V_{n_{1}n_{2}n_{3}n_{4}
}\left(  g_{n_{1}}\frac{\partial}{\partial g_{n_{2}}}\right)  \left(
g_{n_{3}}\frac{\partial}{\partial g_{n_{4}}}\right)  \Phi_{B}(g)\nonumber\\
+\sum\limits_{n_{1}n_{2}}L_{n_{1}n_{2}}\left(  g_{n_{1}}\frac{\partial
}{\partial g_{n_{2}}}\right)  =\mathcal{E}_{B}\Phi_{B}
(g)\,.\label{Schroedinger-g}
\end{gather}

\section{Large-$N_{c}$ limit}

\label{Large-Nc-section}

\setcounter{equation}{0} 

\subsection{Two approaches to the $1/N_{c}$ expansion}

In this section we want to study the large-$N_{c}$ limit in the simplest
models of baryons introduced in Sec. \ref{Simplest-models-section}. As was
mentioned in the introduction, one can use two approaches to the large-$N_{c}
$ limit in fermion systems:

1) approach based on the single-particle \emph{density matrix} obeying the
Hartree equation,

2) approach based on the baryon \emph{wave function} described in terms of the
generating functional $\Phi_{B}(g)$.

In principle these two methods are equivalent. For example, they lead to the
same $1/N_{c}$ expansion for the spectrum of states. The second method based
on the functional $\Phi_{B}(g)$ provides more information. However, one has to
pay some price for the additional information: the large-$N_{c}$ equations
derived for the functional $\Phi_{B}(g)$ are more complicated than the Hartree
equation for the density matrix.

Below we consider both approaches and demonstrate their equivalence.

\subsection{Large $N_{c}$ limit for functionals $\Phi_{B}(g)$}

\label{Phi-B-large-Nc-section}

We want to study the limit of $\Phi_{B}(g)$ at large $N_{c}$. According to Eq.
(\ref{Phi-B-Homogeneous}) we have
\begin{align}
W(\lambda g)  &  =W(g)+\ln\lambda\,,\label{W-rescaling}\\
A_{B}(\lambda g)  &  =A_{B}(g)\,,\label{A-rescaling}
\end{align}
so that
\begin{align}
\sum\limits_{n}g_{n}\frac{\partial}{\partial g_{n}}W(g)  &
=1\,,\label{W-Euler}\\
\sum\limits_{n}g_{n}\frac{\partial}{\partial g_{n}}A_{B}(g)  &  =0\,.
\end{align}

Let us insert Eq. (\ref{Phi-A-W}) into the Schr\"{o}dinger equation
(\ref{Schroedinger-g}). Taking into account the $1/N_{c}$ expansion for the
energy
\begin{equation}
\mathcal{E}_{B}=N_{c}E_{0}+\Delta E_{B}+O(N_{c}^{-1}
)\,,\label{E-B-N-c-expansion}
\end{equation}
we obtain in the leading order
\begin{equation}
\frac{1}{2}\sum\limits_{n_{1}n_{2}n_{3}n_{4}}V_{n_{1}n_{2}n_{3}n_{4}}\left[
g_{n_{1}}\frac{\partial W(g)}{\partial g_{n_{2}}}\right]  \left[  g_{n_{3}
}\frac{\partial W(g)}{\partial g_{n_{4}}}\right]  +\sum\limits_{n_{1}n_{2}
}L_{n_{1}n_{2}}\left[  g_{n_{1}}\frac{\partial W(g)}{\partial g_{n_{2}}
}\right]  =E_{0}\,.\label{W-WKB}
\end{equation}
In the next order of the $1/N_{c}$ expansion we find
\begin{gather}
\sum\limits_{n_{1}n_{2}n_{3}n_{4}}V_{n_{1}n_{2}n_{3}n_{4}}\left[  g_{n_{1}
}\frac{\partial W(g)}{\partial g_{n_{2}}}\right]  \left[  g_{n_{3}
}\frac{\partial\ln A_{B}(g)}{\partial g_{n_{4}}}\right]  +\sum\limits_{n_{1}
n_{2}}L_{n_{1}n_{2}}\left[  g_{n_{1}}\frac{\partial\ln A_{B}(g)}{\partial
g_{n_{2}}}\right] \nonumber\\
+\frac{1}{2}\sum\limits_{n_{1}n_{2}n_{3}n_{4}}V_{n_{1}n_{2}n_{3}n_{4}}\left[
g_{n_{1}}\frac{\partial}{\partial g_{n_{2}}}\left(  g_{n_{3}}\frac{\partial
W(g)}{\partial g_{n_{4}}}\right)  \right]  =\Delta E_{B}\,.\label{A-WKB-0}
\end{gather}

\subsection{Hartree equation}

\label{Hartree-equation-1-section}

As was explained in Sec. \ref{Hartree-general-section}, the energy of the
lowest baryon state can be described in the leading order of the $1/N_{c}$
expansion by the Hartree equations (\ref{h-phi-Hartree-general}),
(\ref{h-phi-Hartree-general-2}). Our case corresponds to $M=1$ in these
equations:
\begin{equation}
\sum\limits_{n_{2}}\left(  \sum\limits_{n_{3}n_{4}}V_{n_{1}n_{2}n_{3}n_{4}
}\phi_{n_{3}}^{\ast}\phi_{n_{4}}+L_{n_{1}n_{2}}\right)  \phi_{n_{2}
}=\varepsilon_{0}\phi_{n_{1}}\,,\label{Hartree-general}
\end{equation}

\begin{equation}
\sum\limits_{n}\phi_{n}^{\ast}\phi_{n}=1\,\,,\label{phi-normalized}
\end{equation}
where $\phi_{n}$ is the single-particle wave function of the level occupied
with $N_{c}$ quarks.

The leading order contribution $N_{c}E_{0}$ to the energy $\mathcal{E}_{B}$
(\ref{E-B-N-c-expansion}) is given by Eq. (\ref{E-0-general})
\begin{equation}
E_{0}=\frac{1}{2}\sum\limits_{n_{1}n_{2}n_{3}n_{4}}V_{n_{1}n_{2}n_{3}n_{4}
}\phi_{n_{1}}^{\ast}\phi_{n_{2}}\phi_{n_{3}}^{\ast}\phi_{n_{4}}+\sum
\limits_{n_{1}n_{2}}L_{n_{1}n_{2}}\phi_{n_{1}}^{\ast}\phi_{n_{2}
}\,.\label{E-Hartree-special}
\end{equation}

\subsection{Agreement between the Hartree equation and the equation for
$W(\phi)$}

\label{Hartree-W-agreement-section}

Now we can compare equation (\ref{W-WKB}) for $W(g)$ with the expression
(\ref{E-Hartree-special}) for the Hartree energy $E_{0}$. We see that the two
expressions agree if we take
\begin{equation}
\phi_{n}=\frac{\partial W(g)}{\partial g_{n}}\,,\quad\phi_{n}^{\ast}
=g_{n}\,.\label{Hartree-point-eq}
\end{equation}
In other words, if we know function $W(g)$, then the Hartree solution
$\phi_{n}$ can be found by solving the equation
\begin{equation}
g_{n}^{\ast}=\frac{\partial W(g)}{\partial g_{n}}\quad\Longrightarrow
\,\phi_{n}=g_{n}^{\ast}\,.\label{W-Hartree}
\end{equation}
Note that the normalization condition (\ref{phi-normalized}) is automatically satisfied
\begin{equation}
\sum\limits_{n}\phi_{n}^{\ast}\phi_{n}=\sum\limits_{n}g_{n}\frac{\partial
W(g)}{\partial g_{n}}=1
\end{equation}
due to the property (\ref{W-Euler}) of the function $W(g)$.

\subsection{Normalization integral}

Let us consider two states (\ref{psi-B-decompos})
\begin{equation}
|\psi^{(k)}\rangle=\sum\limits_{n_{1}n_{2}\ldots n_{N_{c}}}\psi_{n_{1}\ldots
n_{N_{c}}}^{(k)}a_{n_{N_{c}}N_{c}}^{+}\ldots a_{n_{1}1}^{+}|0\rangle
\quad(k=1,2)
\end{equation}
described by functions (\ref{Phi-g-general-def})
\begin{equation}
\Phi^{(k)}(g)=\sum\limits_{n_{1}n_{2}\ldots n_{N_{c}}}\psi_{n_{1}\ldots
n_{N_{c}}}^{(k)}g_{n_{1}}\ldots g_{n_{N_{c}}}\,.
\end{equation}

Then
\begin{align}
\langle\psi^{(1)}|\psi^{(2)}\rangle &  =\sum\limits_{i_{1}i_{2}\ldots
i_{N_{c}}}F_{i_{1}i_{2}\ldots i_{N_{c}}}^{(1)\ast}F_{i_{1}i_{2}\ldots
i_{N_{c}}}^{(2)}=\left(  \frac{1}{N_{c}!}\right)  ^{2}\left(  \frac{\partial
}{\partial g_{i}}\frac{\partial}{\partial g_{i}^{\ast}}\right)  ^{N_{c}
}\left[  \Phi^{(1)}(g)\right]  ^{\ast}\left[  \Phi^{(2)}(g)\right] \nonumber\\
&  =\frac{N_{c}^{N_{c}}}{N_{c}!}\left.  \exp\left(  \frac{1}{N_{c}
}\frac{\partial}{\partial g_{i}}\frac{\partial}{\partial g_{i}^{\ast}}\right)
\left[  \Phi^{(1)}(g)\right]  ^{\ast}\left[  \Phi^{(2)}(g)\right]  \right|
_{g=g^{\ast}=0}\nonumber\\
&  =\frac{N_{c}^{N_{c}}}{N_{c}!}\frac{\int dgdg^{\ast}\exp\left(  -N_{c}
\sum_{i}g_{i}g_{i}^{\ast}\right)  \left[  \Phi^{(1)}(g)\right]  ^{\ast}\left[
\Phi^{(2)}(g)\right]  }{\int dgdg^{\ast}\exp\left(  -N_{c}\sum_{i}g_{i}
g_{i}^{\ast}\right)  }\,.
\end{align}
At large $N_{c}$ we can use the asymptotic expression (\ref{Phi-A-W}) for
$\Phi^{(k)}(g)$. As a result, we find
\begin{equation}
\langle\psi^{(1)}|\psi^{(2)}\rangle\rightarrow\frac{e^{N_{c}}N_{c}^{\nu
_{1}+\nu_{2}}}{\sqrt{2\pi N_{c}}}\frac{\int dgdg^{\ast}\exp\left\{
N_{c}\left\{  -g_{i}g_{i}^{\ast}+\left[  W(g)\right]  ^{\ast}+W(g)\right\}
\right\}  \,\left[  A^{(1)}(g)\right]  ^{\ast}A^{(2)}(g)}{\int dgdg^{\ast}
\exp\left(  -N_{c}\sum_{i}g_{i}g_{i}^{\ast}\right)  \,.}\,.
\end{equation}
This integral can be computed using the saddle point method. The saddle point
equation is nothing else Eq. (\ref{W-Hartree}) with the solution
$g_{n}=\phi_{n}^{\ast}$ given by the Hartree equation (\ref{Hartree-general}). Thus
we find with the exponential accuracy
\begin{equation}
\langle\psi^{(1)}|\psi^{(2)}\rangle\sim\exp\left\{  -N_{c}\left\{  \left[
W(\phi^{\ast})\right]  ^{\ast}+W(\phi^{\ast})\right\}  \right\}
\,.\label{psi-1-psi-2-overlap}
\end{equation}
We cannot use the saddle point method beyond this exponential accuracy because
of the noncommutativity of the limits
\begin{equation}
g_{n}\rightarrow\phi_{n}^{\ast}\,,\quad N_{c}\rightarrow\infty
\end{equation}
which is discussed in Sec. \ref{Two-Nc-limits-section}.

Taking a normalizable state
\begin{align}
\psi^{(1)}  &  =\psi^{(2)}\equiv\psi\,,\\
\langle\psi|\psi\rangle &  =1\,,
\end{align}
we conclude from (\ref{psi-1-psi-2-overlap}) that
\begin{equation}
\mathrm{Re}\,W(\phi^{\ast})=0\,.\label{Re-W-phi-Hartree}
\end{equation}

\section{Large $N_{c}$ limit and RPA equations}

\label{Large-Nc-RPA-section}

\setcounter{equation}{0} 

\subsection{Beyond the leading order of the $1/N_{c}$ expansion}

In the previous section we have established a connection between the
traditional description of the large-$N_{c}$ systems in terms of the Hartree
equation (\ref{Hartree-general}) and our equation (\ref{W-WKB}) for the
functional $W(g)$. As is well known, the Hartree equation determines only the
leading $O(N_{c})$ part of the energy of the states. If one is interested in
the $O(N_{c}^{0})$ corrections, then one has to solve the random phase
approximation (RPA) equations \cite{GLM-65,Pobylitsa-03}.

On the other hand, in the problem of the baryon wave function the analysis of
the next-leading-order of the $1/N_{c}$ expansion is based on Eq.
(\ref{A-WKB-0}) for the functional $A_{B}(g)$. Therefore one can expect that
there must be some connection between the RPA equations and Eq. (\ref{A-WKB-0}).
From our experience with the leading order equations [the Hartree equation
(\ref{Hartree-general}) and Eq. (\ref{W-WKB}) for $W(g)$] we know that this
connection may be rather nontrivial. Indeed, we will see that the RPA equation
really appears in the analysis of the $1/N_{c}$ expansion for the basic
functional $\Phi_{B}(g)$ but in a modified version of the large-$N_{c}$ limit
when the argument $g$ of the functional $\Phi_{B}(g)$ changes with $N_{c}$ in
a special way. This new version of the large-$N_{c}$ limit is discussed in
Sec. \ref{Two-Nc-limits-section}. In Sec. \ref{RPA-equation-subsection} we
turn to the derivation of the RPA equations corresponding to the modified
large-$N_{c}$ limit.

\subsection{Two versions of the large-$N_{c}$ limit}

\label{Two-Nc-limits-section}

Eq. (\ref{Phi-A-W}) describes the asymptotic behavior of $\Phi(g)$ in the
large-$N_{c}$ limit when $g$ is fixed.
\begin{equation}
g=\mathrm{const},\,N_{c}\rightarrow\infty\,.
\end{equation}
But we can also study another limit
\begin{equation}
N_{c}\rightarrow\infty\,,\quad g_{n}=\phi_{n}+\frac{1}{\sqrt{N_{c}}}\Delta
_{n}\,,\quad\Delta_{n}=\mathrm{const}\,.\label{Nc-limit-2}
\end{equation}
when $g_{n}$ approaches the Hartree solution $\phi_{n}$ in the limit
$N_{c}\rightarrow\infty$. In addition let us impose the condition on
$\Delta_{n}$
\begin{equation}
\sum\limits_{n}\phi_{n}^{\ast}\Delta_{n}=0\,.\label{phi-Delta-ortho}
\end{equation}
In the limit (\ref{Nc-limit-2}) we have the asymptotic behavior
\begin{equation}
\Phi_{B}\left(  \phi_{n}+\frac{1}{\sqrt{N_{c}}}\Delta_{n}\right)
\overset{N_{c}\rightarrow\infty}{\longrightarrow}N_{c}^{\nu_{B}}C_{B}
(\Delta_{n})\exp\left[  N_{c}W(\phi)\right] \label{Phi-B-large-Nc-limit-2}
\end{equation}
different from Eq. (\ref{Phi-A-W}). Taking $\Delta_{n}=0$ and comparing the
asymptotic expression (\ref{Phi-B-large-Nc-limit-2}) with Eq. (\ref{Phi-A-W}),
we obtain
\begin{equation}
C_{B}(0)=A_{B}(\phi)\,.
\end{equation}
Replacing $\Delta_{n}\rightarrow\lambda^{-1}\Delta_{n}$ in Eq.
(\ref{Phi-B-large-Nc-limit-2}) and using relation (\ref{Phi-B-Homogeneous}),
we find
\begin{gather}
\Phi_{B}\left(  \lambda\phi_{n}+\frac{1}{\sqrt{N_{c}}}\Delta_{n}\right)
=\lambda^{N_{c}}\Phi_{B}\left(  \phi_{n}+\frac{1}{\sqrt{N_{c}}}\lambda
^{-1}\Delta_{n}\right) \nonumber\\
\overset{N_{c}\rightarrow\infty}{\longrightarrow}N_{c}^{\nu_{B}}C_{B}
(\lambda^{-1}\Delta_{n})\exp\left\{  N_{c}\left[  W(\phi)+\ln\lambda\right]
\right\}  \,.\label{Phi-B-large-Nc-limit-2b}
\end{gather}

\subsection{Special choice of basis}

\label{Special-basis-section}

It is convenient to choose the basis diagonalizing the Hartree Hamiltonian
such that
\begin{equation}
\phi_{n}^{s}=\delta_{n}^{s}\,.\label{basis-choice}
\end{equation}
In models with one occupied level we will label this level with the index
$s=0$ and use the short notation
\begin{equation}
\phi_{n}\equiv\phi_{n}^{0}=\delta_{n}^{0}\,.
\end{equation}

Then the Hartree Hamiltonian (\ref{h-phi-Hartree-general-2}) becomes
\begin{equation}
h_{mn}=V_{mn00}+L_{mn}
\end{equation}
and the Hartree equation (\ref{Hartree-general}) takes the form
\begin{equation}
V_{mn00}+L_{mn}=\varepsilon_{m}\delta_{mn}\,.\label{Hartree-special-basis}
\end{equation}
In particular,
\begin{equation}
\varepsilon_{0}=V_{0000}+L_{00}\,.\label{Hartree-special-basis-2}
\end{equation}

The Hartree energy is
\begin{equation}
E_{0}=\frac{1}{2}V_{0000}+L_{00}\,.\label{E0-Hartree}
\end{equation}
We introduce a special notation for the components $g_{n}$ with $n>0$:
\begin{equation}
\tilde{g}_{n}\overset{n>0}{=}g_{n}\,,\quad\tilde{g}_{0}=0
\end{equation}
so that
\begin{equation}
g_{n}=g_{0}\phi_{n}+\tilde{g}_{n}\,,
\end{equation}
\begin{equation}
\Phi_{B}\left(  g\right)  =\Phi_{B}\left(  g_{0},\tilde{g}\right)  \,.
\end{equation}
Now we find from Eq. (\ref{Phi-B-large-Nc-limit-2b})
\begin{equation}
\left[  \Phi_{B}\left(  g\right)  \right]  _{\tilde{g}_{n}=\Delta_{n}
/\sqrt{N_{c}}}=\Phi_{B}\left(  g_{0}\phi_{n}+\frac{\Delta_{n}}{\sqrt{N_{c}}
}\right)  \overset{N_{c}\rightarrow\infty,\Delta_{n}=\mathrm{const}
}{\longrightarrow}N_{c}^{\nu_{B}}C_{B}(g_{0}^{-1}\Delta_{n})\exp\left\{
N_{c}\left[  W(\phi)+\ln g_{0}\right]  \right\}  \,.
\end{equation}
Therefore
\begin{gather}
g_{0}\frac{\partial}{\partial g_{0}}\left[  \Phi_{B}\left(  g\right)  \right]
_{\tilde{g}_{n}=\Delta_{n}/\sqrt{N_{c}}}\overset{N_{c}\rightarrow\infty
,\Delta_{n}=\mathrm{const}}{\longrightarrow}g_{0}\frac{\partial}{\partial
g_{0}}N_{c}^{\nu_{B}}C_{B}(g_{0}^{-1}\Delta_{n})\exp\left\{  N_{c}\left[
W(\phi)+\ln g_{0}\right]  \right\} \nonumber\\
=\left\{  N_{c}^{\nu_{B}}\exp\left\{  N_{c}\left[  W(\phi)+\ln g_{0}\right]
\right\}  \right\}  \left.  N_{c}\left(  1-\frac{1}{N_{c}}\sum\limits_{n>0}
\tilde{\Delta}_{n}\frac{\partial}{\partial\tilde{\Delta}_{n}}\right)
C_{B}(\tilde{\Delta}_{n})\right|  _{\tilde{\Delta}_{n}=g_{0}^{-1}\Delta_{n}
}\,.\label{g0-dg0-a}
\end{gather}
On the LHS the differential operator $g_{0}\partial/\partial g_{0}$ acts on
$\Phi_{B}\left(  g\right)  $ whereas on the RHS we have the corresponding
operator acting on $C_{B}(\tilde{\Delta}_{n})$. Equation (\ref{g0-dg0-a})
means that the following $\tilde{\Delta}$ representation is valid for the
operator $g_{0}\partial/\partial g_{0}$:
\begin{equation}
g_{0}\frac{\partial}{\partial g_{0}}\rightarrow N_{c}\left(  1-\frac{1}{N_{c}
}\sum\limits_{n>0}\tilde{\Delta}_{n}\frac{\partial}{\partial\tilde{\Delta}
_{n}}\right)  \,.\label{g0-dg0}
\end{equation}
Similarly we find
\begin{align}
\tilde{g}_{m}\frac{\partial}{\partial g_{0}}  &  \rightarrow\sqrt{N_{c}}
\tilde{\Delta}_{m}\left(  1-\frac{1}{N_{c}}\sum\limits_{n>0}\tilde{\Delta}
_{n}\frac{\partial}{\partial\tilde{\Delta}_{n}}\right)  \,,\\
g_{0}\frac{\partial}{\partial\tilde{g}_{m}}  &  \rightarrow\sqrt{N_{c}
}\frac{\partial}{\partial\tilde{\Delta}_{m}}\,,\\
g_{n}\frac{\partial}{\partial\tilde{g}_{m}}  &  \rightarrow\tilde{\Delta}
_{n}\frac{\partial}{\partial\tilde{\Delta}_{m}}\,.\label{gn-dgm}
\end{align}

\subsection{RPA equations from the large-$N_{c}$ limit}

\label{RPA-equation-subsection}

Now we can apply the results (\ref{g0-dg0})--(\ref{gn-dgm}) to the
Schr\"{o}dinger equation (\ref{Schroedinger-g}). The terms of order $N_{c}$
generated by $g_{0}\partial/\partial g_{0}$ give the Hartree energy
(\ref{E0-Hartree}). The terms of order $\sqrt{N_{c}}$ cancel because of the
equation (\ref{Hartree-special-basis}). The terms of order $N_{c}^{0}$ lead to
the equation
\begin{gather}
\sum\limits_{m,n>0}\left[  V_{00nm}\left(  \tilde{\Delta}_{n}\frac{\partial
}{\partial\tilde{\Delta}_{m}}\right)  +\frac{1}{2}V_{m00n}\left(
\tilde{\Delta}_{m}\frac{\partial}{\partial\tilde{\Delta}_{n}}+\frac{\partial
}{\partial\tilde{\Delta}_{n}}\tilde{\Delta}_{m}\right)  +\frac{1}{2}
V_{0m0n}\frac{\partial}{\partial\tilde{\Delta}_{m}}\frac{\partial}
{\partial\tilde{\Delta}_{m}}\right. \nonumber\\
\left.  +\frac{1}{2}V_{m0n0}\tilde{\Delta}_{m}\tilde{\Delta}_{n}+L_{mn}
\tilde{\Delta}_{m}\frac{\partial}{\partial\tilde{\Delta}_{n}}-\left(
V_{0000}+L_{00}\right)  \delta_{mn}\tilde{\Delta}_{m}\frac{\partial}
{\partial\tilde{\Delta}_{n}}\right]  C_{B}(\tilde{\Delta})\nonumber\\
=\Delta E_{B}C_{B}(\tilde{\Delta})\,.
\end{gather}
This equation can be simplified using Eq. (\ref{Hartree-special-basis-2}):
\begin{gather}
\sum\limits_{m,n>0}\left[  V_{00nm}\left(  \tilde{\Delta}_{n}\frac{\partial
}{\partial\tilde{\Delta}_{m}}\right)  +\frac{1}{2}V_{m00n}\left(
\tilde{\Delta}_{m}\frac{\partial}{\partial\tilde{\Delta}_{n}}+\frac{\partial
}{\partial\tilde{\Delta}_{n}}\tilde{\Delta}_{m}\right)  +\frac{1}{2}
V_{0m0n}\frac{\partial}{\partial\tilde{\Delta}_{m}}\frac{\partial}
{\partial\tilde{\Delta}_{m}}\right. \nonumber\\
\left.  +\frac{1}{2}V_{m0n0}\tilde{\Delta}_{m}\tilde{\Delta}_{n}+L_{mn}
\tilde{\Delta}_{m}\frac{\partial}{\partial\tilde{\Delta}_{n}}-\varepsilon
_{0}\delta_{mn}\tilde{\Delta}_{m}\frac{\partial}{\partial\tilde{\Delta}_{n}
}\right]  C_{B}(\tilde{\Delta})=\Delta E_{B}C_{B}(\tilde{\Delta})\,.
\end{gather}

Let us introduce operators
\begin{align}
b_{m} &  =\frac{\partial}{\partial\tilde{\Delta}_{m}}\,,\label{a-Delta}\\
b_{m}^{+} &  =\tilde{\Delta}_{m}\,\,,\label{a-dagger-Delta}
\end{align}
\begin{equation}
\lbrack b_{m},b_{n}^{+}]=\delta_{mn}\,.
\end{equation}
Then
\begin{gather}
\sum\limits_{m,n>0}\left[  V_{00nm}b_{n}^{+}b_{m}+\frac{1}{2}V_{m00n}\left(
b_{m}^{+}b_{n}+b_{n}b_{m}^{+}\right)  +\frac{1}{2}V_{0m0n}b_{m}^{+}
b_{n}\right. \nonumber\\
\left.  +\frac{1}{2}V_{m0n0}b_{m}^{+}b_{n}^{+}+L_{mn}b_{m}^{+}b_{n}
-\varepsilon_{0}\delta_{mn}b_{m}^{+}b_{n}\right]  C_{B}=\Delta E_{B}C_{B}\,.
\end{gather}
Note that the $b_{n}$ vacuum $|0\rangle$
\begin{equation}
b_{m}|0\rangle=0\label{a-vac}
\end{equation}
corresponds in the $\tilde{\Delta}$ representation (\ref{a-Delta}),
(\ref{a-dagger-Delta}) to the wave function
\begin{equation}
\Psi(\tilde{\Delta})\equiv1\,.
\end{equation}
Thus we have the Hamiltonian
\begin{align}
H_{\mathrm{RPA}} &  \equiv\sum\limits_{m,n>0}\left[  V_{00nm}b_{n}^{+}
b_{m}+\frac{1}{2}V_{m00n}\left(  b_{m}^{+}b_{n}+b_{n}b_{m}^{+}\right)
+\frac{1}{2}V_{0m0n}b_{m}b_{n}\right. \nonumber\\
&  \left.  +\frac{1}{2}V_{m0n0}b_{m}^{+}b_{n}^{+}+L_{mn}b_{m}^{+}
b_{n}-\varepsilon_{0}\delta_{mn}b_{m}^{+}b_{n}\right] \nonumber\\
&  =\frac{1}{2}\sum\limits_{m,n>0}\left(
\begin{array}
[c]{c}
b_{m}\\
b_{m}^{+}
\end{array}
\right)  ^{T}\left(
\begin{array}
[c]{cc}
V_{0m0n} & V_{n00m}+V_{00nm}+L_{nm}-\varepsilon_{0}\delta_{mn}\\
V_{m00n}+V_{00mn}+L_{mn}-\varepsilon_{0}\delta_{mn} & V_{m0n0}
\end{array}
\right)  \left(
\begin{array}
[c]{c}
b_{n}\\
b_{n}^{+}
\end{array}
\right) \nonumber\\
&  -\frac{1}{2}\sum\limits_{m>0}\left[  \left(  V_{00mm}+L_{mm}\right)
-\varepsilon_{0}\right]  \,.
\end{align}
According to Eqs. (\ref{V-sym}) and (\ref{Hartree-special-basis}) we have
\begin{equation}
V_{00nm}+L_{nm}=\varepsilon_{m}\delta_{nm}\,,
\end{equation}
\begin{equation}
\sum\limits_{m>0}\left(  V_{00mm}+L_{mm}\right)  =\sum\limits_{m>0}
\varepsilon_{m}\,.
\end{equation}
Now we find
\begin{equation}
H_{\mathrm{RPA}}=\frac{1}{2}\sum\limits_{m,n>0}\left(
\begin{array}
[c]{c}
b_{m}\\
b_{m}^{+}
\end{array}
\right)  ^{T}\mathcal{R}\left(
\begin{array}
[c]{c}
b_{n}\\
b_{n}^{+}
\end{array}
\right)  -\frac{1}{2}\sum\limits_{m>0}\left(  \varepsilon_{m}-\varepsilon
_{0}\right)  \,,\label{H-RPA-def}
\end{equation}
where
\begin{equation}
\mathcal{R}=\left(
\begin{array}
[c]{cc}
V_{0m0n} & V_{n00m}+\left(  \varepsilon_{m}-\varepsilon_{0}\right)
\delta_{mn}\\
V_{m00n}+\left(  \varepsilon_{m}-\varepsilon_{0}\right)  \delta_{mn} &
V_{m0n0}
\end{array}
\right)  \,.\label{RPA-matrix-def}
\end{equation}
This is nothing else but the large-$N_{c}$ version of the RPA Hamiltonian.

\subsection{Diagonalization of the RPA Hamiltonian}

We can diagonalize the RPA Hamiltonian (\ref{H-RPA-def}) using the Bogolyubov
transformation
\begin{align}
Ub_{m}U^{-1} &  =\sum\limits_{n>0}\left(  \alpha_{mn}b_{n}+\beta_{mn}b_{n}
^{+}\right)  \,,\label{U-alpha-beta}\\
Ub_{m}^{+}U^{-1} &  =\sum\limits_{n>0}\left(  \beta_{mn}^{\ast}b_{n}
+\alpha_{mn}^{\ast}b_{n}^{+}\right)  \,,\label{U-alpha-beta-2}
\end{align}
\begin{align}
U^{-1}b_{m}U &  =\sum\limits_{n>0}\left(  \alpha_{nm}^{\ast}b_{n}-\beta
_{nm}b_{n}^{+}\,\right)  ,\label{U-inv-1}\\
U^{-1}b_{m}^{+}U &  =\sum\limits_{n>0}\left(  -\beta_{nm}^{\ast}b_{n}
+\alpha_{nm}b_{n}\right)  \,.\label{U-inv-2}
\end{align}
Matrices $\alpha,\beta$ obey the conditions
\begin{gather}
\alpha\alpha^{+}-\beta\beta^{+}=1\,,\label{alpha-beta-1}\\
\alpha\beta^{T}=\beta\alpha^{T}\,,\label{alpha-beta-2}
\end{gather}
\begin{gather}
\alpha^{+}\alpha-\beta^{T}\beta^{\ast}=1\,,\label{alpha-beta-3}\\
\alpha^{+}\beta=\beta^{T}\alpha^{\ast}\,.\label{alpha-beta-4}
\end{gather}
Eqs. (\ref{alpha-beta-1}) and (\ref{alpha-beta-2}) can be rewritten in the
form
\begin{equation}
\left(
\begin{array}
[c]{cc}
\alpha & \beta\\
\beta^{\ast} & \alpha^{\ast}
\end{array}
\right)  \left(
\begin{array}
[c]{cc}
\alpha^{+} & -\beta^{T}\\
-\beta^{+} & \alpha^{T}
\end{array}
\right)  =1\,.\label{alpha-beta-restriction}
\end{equation}
Introducing the notation
\begin{equation}
\mathcal{S}=\left(
\begin{array}
[c]{cc}
\alpha & \beta\\
\beta^{\ast} & \alpha^{\ast}
\end{array}
\right)  \,,
\end{equation}
we can write
\begin{equation}
U\left(
\begin{array}
[c]{c}
b_{m}\\
b_{m}^{+}
\end{array}
\right)  U^{-1}=\left[  \mathcal{S}\left(
\begin{array}
[c]{c}
b\\
b^{+}
\end{array}
\right)  \right]  _{m}\,.\label{S-RPA-explicit}
\end{equation}
Now we can diagonalize the RPA Hamiltonian (\ref{H-RPA-def}) using Eq.
(\ref{S-RPA-explicit})
\begin{equation}
UH_{\mathrm{RPA}}U^{-1}=\frac{1}{2}\sum\limits_{m,n>0}\left(
\begin{array}
[c]{c}
b_{m}\\
b_{m}^{+}
\end{array}
\right)  ^{T}\left(  \mathcal{S}^{T}\mathcal{RS}\right)  _{mn}\left(
\begin{array}
[c]{c}
b_{n}\\
b_{n}^{+}
\end{array}
\right)  -\frac{1}{2}\sum\limits_{m>0}\left(  \varepsilon_{m}-\varepsilon
_{0}\right)  \,.
\end{equation}
Let us choose $\mathcal{S}$ so that
\begin{equation}
\mathcal{R}=\mathcal{S}^{T}\left(
\begin{array}
[c]{cc}
0 & \Omega\\
\Omega & 0
\end{array}
\right)  \mathcal{S}\,,
\end{equation}
where $\Omega$ is a diagonal matrix
\begin{equation}
\Omega=\mathrm{diag}(\Omega_{m})\,.
\end{equation}
Then
\begin{equation}
UH_{\mathrm{RPA}}U^{-1}=\frac{1}{2}\sum\limits_{m>0}\left[  \Omega_{m}\left(
b_{m}^{+}b_{m}+b_{m}b_{m}^{+}\right)  -\left(  \varepsilon_{m}-\varepsilon
_{0}\right)  \right]  \,.\label{H-U-diag}
\end{equation}
The spectrum of this Hamiltonian gives the $O(N_{c}^{0})$ contribution $\Delta
E_{B}$ to the $1/N_{c}$ expansion of the total energy (\ref{E-B-N-c-expansion}):
\begin{equation}
\Delta E_{B}=\sum\limits_{m>0}\left[  \Omega_{m}\left(  n_{m}+\frac{1}
{2}\right)  -\frac{1}{2}\left(  \varepsilon_{m}-\varepsilon_{0}\right)
\,\right]  .
\end{equation}

The corresponding eigenstates of $H_{\mathrm{RPA}}$ are according to Eqs.
(\ref{U-inv-2}) and (\ref{H-U-diag})
\begin{equation}
U^{-1}\left[  \prod\limits_{k}\left(  b_{k}^{+}\right)  ^{n_{k}}\right]
|0\rangle=\left[  \prod\limits_{k}\left(  -\sum\limits_{m>0}\beta_{mk}^{\ast
}b_{m}+\alpha_{mk}b_{m}^{+}\right)  ^{n_{k}}\right]  U^{-1}|0\rangle
\,.\label{U-a-eigenstates}
\end{equation}
Here $|0\rangle$ is the $b_{m}$ vacuum (\ref{a-vac}) and the ground state of
$H_{\mathrm{RPA}}$ is
\begin{equation}
|0_{\mathrm{RPA}}\rangle=U^{-1}|0\rangle\,.\label{vac-RPA}
\end{equation}

Using Eqs. (\ref{U-inv-1}) and (\ref{a-vac}), we find
\begin{equation}
0=U^{-1}b_{m}|0\rangle=\left(  U^{-1}b_{m}U\right)  U^{-1}|0\rangle
=\sum\limits_{n>0}\left(  \alpha_{nm}^{\ast}b_{n}-\beta_{nm}b_{n}^{+}\right)
U^{-1}|0\rangle\,.\label{U-0-eq}
\end{equation}
Combining Eqs. (\ref{vac-RPA}) and (\ref{U-0-eq}), we obtain
\begin{equation}
\sum\limits_{n>0}\left(  \alpha_{nm}^{\ast}b_{n}-\beta_{nm}b_{n}^{+}\right)
|0_{\mathrm{RPA}}\rangle=0\,.\label{U-0-eq-2}
\end{equation}

\subsection{Calculation of functions $C_{B}(\tilde{\Delta})$}

Functions $C_{B}(\tilde{\Delta})$ were defined via the asymptotic expression
(\ref{Phi-B-large-Nc-limit-2}). Now let us compute these functions. We start
from the function $C_{0}(\tilde{\Delta})$ corresponding to the ground state
$|0_{\mathrm{RPA}}\rangle$. The ground state $|0_{\mathrm{RPA}}\rangle$ obeys
Eq. (\ref{U-0-eq-2}). In the $\tilde{\Delta}$ representation (\ref{a-Delta}),
(\ref{a-dagger-Delta}), Eq. (\ref{U-0-eq-2}) takes the form
\begin{equation}
\sum\limits_{n>0}\left(  \alpha_{nm}^{\ast}\frac{\partial}{\partial
\tilde{\Delta}_{n}}-\beta_{nm}\tilde{\Delta}_{n}\right)  C_{0}(\tilde{\Delta
})=0\,.
\end{equation}
The solution of this equation is
\begin{equation}
C_{0}(\tilde{\Delta})=\mathcal{N}\exp\left[  \frac{1}{2}(\tilde{\Delta}
X\tilde{\Delta})\right]  \,,\label{C0-ansatz}
\end{equation}
where
\begin{equation}
(\tilde{\Delta}X\tilde{\Delta})=\sum\limits_{m,n>0}\tilde{\Delta}_{m}
X_{mn}\tilde{\Delta}_{n}\,,
\end{equation}
\begin{equation}
X_{mn}=\left[  \left(  \alpha^{\ast T}\right)  ^{-1}\beta^{T}\right]
_{mn}=\left[  \left(  \alpha^{+}\right)  ^{-1}\beta^{T}\right]  _{mn}
\label{X-def-0}
\end{equation}
and $\mathcal{N}$ is a normalization factor.

According to Eq. (\ref{alpha-beta-4}) we have
\begin{equation}
\beta\left(  \alpha^{\ast}\right)  ^{-1}=\left(  \alpha^{+}\right)  ^{-1}
\beta^{T}\,.
\end{equation}
This means that matrix
\begin{equation}
X_{mn}=\left[  \left(  \alpha^{+}\right)  ^{-1}\beta^{T}\right]  _{mn}=\left[
\beta\left(  \alpha^{\ast}\right)  ^{-1}\right]  _{mn}\label{X-def}
\end{equation}
is symmetric
\begin{equation}
X_{mn}=X_{nm}\,.\label{X-symmetric}
\end{equation}

The excited states (\ref{U-a-eigenstates}) are described in the $\tilde
{\Delta}$ representation (\ref{a-Delta}), (\ref{a-dagger-Delta}) by the wave
functions
\begin{equation}
C_{B}(\tilde{\Delta})=\mathcal{N}\prod\limits_{k}\left(  \sum\limits_{m>0}
-\beta_{mk}^{\ast}\frac{\partial}{\partial\tilde{\Delta}_{m}}+\alpha
_{mk}\tilde{\Delta}_{m}\right)  ^{n_{k}}\exp\left[  \frac{1}{2}(\tilde{\Delta
}X\tilde{\Delta})\right]  =\mathcal{N}P_{B}(\tilde{\Delta})\exp\left[
\frac{1}{2}(\tilde{\Delta}X\tilde{\Delta})\right]  \,.\label{C-B-P-B-exp}
\end{equation}
Here $P_{B}(\tilde{\Delta})$ is a polynomial of degree
\begin{equation}
N_{B}=\sum\limits_{k}n_{k}\,.
\end{equation}
At large $\tilde{\Delta}_{n}$ we have according to Eq. (\ref{C-B-P-B-exp})
\begin{equation}
P_{B}(\tilde{\Delta})\overset{\tilde{\Delta}\rightarrow\infty}{\longrightarrow
}\prod\limits_{k}\left[  \left(  -\beta^{+}X\tilde{\Delta}\right)
_{k}\right]  ^{n_{k}}\,,\label{P-Delta-large}
\end{equation}
where
\begin{equation}
\left(  -\beta^{+}X\tilde{\Delta}\right)  _{k}=-\sum\limits_{mn}\beta
_{mk}^{\ast}X_{mn}\tilde{\Delta}_{n}\,.
\end{equation}

Using Eqs. (\ref{X-def}) and (\ref{alpha-beta-3}), we find
\begin{equation}
\beta^{+}X=\beta^{+}\beta\left(  \alpha^{\ast}\right)  ^{-1}=\left(
\alpha^{T}\alpha^{\ast}-1\right)  \left(  \alpha^{\ast}\right)  ^{-1}
=\alpha^{T}-\left(  \alpha^{\ast}\right)  ^{-1}\,.
\end{equation}

\subsection{Matching two types of the $1/N_{c}$ expansion}

In Sec. \ref{Two-Nc-limits-section} we have described two different asymptotic
expressions for the large-$N_{c}$ limit. These two expressions are valid in
different regions. However, we can match the two expressions in their common
overlap area.

Let us insert Eq. (\ref{C-B-P-B-exp}) into Eq. (\ref{Phi-B-large-Nc-limit-2})

\begin{equation}
\Phi_{B}\left(  \phi+\frac{\tilde{\Delta}}{\sqrt{N_{c}}}\right)
=\mathcal{N}P_{B}(\tilde{\Delta})\exp\left[  \frac{1}{2}(\tilde{\Delta}
X\tilde{\Delta})\right]  \exp\left[  N_{c}W(\phi)\right]  \quad\left[
\tilde{\Delta}_{n}=O(N_{c}^{0})\right]  \,.\label{Phi-as-1}
\end{equation}
This expression must agree with with Eq. (\ref{Phi-A-W})
\begin{equation}
\Phi_{B}\left(  \phi+\frac{\tilde{\Delta}}{\sqrt{N_{c}}}\right)  =N_{c}
^{\nu_{B}}A_{B}\left(  \phi+\frac{\tilde{\Delta}}{\sqrt{N_{c}}}\right)
\exp\left[  N_{c}W\left(  \phi+\frac{\tilde{\Delta}}{\sqrt{N_{c}}}\right)
\right]  \,,\quad\tilde{\Delta}_{n}=O(\sqrt{N_{c}})\label{Phi-A-W-1}
\end{equation}
in the region
\begin{equation}
1\ll\tilde{\Delta}\ll\sqrt{N_{c}}\,.
\end{equation}
In this region only the leading term of degree (\ref{P-Delta-large}) survives
in the polynomial $P_{B}(\tilde{\Delta})$ so that Eq. (\ref{Phi-as-1}) becomes
\begin{equation}
\Phi_{B}\left(  \phi+\frac{\tilde{\Delta}}{\sqrt{N_{c}}}\right)
=\mathcal{N}\left\{  \prod\limits_{k}\left[  \left(  -\beta^{+}X\tilde{\Delta
}\right)  _{k}\right]  ^{n_{k}}\right\}  \exp\left[  \frac{1}{2}(\tilde
{\Delta}X\tilde{\Delta})\right]  \exp\left[  N_{c}W(\phi)\right]
\,.\label{RPA-matching}
\end{equation}

Now we want to check the general factorization relation
(\ref{A-factorization-1})
\begin{equation}
A_{B}\left(  g\right)  =A^{(0)}(g)\prod\limits_{k}\left[  A_{k}(g)\right]
^{n_{k}}=\tilde{A}^{(0)}(\tilde{g})\prod\limits_{k}\left[  \tilde{A}_{k}
(\tilde{g})\right]  ^{n_{k}}\,.\label{A-factorization-2}
\end{equation}
If this factorization holds, then Eq. (\ref{Phi-A-W-1}) takes the form
\begin{gather}
\Phi_{B}\left(  \phi+\frac{\tilde{\Delta}}{\sqrt{N_{c}}}\right)  =N_{c}
^{\nu_{B}}\tilde{A}^{(0)}\left(  \frac{\tilde{\Delta}}{\sqrt{N_{c}}}\right)
\prod\limits_{k}\left[  \tilde{A}_{k}\left(  \frac{\tilde{\Delta}}{\sqrt
{N_{c}}}\right)  \right]  ^{n_{k}}\nonumber\\
\times\,\exp\left[  N_{c}W\left(  \phi+\frac{\tilde{\Delta}}{\sqrt{N_{c}}
}\right)  \right]  \,,\quad\tilde{\Delta}_{n}=O(\sqrt{N_{c}})
\end{gather}

Due to Eq. (\ref{Hartree-point-eq}) and condition (\ref{phi-Delta-ortho}),
\begin{equation}
\left(  \phi^{\ast}\delta\tilde{\phi}\right)
=0\,,\label{delta-phi-orthogonal}
\end{equation}
the variation $W\left(  \phi+\delta\tilde{\phi}\right)  $ has no linear term:
\begin{equation}
W\left(  \phi+\delta\tilde{\phi}\right)  =W(\phi)+\frac{1}{2}W^{(2)}
(\delta\tilde{\phi},\delta\tilde{\phi})\,.\label{W-phi-expansion}
\end{equation}
Thus
\begin{equation}
\Phi_{B}\left(  \phi+\frac{\tilde{\Delta}}{\sqrt{N_{c}}}\right)
\approx\tilde{A}^{(0)}(0)\prod_{k}\left[  A_{k}\left(  \frac{\tilde{\Delta}
}{\sqrt{N_{c}}}\right)  \right]  ^{n_{k}}\exp\left[  N_{c}W(\phi)+\frac{1}
{2}W^{(2)}(\delta\tilde{\Delta},\delta\tilde{\Delta})\right]  \,.
\end{equation}
Inserting this into Eq. (\ref{RPA-matching}), we see that
\begin{align}
&  \mathcal{N}\left\{  \prod\limits_{k}\left[  \left(  -\beta^{+}
X\tilde{\Delta}\right)  _{k}\right]  ^{n_{k}}\right\}  \exp\left[  \frac{1}
{2}(\tilde{\Delta}X\tilde{\Delta})\right]  \exp\left[  N_{c}W(\phi)\right]
\nonumber\\
&  =N_{c}^{\nu_{B}}\tilde{A}^{(0)}(0)\prod_{k}\left[  \tilde{A}_{k}\left(
\frac{\tilde{\Delta}}{\sqrt{N_{c}}}\right)  \right]  ^{n_{k}}\exp\left[
N_{c}W(\phi)+\frac{1}{2}W^{(2)}(\delta\tilde{\Delta},\delta\tilde{\Delta
})\right]  \,.
\end{align}
Now we see that
\begin{equation}
\tilde{A}^{(0)}(0)=\mathcal{N}N_{c}^{-\nu_{B}+\sum_{k}n_{k}/2}\,,
\end{equation}
\begin{equation}
A_{k}\left(  \phi+\delta\tilde{\phi}\right)  =\left(  -\beta^{+}
X\,\delta\tilde{\phi}\right)  _{k}\,,\label{A-k-delta-phi}
\end{equation}
\begin{equation}
W^{(2)}(\delta\tilde{\phi},\delta\tilde{\phi})=\left(  \delta\tilde{\phi
}\,X\,\delta\tilde{\phi}\right)  \,.\label{W2-res}
\end{equation}

\subsection{Asymptotic behavior in the vicinity of the Hartree solution}

Inserting Eq. (\ref{W2-res}) into Eq. (\ref{W-phi-expansion}), we find
\begin{equation}
W\left(  \phi+\delta\tilde{\phi}\right)  =W(\phi)+\frac{1}{2}\left(
\delta\tilde{\phi}\,X\,\delta\tilde{\phi}\right)  +O\left(  \delta\tilde{\phi
}^{3}\right)  \,.
\end{equation}
Combining this result with Eq. (\ref{W-rescaling}), we obtain
\begin{equation}
W\left(  \lambda\phi+\delta\tilde{\phi}\right)  =W\left(  \phi+\lambda
^{-1}\delta\tilde{\phi}\right)  +\ln\lambda=W(\phi)+\ln\lambda+\frac{1}
{2\lambda^{2}}\left(  \delta\tilde{\phi}\,X\,\delta\tilde{\phi}\right)
+O\left(  \delta\tilde{\phi}^{3}\right)  \,.
\end{equation}
Taking into account Eq. (\ref{basis-choice}), we conclude that if
\begin{equation}
|g_{m}|\ll g_{0}\quad(m>0)
\end{equation}
then
\begin{equation}
W\left(  g\right)  =W(\phi)+\ln g_{0}+\frac{1}{2\left(  g_{0}\right)  ^{2}
}\sum\limits_{m,n>0}g_{m}X_{mn}g_{n}+O\left(  \left(  \frac{g_{m>0}}{g_{0}
}\right)  ^{3}\right)  \,.\label{W-g-near-Hartree}
\end{equation}

In Sec. \ref{Phi-B-large-Nc-section} we derived Eq. (\ref{W-WKB}) for the
function $W\left(  g\right)  $. This is a partial differential equation which
has many solutions. The asymptotic expansion (\ref{W-g-near-Hartree}) plays
the role of the boundary condition for this differential equation. The matrix
$X_{mn}$ appearing in this boundary condition is given by Eq. (\ref{X-def}).

In the same way, using Eq. (\ref{A-rescaling}), we can rewrite relation
(\ref{A-k-delta-phi}) in the form
\begin{equation}
A_{k}\left(  g\right)  =-\frac{1}{g_{0}}\sum\limits_{m,n>0}\beta_{mk}^{\ast
}X_{mn}g_{n}+O\left(  \left(  \frac{g_{m>0}}{g_{0}}\right)  ^{2}\right)  \,.
\end{equation}
This asymptotic behavior of $A_{k}\left(  g\right)  $ should be used as the
boundary condition for the differential equation (\ref{A-WKB-0}).

\section{Classical dynamics of large-$N_{c}$ systems}

\label{Classical-dynamics-section}

\setcounter{equation}{0} 

\subsection{Hamilton--Jacobi equation}

According to our previous results, function $W(g)$ can be computed by solving
the differential equation (\ref{W-WKB}) with the boundary condition
(\ref{W-g-near-Hartree}). This representation for $W(g)$ is not quite
convenient since one has to deal with the \emph{partial} differential equation
(\ref{W-WKB}). In this section we show how the problem can be reduced to the
analysis of \emph{ordinary} differential equations. This reduction can be done
using the well-known Hamilton--Jacobi method.

Eq. (\ref{W-WKB}) has the structure of the classical Hamilton--Jacobi
equation. In order to make this structure explicit, we change the notation
\begin{equation}
g\rightarrow q\,,\label{g-to-q}
\end{equation}
\begin{equation}
W(g)\rightarrow S(q)\,.\label{W-to-S}
\end{equation}
Then Eq. (\ref{W-WKB}) takes the form
\begin{equation}
H\left(  \frac{\partial S(q)}{\partial q},q\right)  =E_{0}\,,\label{S-HJ}
\end{equation}
where the Hamiltonian is
\begin{equation}
H(p,q)=\frac{1}{2}\sum\limits_{n_{1}n_{2}n_{3}n_{4}}V_{n_{1}n_{2}n_{3}n_{4}
}\left(  q_{n_{1}}p_{n_{2}}\right)  \left(  q_{n_{3}}p_{n_{4}}\right)
+\sum\limits_{n_{1}n_{2}}L_{n_{1}n_{2}}\left(  q_{n_{1}}p_{n_{2}}\right)
\,.\label{H-V-L}
\end{equation}

One should keep in mind that $g$ can be complex and $W(g)$ is an analytical
function of $g$ (which may have singular and branch points). Therefore our
classical mechanics is complex.

As is well known, the solutions of the Hamilton--Jacobi equation can be
constructed in terms of the action for the trajectories obeying the
Hamiltonian equations
\begin{equation}
\frac{dp_{n}}{dt}=-\frac{\partial H(p,q)}{\partial q_{n}},\quad\frac{dq_{n}
}{dt}=\frac{\partial H(p,q)}{\partial p_{n}}\,.\label{pq-Hamilton-eq}
\end{equation}
Solution $S(q)$ of the Hamilton--Jacobi equation (\ref{S-HJ}) describes the
set of trajectories covering the coordinate space. All these trajectories have
the same energy $E_{0}$ and obey the condition
\begin{equation}
p_{n}=\frac{\partial S(q)}{\partial q_{n}}\,.\label{p-dS-dq}
\end{equation}

\subsection{Static and time-dependent Hartree equations}

\label{TDHE-interpretation-section}

If we change the notation
\begin{equation}
\phi_{n}\rightarrow p_{n}\,,\quad\phi_{n}^{\ast}\rightarrow q_{n}
\label{phi-p-q}
\end{equation}
in the Hartree equation (\ref{Hartree-general}) and in the normalization
condition (\ref{phi-normalized}) then we obtain
\begin{align}
\frac{\partial}{\partial p_{n}}H(p,q) &  =\varepsilon_{0}q_{n}
\,,\label{Hartree-pq-1}\\
\frac{\partial}{\partial q_{n}}H(p,q) &  =\varepsilon_{0}p_{n}\,,\\
\sum\limits_{n}p_{n}q_{n} &  =1\,,\label{Hartree-pq-3}\\
p_{n} &  =q_{n}^{\ast}\,.\label{Hartree-pq-4}
\end{align}
Let us stress that we consider $p_{n}$ and $q_{n}$ as independent complex
variables of our phase space. Conditions (\ref{Hartree-pq-1})--(\ref{Hartree-pq-4})
define a point in this space. At this point the
variables $p_{n}$ and $q_{n}$ are complex conjugate. But generally speaking
$p_{n}\neq q_{n}^{\ast}$.

Strictly speaking, Eqs. (\ref{Hartree-pq-1})--(\ref{Hartree-pq-4}) have more
than one solution. For example, these equations are invariant under the phase
transformation
\begin{equation}
p_{n}\rightarrow e^{i\alpha}p_{n}\,,\quad q_{n}\rightarrow e^{-i\alpha}
q_{n}\,.
\end{equation}
In principle, there can be additional symmetries which increase the number of solutions.

If one chooses the basis (\ref{basis-choice}), then the solution of the
Hartree equations (\ref{Hartree-pq-1})--(\ref{Hartree-pq-4}) is
\begin{equation}
p_{n}=q_{n}=\delta_{n0}\,.
\end{equation}
Parameter $\varepsilon_{0}$ is given by Eq. (\ref{Hartree-special-basis-2}).

The change of notation (\ref{phi-p-q}) in the Hamilton equations
(\ref{pq-Hamilton-eq}) transforms them into the form
\begin{equation}
\frac{d\phi_{m}}{dt}=-\sum\limits_{n}h_{mn}\phi_{n}\,,\quad\frac{d\phi
_{n}^{\ast}}{dt}=\sum\limits_{m}\phi_{m}^{\ast}h_{mn}
\,,\label{phi-Euclid-TDHE}
\end{equation}
where
\begin{equation}
h_{mn}=\sum\limits_{ij}V_{mnij}\phi_{i}^{\ast}\phi_{j}+L_{mn}\,.
\end{equation}
These equations can be interpreted as a Euclidean version of TDHE
(\ref{TDHE-1}), (\ref{TDHE-2}). It should be stressed that the variables
$\phi_{n} $ and $\phi_{n}^{\ast}$ are not complex conjugate in equations
(\ref{phi-Euclid-TDHE}) in contrast to the standard TDHE.

\subsection{Asymptotic conditions}

In the vicinity of the ``Hartree point''
(\ref{Hartree-pq-1})--(\ref{Hartree-pq-4}), the Hamilton equations for trajectories can be
approximated by
\begin{align}
\frac{dq_{n}}{dt}  &  =\frac{\partial}{\partial p_{n}}H(p,q)=\varepsilon
_{0}q_{n}\,,\\
\frac{dp_{n}}{dt}  &  =-\frac{\partial}{\partial q_{n}}H(p,q)\approx
-\varepsilon_{0}p_{n}
\end{align}
with the asymptotic solutions at $t\rightarrow-\infty$
\begin{align}
&  q_{n}(t)\overset{t\rightarrow-\infty}{=}\delta_{n0}\exp\left[
\varepsilon_{0}\left(  t-\tau\right)  \right]  \,,\label{q-asymptotitc}\\
&  p_{n}(t)\overset{t\rightarrow-\infty}{=}\delta_{n0}\left[  -\varepsilon
_{0}\left(  t-\tau\right)  \right]  \,.\label{p-asymptotitc}
\end{align}
According to Eq. (\ref{H-V-L}) this asymptotic behavior corresponds to the
Hartree energy $E_{0}$ (\ref{E0-Hartree})
\begin{equation}
H(p,q)=E_{0}\,.\label{H-pq-E0}
\end{equation}

\subsection{Integral of motion $\sum_{n}p_{n}q_{n}$}

The Hamiltonian (\ref{H-V-L}) and equations of motion (\ref{pq-Hamilton-eq})
are invariant under the transformations
\begin{equation}
p_{n}\rightarrow\lambda^{-1}p_{n}\,,\quad q_{n}\rightarrow\lambda q_{n}\,.
\end{equation}
Therefore
\begin{equation}
\left(  p_{n}\frac{\partial}{\partial p_{n}}-q_{n}\frac{\partial}{\partial
q_{n}}\right)  H=0\,.
\end{equation}
This identity can be rewritten in terms of the Poisson bracket
\begin{equation}
\left\{  \sum\limits_{n}p_{n}q_{n},H\right\}  =0\,.
\end{equation}
Thus we have the integral of motion
\begin{equation}
\sum\limits_{n}p_{n}q_{n}=\mathrm{const}\,.\label{pq-integral}
\end{equation}
For the trajectories with the asymptotic behavior (\ref{q-asymptotitc}),
(\ref{p-asymptotitc}) we have
\begin{equation}
\sum\limits_{n}p_{n}q_{n}=1\,.\label{pq-one}
\end{equation}

\subsection{Boundary condition}

Using notation (\ref{g-to-q}), (\ref{W-to-S}), we can rewrite the asymptotic
expansion (\ref{W-g-near-Hartree}) in the form
\begin{equation}
S\left(  q\right)  =\left.  S(q^{H})\right|  _{q^{H}=\delta_{n0}}+\ln
q_{0}+\frac{1}{2}\sum\limits_{m,n>0}X_{mn}\frac{q_{m}}{q_{0}}\frac{q_{n}
}{q_{0}}+O\left[  \left|  \frac{q_{m>0}}{q_{0}}\right|  ^{3}\right]
\,.\label{S-q-near-Hartree}
\end{equation}
This expression is valid for
\begin{equation}
|q_{m}|\ll q_{0}\quad(m>0)\,.
\end{equation}
Inserting this decomposition into Eq. (\ref{p-dS-dq}), we find
\begin{equation}
p_{n}(q)=\frac{\partial S(q)}{\partial q_{n}}=\frac{1}{q_{0}}\delta_{n0}
\quad\mathrm{if}\quad|q_{m}|\ll q_{0}\quad(m>0)\,\,.\label{p-q-near-Hartree}
\end{equation}
This agrees with the asymptotic behavior of trajectories
(\ref{q-asymptotitc}), (\ref{p-asymptotitc}) at $t\rightarrow-\infty$.

Now we understand that the action $S(q)$ is associated with the configuration
of trajectories $p(t)$, $q(t)$ which start at $t\rightarrow-\infty$ at points
$p,q$ obeying Eqs. (\ref{Hartree-pq-1})--(\ref{Hartree-pq-3}) with the
asymptotic behavior (\ref{q-asymptotitc}), (\ref{p-asymptotitc}). All these
trajectories have the same energy $E_{0}$ and the same integral of motion
(\ref{pq-one}).

Let $p(t)$, $q(t)$ be one of these trajectories with energy $E_{0}$:
\begin{equation}
H\left[  p(t),q(t)\right]  =E_{0}\,.\label{H-E0}
\end{equation}
Let us assume that at $t=0$ this trajectory passes through the point
$p^{(0)},q^{(0)}$:
\begin{equation}
p(0)=p^{(0)},\quad q(0)=q^{(0)}\,,\label{pq-t-0}
\end{equation}
\begin{equation}
H\left(  p^{(0)},q^{(0)}\right)  =E_{0}\,\label{H-p0-q0}
\end{equation}
and at $t\rightarrow-\infty$ it has the asymptotic behavior
(\ref{q-asymptotitc}), (\ref{p-asymptotitc}) with some
\begin{equation}
\tau=\tau(q^{(0)})
\end{equation}
depending on $q_{0}$:
\begin{align}
&  q_{n}(t)\overset{t\rightarrow-\infty}{=}\delta_{n0}\exp\left\{
\varepsilon_{0}\left[  t-\tau(q^{(0)})\right]  \right\}
\,,\label{q-asymptotitc-tau}\\
&  p_{n}(t)\overset{t\rightarrow-\infty}{=}\delta_{n0}\left\{  -\varepsilon
_{0}\left[  t-\tau(q^{(0)})\right]  \right\}  \,.\label{p-asymptotitc-tau}
\end{align}

Note that at $t\rightarrow-\infty$ we have along this trajectory
\begin{equation}
L=\sum\limits_{n}p_{n}\dot{q}_{n}-H(p,q)\overset{t\rightarrow-\infty
}{\rightarrow}\varepsilon_{0}-E_{0}\,.
\end{equation}
This means that the integral
\begin{equation}
\int_{-\infty}^{0}dt\left(  L+E_{0}-\varepsilon_{0}\right)
\end{equation}
is convergent at $t\rightarrow-\infty$. Now let us define the function
\begin{equation}
S(q^{(0)})=\int_{-\infty}^{0}dt\left(  L+E_{0}-\varepsilon_{0}\right)
-\varepsilon_{0}\tau(q^{(0)})\label{S-def-int}
\end{equation}
and show that that this function obeys the differential equation (\ref{S-HJ}).

According to Eq. (\ref{H-E0}) we have
\begin{equation}
L+E_{0}=L+H=\sum\limits_{n}p_{n}\dot{q}_{n}\,.
\end{equation}
Therefore
\begin{equation}
S(q^{(0)})=\int_{-\infty}^{0}dt\left[  \sum\limits_{n}p_{n}(t)\dot{q}
_{n}(t)-\varepsilon_{0}\right]  -\varepsilon_{0}\tau(q^{(0)})\label{S-int-2}
\end{equation}
Hence
\begin{align}
S(q^{(0)})  &  =\lim_{T\rightarrow-\infty}\int_{T+\tau(q^{(0)})}^{0}
dt\sum\limits_{n}p_{n}(t)\dot{q}_{n}(t)+\varepsilon_{0}T\nonumber\\
&  =\lim_{T\rightarrow-\infty}\int_{\delta_{m0}\exp\left(  \varepsilon
_{0}T\right)  }^{q_{m}^{(0)}}\sum\limits_{n}p_{n}dq_{n}+\varepsilon
_{0}T\,.\label{S-int-3}
\end{align}
Taking some large but fixed $|T|$ on the RHS, we see that this definition of
$S(q^{(0)})$ has the standard property
\begin{equation}
\frac{\partial S(q^{(0)})}{\partial q_{n}^{(0)}}=p_{n}^{(0)}\,,\label{dS-dq0}
\end{equation}
where $p_{n}^{(0)}$ is the momentum $p_{n}(t)$ taken at $t=0$ according to Eq.
(\ref{pq-t-0}). Inserting Eq. (\ref{dS-dq0}) into Eq. (\ref{H-p0-q0}) and
taking into account that the point $q^{(0)}$ is arbitrary, we find that the
function $S(q)$ defined by Eq. (\ref{S-def-int}) obeys the Hamilton--Jacobi
equation (\ref{S-HJ}).

Since the definition (\ref{S-def-int}) of $S(q)$ is based on the trajectories
with the asymptotic behavior (\ref{q-asymptotitc-tau}),
(\ref{p-asymptotitc-tau}), we automatically have the property
(\ref{p-q-near-Hartree}). Therefore function $S(q)$ defined by Eq.
(\ref{S-def-int}) also obeys the boundary condition (\ref{S-q-near-Hartree}).
Since both differential equation (\ref{S-HJ}) and the boundary condition
(\ref{S-q-near-Hartree}) are satisfied, we conclude that Eq. (\ref{S-def-int})
gives a correct representation for the functional
\begin{equation}
W(g)\equiv S(q^{(0)})\,,\quad g\equiv q^{(0)}\label{W-S-id}
\end{equation}
which determines the large-$N_{c}$ behavior (\ref{Phi-A-W}).

The construction described in this section leads us to the following algorithm
of the calculation of $W(g)$.

1) Solve equations of motion (\ref{pq-Hamilton-eq}) imposing boundary
condition
\begin{equation}
q(0)=g
\end{equation}
with given $g$ (\ref{q-0-g}) and boundary conditions
(\ref{q-asymptotitc-tau}), (\ref{p-asymptotitc-tau})
\begin{align}
&  q_{n}(t)\overset{t\rightarrow-\infty}{=}\delta_{n0}\exp\left\{
\varepsilon_{0}\left[  t-\tau(g)\right]  \right\}  \,,\\
&  p_{n}(t)\overset{t\rightarrow-\infty}{=}\delta_{n0}\left\{  -\varepsilon
_{0}\left[  t-\tau(g)\right]  \right\}  \,.
\end{align}
with \emph{unknown} $\tau(g)$. In this way one finds the trajectory and
$\tau(g)$.

2) Now $W(g)$ is given by Eqs. (\ref{S-int-2}), (\ref{W-S-id}):
\begin{equation}
W(g)=\int_{-\infty}^{0}dt\left[  \sum\limits_{n}p_{n}(t)\dot{q}_{n}
(t)-\varepsilon_{0}\right]  -\varepsilon_{0}\tau(g)\,.
\end{equation}

\subsection{Case $L_{mn}=0$}

\label{L-0-case-section}

Let us consider the special case when
\begin{equation}
L_{mn}=0\label{L-0}
\end{equation}
in the Hamiltonian (\ref{H-V-L}). In this case the Hartree equations
(\ref{Hartree-special-basis}), (\ref{E0-Hartree}) lead to the simple relation
\begin{equation}
E_{0}=\frac{1}{2}\varepsilon_{0}\,.\label{E0-eps0}
\end{equation}
In the case (\ref{L-0}) the Hamiltonian (\ref{H-V-L}) is quadratic in $p$ so
that
\begin{equation}
\sum\limits_{n}p_{n}\frac{\partial H(p,q)}{\partial p_{n}}=2H(p,q)\,.
\end{equation}
Combining this with Eq. (\ref{H-E0}), we find
\begin{equation}
\sum\limits_{n}p_{n}\dot{q}_{n}=\sum\limits_{n}p_{n}\frac{\partial H}{\partial
p_{n}}=2H(p,q)=2E_{0}\,.
\end{equation}
Taking into account Eq. (\ref{E0-eps0}), we arrive at
\begin{equation}
\sum\limits_{n}p_{n}\dot{q}_{n}-\varepsilon_{0}=2E_{0}-\varepsilon_{0}=0\,.
\end{equation}
Now we insert this into Eq. (\ref{S-int-2})
\begin{equation}
S(q^{(0)})=-\varepsilon_{0}\tau(q^{(0)})\,.
\end{equation}
This allows us to rewrite Eqs. (\ref{q-asymptotitc-tau}) and
(\ref{p-asymptotitc-tau}) in the form
\begin{align}
&  q_{n}(t)\overset{t\rightarrow-\infty}{=}\delta_{n0}\exp\left[
\varepsilon_{0}t+S(q^{(0)})\right]  \,,\label{q-S-q0}\\
&  p_{n}(t)\overset{t\rightarrow-\infty}{=}\delta_{n0}\left[  -\varepsilon
_{0}t-S(q^{(0)})\right]  \,.\label{p-S-q0}
\end{align}
Thus in the case (\ref{L-0}) the action can be directly read from the
asymptotic behavior of $q_{n}(t)$, $p_{n}(t)$ at $t\rightarrow-\infty$.
Remember that these trajectories are fixed by extra conditions (\ref{pq-t-0})
\begin{equation}
q(0)=q^{(0)}
\end{equation}
and (\ref{H-pq-E0}), (\ref{E0-eps0})
\begin{equation}
H\left(  p^{(0)},q^{(0)}\right)  =E_{0}=\frac{1}{2}\varepsilon_{0}\,.
\end{equation}

Now we conclude from Eqs. (\ref{W-S-id}), (\ref{q-S-q0}), (\ref{p-S-q0}) that
in the case $L_{mn}=0$ the functional $W(g)$ is given by the equation
\begin{equation}
W(g)=\ln I(g)\quad\left(  \mathrm{if}\,L_{mn}=0\right)  \,,\label{W-ln-I-0}
\end{equation}
where $I(g)$ is the parameter of the trajectories obeying the asymptotic
conditions
\begin{gather}
q_{n}(t)\overset{t\rightarrow-\infty}{=}\delta_{n0}I(g)e^{\varepsilon_{0}
t}\,,\label{q-I}\\
p_{n}(t)\overset{t\rightarrow-\infty}{=}\frac{\delta_{n0}}{I(g)}
e^{-\varepsilon_{0}t}\,,\label{p-I}\\
q_{n}(0)=g_{n}\,.\label{q-0-g}
\end{gather}

Let us summarize. In the case $L_{mn}=0$ the calculation of the functional
$W(g)$ reduces to the following steps:

1) Solve equations of motion (\ref{pq-Hamilton-eq}) imposing boundary
conditions (\ref{q-I}) -- (\ref{q-0-g}) with given $g$ and with \emph{unknown}
$I(g)$. In this way one finds $I(g)$.

2) Express $W(g)$ via $I(g)$ according to Eq. (\ref{W-ln-I-0}).

\section{Example: asymmetric rotator as a large-$N_{c}$ system}

\label{Asymmetric-rotator-section}

\setcounter{equation}{0} 

\subsection{Model}

In the above sections we have considered the general theory of fermionic
systems described by operators $a_{nc}$, $a_{nc}^{+}$
(\ref{a-a-dagger-anticommutator}) with $c=1,2,\ldots,N_{c}$. Now we want to
consider the simplest version of these models with only two values for the
index $n$ of $a_{nc}$. It is convenient to interpret this two-valued index as
the projection of spin $1/2$. In other words, we want to consider the model
described by the Hamiltonian
\begin{equation}
H=\frac{1}{2N_{c}}\sum\limits_{ab}K_{ab}J_{a}J_{b}+\sum\limits_{a}Y_{a}
J_{a}\,,\label{H-J}
\end{equation}
where
\begin{equation}
J_{b}=\sum\limits_{j,j^{\prime}=\pm\frac{1}{2}}\sum\limits_{c=1}^{N_{c}
}\frac{1}{2}a_{j^{\prime}c}^{+}\left(  \tau_{b}\right)  _{j^{\prime}j}
a_{jc}\,.
\end{equation}
Obviously this Hamiltonian belongs to the class (\ref{H-general}) with
parameters
\begin{align}
V_{n_{1}n_{2}n_{3}n_{4}}  &  =\frac{1}{4}\sum\limits_{ab}K_{ab}\left(
\tau_{a}\right)  _{n_{1}n_{2}}\left(  \tau_{b}\right)  _{n_{3}n_{4}
}\,,\label{V-rotator}\\
L_{n_{1}n_{2}}  &  =\frac{1}{2}\sum\limits_{a}Y_{a}\left(  \tau_{a}\right)
_{n_{1}n_{2}}\,.
\end{align}

On the other hand, Hamiltonian (\ref{H-J}) is nothing else but the rotator
described by the angular momentum $J_{a}$
\begin{equation}
\left[  J_{a},J_{b}\right]  =i\varepsilon_{abc}J_{c}\,.
\end{equation}
We will consider the case of the asymmetric rotator ($I_{ab}\neq I\delta_{ab}
$) but the quantum number $\mathbf{J}^{2}$ is still conserved:
\begin{equation}
\left[  \mathbf{J}^{2},H\right]  =0
\end{equation}
and has the standard eigenvalues
\begin{equation}
\mathbf{J}^{2}=J(J+1)\,.
\end{equation}

We want to study the states made of $N_{c}$ quarks and antisymmetric in color.
The Fermi statistics and the color antisymmetry lead to the complete symmetry
of the spin wave function so that we deal with
\begin{equation}
J=\frac{N_{c}}{2}\,.\label{J-N-constraint}
\end{equation}
Thus the algebraic formulation of the problem of the baryon is obvious: take
the sector with $J=N_{c}/2$ and diagonalize the Hamiltonian (\ref{H-J}) in
this sector.

\subsection{Direct semiclassical approach}

In principle, we can find the spectrum of low-lying states of the Hamiltonian
(\ref{H-J}) at large $N_{c}$ solving the general Hartree equation
(\ref{Hartree-general}) and diagonalizing the RPA Hamiltonian
(\ref{H-RPA-def}). However, the same results can be obtained by
directly applying the
semiclassical approximation to the large momentum $J$ (\ref{J-N-constraint})
in the Hamiltonian (\ref{H-J}).

Instead of solving the Hartree equation (\ref{Hartree-general}) we can simply
minimize the Hamiltonian (\ref{H-J}) considering it as a function of the
classical momentum $\mathbf{J}$. This minimization should be performed
assuming the classical version of the constraint (\ref{J-N-constraint}).
Introducing the Lagrange multiplier $\lambda$, we arrive at the extremum
problem for the ground state energy (\ref{E-B-N-c-expansion}) in the leading
order of the $1/N_{c}$ expansion:
\begin{equation}
N_{c}E_{0}=\min_{\mathbf{J}}\left[  \frac{1}{2N_{c}}\sum\limits_{ab}\left(
K_{ab}-\lambda\delta_{ab}\right)  J_{a}J_{b}+\sum\limits_{a}Y_{a}J_{a}\right]
\,,
\end{equation}
This leads to the equation
\begin{equation}
\frac{1}{N_{c}}\sum\limits_{b}\left(  K_{ab}-\lambda\delta_{ab}\right)
J_{b}+Y_{a}=0\,.\label{lambda-J}
\end{equation}
If $K-\lambda$ is not degenerate, then
\begin{equation}
J_{a}=-N_{c}\sum\limits_{b}\left[  (K-\lambda)^{-1}\right]  _{ab}
Y_{b}\,.\label{J-via-c}
\end{equation}
Inserting this into Eq. (\ref{J-N-constraint}), we find
\begin{equation}
\left(  Y,(K-\lambda)^{-2}Y\right)  =\frac{1}{4}\,.\label{lambda-ground-state}
\end{equation}
This equation determines $\lambda$. Knowing $\lambda$, we can find $J$ from
Eq. (\ref{J-via-c}).

The leading order of the $1/N_{c}$ expansion (\ref{E-B-N-c-expansion}) for the
energy of the lowest states is determined by the parameter
\begin{align}
E_{0} &  =\frac{1}{N_{c}}\left[  \frac{1}{2N_{c}}K_{ab}J_{a}J_{b}+Y_{a}
J_{a}\right]  =\frac{1}{2}Y(K-\lambda)^{-1}K(K-\lambda)^{-1}Y-Y(K-\lambda
)^{-1}Y\nonumber\\
&  =-\frac{1}{2}Y\frac{K-2\lambda}{(K-\lambda)^{2}}
Y\,.\label{E0-rotator-general}
\end{align}
In principle, we may also have other solutions of Eq. (\ref{lambda-J})
corresponding to
\begin{equation}
\det(K-\lambda)=0\,.\label{lambda-special}
\end{equation}

\subsection{Special case}

Below we concentrate on the case when matrix $K_{ab}$ is diagonal
\begin{equation}
K_{ab}=K_{a}\delta_{ab}\label{K-diagonal}
\end{equation}
and
\begin{equation}
Y_{1}=Y_{2}=0\,.\label{B-12-0}
\end{equation}
We also assume that
\begin{equation}
Y_{3}>0\,,\label{Y-3-positive}
\end{equation}
\begin{align}
K_{1}-K_{3}+2Y_{3}  &  >0\,,\label{case-1-stability-1}\\
K_{2}-K_{3}+2Y_{3}  &  >0\,.\label{case-1-stability-2}
\end{align}

In this case Eq. (\ref{lambda-ground-state}) yields
\begin{equation}
K_{3}-\lambda=\pm2Y_{3}
\end{equation}
and Eq. (\ref{E0-rotator-general}) results in
\begin{equation}
E_{0}=-\frac{1}{8}\left(  K_{3}-2\lambda\right)  =-\frac{1}{8}\left[
K_{3}-2\left(  K_{3}\mp2Y_{3}\right)  \right]  \,.
\end{equation}
According to Eq. (\ref{Y-3-positive}) the minimal energy corresponds to the
upper sign. Thus
\begin{equation}
E_{0}=\frac{1}{8}\left(  K_{3}-4Y_{3}\right)  \,.\label{E0-KB}
\end{equation}

We find from Eq. (\ref{J-via-c})
\begin{equation}
J_{1}=J_{2}=0\,,\quad J_{3}=-\frac{N_{c}}{2}\,.\label{J-clasical-res}
\end{equation}
One can show that under conditions
(\ref{Y-3-positive})--(\ref{case-1-stability-2}) this solution gives the
true semiclassical
ground state and solutions of Eq. (\ref{lambda-special}) can be ignored.

\subsection{Hartree equation}

The semiclassical spin $J_{a}$ has the following interpretation in terms of
the solution of the Hartree equation (\ref{Hartree-general})
\begin{equation}
J_{a}=\frac{N_{c}}{2}\sum\limits_{mn}\phi_{m}^{\ast}\left(  \tau_{a}\right)
_{mn}\phi_{m}\,.
\end{equation}
Therefore the classical solution (\ref{J-clasical-res}) corresponds to
\begin{equation}
\phi_{m}=\delta_{m2}=\left(
\begin{array}
[c]{c}
0\\
1
\end{array}
\right)  \,.\label{phi-Hartree-rotator}
\end{equation}
In the Hartree equation (\ref{Hartree-general}) we have the single-particle
Hamiltonian
\begin{equation}
h_{n_{1}n_{2}}=\sum\limits_{n_{3}n_{4}}V_{n_{1}n_{2}n_{3}n_{4}}\phi_{n_{3}
}^{\ast}\phi_{n_{4}}+L_{n_{1}n_{2}}=\frac{1}{4}\left(  -K_{3}+2Y_{3}\right)
\left(  \tau_{3}\right)  _{n_{1}n_{2}}\,.
\end{equation}
Hence the single-particle energy $\varepsilon_{0}$ is
\begin{equation}
\varepsilon_{0}=\frac{1}{4}\left(  K_{3}-2Y_{3}\right)  \,.
\end{equation}
It is easy to see that the general expression for the Hartree energy $E_{0}$
(\ref{E-Hartree-special}) reproduces the above result (\ref{E0-KB}). The
single nonoccupied Hartree state is
\begin{equation}
\phi_{m}^{(1)}=\left(
\begin{array}
[c]{c}
1\\
0
\end{array}
\right) \label{phi-nonocc}
\end{equation}
and its energy is
\begin{equation}
\varepsilon_{1}=-\varepsilon_{0}=-\frac{1}{4}\left(  K_{3}-2Y_{3}\right)
\,.\label{eps-1-0}
\end{equation}

\subsection{RPA equation}

In the general RPA matrix $\mathcal{R}$ (\ref{RPA-matrix-def}), the indices
$m,n$ correspond to the nonoccupied eigenstates of the Hartree equation. In
our case we have only one nonoccupied state (\ref{phi-nonocc}). We label the
occupied Hartree state with $m=0$ whereas the nonoccupied state is labeled by
$m=1$. As a result, matrix $\mathcal{R}$ (\ref{RPA-matrix-def}) takes the form
\begin{equation}
\mathcal{R}=\left(
\begin{array}
[c]{cc}
V_{0101} & V_{1001}+\varepsilon_{1}-\varepsilon_{0}\\
V_{1001}+\varepsilon_{1}-\varepsilon_{0} & V_{1010}
\end{array}
\right)  \,.
\end{equation}
Using Eq. (\ref{V-rotator}), we find for the diagonal tensor $K_{ab}$
(\ref{K-diagonal})
\begin{align}
V_{0101} &  =V_{1010}=\frac{1}{4}\left(  K_{1}-K_{2}\right)  \,,\\
V_{1001} &  =\frac{1}{4}\left(  K_{1}+K_{2}\right)  \,.
\end{align}
With these coefficients $V_{n_{1}n_{2}n_{3}n_{4}}$ and with expression
(\ref{eps-1-0}) for $\varepsilon_{1}-\varepsilon_{0}$, we obtain
\begin{equation}
\mathcal{R}=\frac{1}{4}\left(
\begin{array}
[c]{cc}
K_{1}-K_{2} & K_{1}+K_{2}+2\left(  -K_{3}+2Y_{3}\right) \\
K_{1}+K_{2}+2\left(  -K_{3}+2Y_{3}\right)  & K_{1}-K_{2}
\end{array}
\right)  \,.
\end{equation}
This leads to the Hamiltonian (\ref{H-RPA-def})
\begin{equation}
H_{\mathrm{RPA}}=\frac{1}{8}\left[  (K_{1}-K_{2})\left(  aa+a^{+}a^{+}\right)
+\left[  K_{1}+K_{2}+2\left(  -K_{3}+2Y_{3}\right)  \right]  \left(
a^{+}a+aa^{+}\right)  \right]  -\frac{1}{4}\left(  -K_{3}+2Y_{3}\right)  \,.
\end{equation}
Performing Bogolyubov transformation
\begin{equation}
a=\frac{\kappa^{1/2}+\kappa^{-1/2}}{2}b+\frac{\kappa^{1/2}-\kappa^{-1/2}}
{2}b^{+}\label{a-via-b}
\end{equation}
with
\begin{equation}
\kappa=\sqrt{\frac{K_{2}-K_{3}+2Y_{3}}{K_{1}-K_{3}+2Y_{3}}}
\,,\label{kappa-via-K}
\end{equation}
we find
\begin{equation}
H_{\mathrm{RPA}}=\frac{\omega}{2}\left(  b^{+}b+bb^{+}\right)  +\frac{1}
{4}\left(  K_{3}-2Y_{3}\right)  \,,
\end{equation}
where
\begin{equation}
\omega=\frac{1}{2}\sqrt{\left(  K_{1}-K_{3}+2Y_{3}\right)  \left(  K_{2}
-K_{3}+2Y_{3}\right)  }\,.\label{omega-via-K}
\end{equation}
According to Eqs. (\ref{kappa-via-K}) and (\ref{omega-via-K}) we have
\begin{align}
K_{1} &  =K_{3}-2Y_{3}+2\kappa^{-1}\omega\,,\label{K1-kappa-omega}\\
K_{2} &  =K_{3}-2Y_{3}+2\kappa\omega\,.\label{K2-kappa-omega}
\end{align}

Comparing Eq. (\ref{H-U-diag}) with Eq. (\ref{a-via-b}), we see that
\begin{equation}
\mathcal{S}=\frac{1}{2}\left(
\begin{array}
[c]{cc}
\kappa^{1/2}+\kappa^{-1/2} & \kappa^{1/2}-\kappa^{-1/2}\\
\kappa^{1/2}-\kappa^{-1/2} & \kappa^{1/2}+\kappa^{-1/2}
\end{array}
\right) \label{S-rotator}
\end{equation}
and the only RPA excitation energy is
\begin{equation}
\Omega_{1}=\omega\,.
\end{equation}
Comparing our result (\ref{S-rotator}) for $\mathcal{S}$ with Eq.
(\ref{S-RPA-explicit}), we obtain
\begin{align}
\alpha_{11}  &  =\kappa^{1/2}+\kappa^{-1/2}\,,\\
\beta_{11}  &  =\kappa^{1/2}-\kappa^{-1/2}\,.
\end{align}
Using Eq. (\ref{X-def}), we find
\begin{equation}
X_{11}=\left[  \beta\left(  \alpha^{\ast}\right)  ^{-1}\right]  _{11}
=\frac{\kappa^{1/2}-\kappa^{-1/2}}{\kappa^{1/2}+\kappa^{-1/2}}
\,.\label{X11-res}
\end{equation}

\subsection{Method of generating functionals}

In the case of the Hamiltonian (\ref{H-J}), Eq. (\ref{W-WKB}) for the
functional $W(g)$ takes the form
\begin{equation}
\sum\limits_{ab}\frac{1}{2}K_{ab}\mathcal{J}_{a}\mathcal{J}_{b}+\sum
\limits_{a}Y_{a}\mathcal{J}_{a}=E_{0}\,,\label{f-Schroedinger}
\end{equation}
where
\begin{equation}
\mathcal{J}_{a}=g_{i}\left(  \frac{\tau_{a}}{2}\right)  _{ij}\frac{\partial
W(g)}{\partial g_{j}}\,.\label{I-cal-def}
\end{equation}

According to Eq. (\ref{W-rescaling}) function $W(g_{1},g_{2})$ has a
nontrivial dependence only on the ration $g_{1}/g_{2}$ so that we can
represent $W(g_{1},g_{2})$ in the form
\begin{equation}
W(g_{1},g_{2})=f\left(  \frac{g_{1}}{g_{2}}\right)  +\frac{1}{2}\ln\left(
g_{1}g_{2}\right) \label{W-R-0}
\end{equation}
with some function $f$. It is easy to see that Eq. (\ref{f-Schroedinger})
leads to an ordinary differential equation for the function $f$. In order to
solve this equation it is convenient to change the variables:
\begin{equation}
\zeta=\left(  \frac{g_{1}}{g_{2}}\right)  ^{2}\,,\quad w=\ln\left(  g_{1}
g_{2}\right)  \,,\quad R=4f\,.\label{zeta-w-def}
\end{equation}
In terms of these variables representation (\ref{W-R-0}) becomes
\begin{equation}
W(g_{1},g_{2})=\frac{1}{4}R(\zeta)+\frac{1}{2}w\,.\label{W-R-def}
\end{equation}

A straightforward calculation of expressions (\ref{I-cal-def}) shows that
\begin{align}
\mathcal{J}_{1} &  =\frac{1}{4\sqrt{\zeta}}\left[  (1+\zeta)-(\zeta
-1)\frac{dR(\zeta)}{d\zeta}\right]  \,,\label{I-cal-1}\\
\mathcal{J}_{2} &  =\frac{i}{4\sqrt{\zeta}}\left[  (1-\zeta)+(\zeta
+1)\frac{dR(\zeta)}{d\zeta}\right]  \,,\\
\mathcal{J}_{3} &  =\frac{1}{2}\zeta\frac{\partial R(\zeta)}{\partial\zeta
}\,.\label{I-cal-3}
\end{align}
In the case (\ref{K-diagonal}), (\ref{B-12-0}) we find from Eq.
(\ref{f-Schroedinger})
\begin{equation}
\frac{1}{2}\sum\limits_{a=1}^{3}K_{a}\left(  \mathcal{J}_{a}\right)
^{2}+Y_{3}\mathcal{J}_{3}=E_{0}\,.
\end{equation}
Inserting Eqs. (\ref{I-cal-1})--(\ref{I-cal-3}), we obtain
\begin{gather}
\frac{K_{1}}{2\zeta}\left[  (1+\zeta)-(\zeta-1)\zeta\frac{dR(\zeta)}{d\zeta
}\right]  ^{2}-\frac{K_{2}}{2\zeta}\left[  (1-\zeta)+(\zeta+1)\zeta
\frac{dR(\zeta)}{d\zeta}\right]  ^{2}\nonumber\\
+2K_{3}\left[  \zeta\frac{dR(\zeta)}{d\zeta}\right]  ^{2}+8Y_{3}
\zeta\frac{dR(\zeta)}{d\zeta}=16E_{0}\,.\label{dr-d-zeta-eq}
\end{gather}

\subsection{Agreement with the Hartree equation}

\label{Hartree-agrement-section}

At the point $g_{n}=\phi_{n}^{\ast}$ (\ref{Hartree-point-eq}) corresponding to
the solution of the Hartree equation we must have according to Eq.
(\ref{W-R-0})
\begin{equation}
\phi_{n}=\frac{\partial}{\partial g_{n}}\left[  \frac{1}{4}R\left(  \left(
\frac{g_{1}}{g_{2}}\right)  ^{2}\right)  +\frac{1}{2}\ln\left(  g_{1}
g_{2}\right)  \,\right]  \,\,.
\end{equation}
Taking into account Eq. (\ref{phi-Hartree-rotator}), we see that this imposes
the following constraint on the behavior of $R(\zeta)$ at small $\zeta$
\begin{equation}
R(\zeta)\overset{\zeta\rightarrow0}{=}\mathrm{const}-\ln\zeta+\ldots
\end{equation}

Let us find the next term of this small $\zeta$ expansion. Applying Eq.
(\ref{W-g-near-Hartree}) to our case and taking $X_{11}$ from Eq.
(\ref{X11-res}), we obtain
\begin{equation}
W\left(  g\right)  =W(\phi)+\ln g_{2}+\frac{1}{2}\frac{\kappa^{1/2}
-\kappa^{-1/2}}{\kappa^{1/2}+\kappa^{-1/2}}\left(  \frac{g_{1}}{g_{2}}\right)
^{2}+O\left(  \left(  \frac{g_{1}}{g_{2}}\right)  ^{4}\right)  \,.
\end{equation}
According to Eq. (\ref{Re-W-phi-Hartree}) we have for normalizable states
(with the appropriately chosen phase)
\begin{equation}
W(\phi)=0\,.
\end{equation}
Therefore
\begin{equation}
W\left(  g\right)  =\ln g_{2}+\frac{1}{2}\frac{\kappa^{1/2}-\kappa^{-1/2}
}{\kappa^{1/2}+\kappa^{-1/2}}\left(  \frac{g_{1}}{g_{2}}\right)  ^{2}+O\left(
\left(  \frac{g_{1}}{g_{2}}\right)  ^{4}\right)  \,.
\end{equation}

Combining this with Eq. (\ref{W-R-def}), we find
\begin{equation}
R(\zeta)\overset{\zeta\rightarrow0}{=}-\ln\zeta+2\frac{\kappa^{1/2}
-\kappa^{-1/2}}{\kappa^{1/2}+\kappa^{-1/2}}\zeta+O\left(  \zeta^{2}\right)
\,.\label{R-small-zeta}
\end{equation}

\subsection{Calculation of $W(g)$}

Eq. (\ref{dr-d-zeta-eq}) is a quadratic equation with respect to $\partial
R/\partial z$ which can be easily solved. Using expression (\ref{E0-KB}) for
$E_{0}$, we find from Eq. (\ref{dr-d-zeta-eq})
\begin{equation}
\zeta\frac{\partial R}{\partial\zeta}=\frac{-\left[  \eta(1-\zeta^{2}
)+C\zeta\right]  \pm2\zeta\sqrt{\left(  \xi^{2}-\eta^{2}\right)  -\eta C\zeta
}}{\eta(1+\zeta^{2})-\left(  2\xi-C\right)  \zeta}\,.\label{zeta-d-zeta-R}
\end{equation}
Here we have introduced the compact notation
\begin{equation}
C=4Y_{3}\,,\label{C-via-Y3}
\end{equation}
\begin{align}
\xi &  =\left(  \kappa^{-1}+\kappa\right)  \omega=\frac{1}{2}(K_{1}
+K_{2})-K_{3}+2Y_{3}\,,\label{alpha-AB-kappa-omega}\\
\eta &  =\left(  \kappa^{-1}-\kappa\right)  \omega=\frac{1}{2}(K_{1}
-K_{2})\,.\label{beta-AB-kappa-omega}
\end{align}
In order to fix the sign uncertainty on the RHS of (\ref{zeta-d-zeta-R}), let
us consider the limit $\zeta\rightarrow0$
\begin{equation}
\zeta\frac{\partial R}{\partial\zeta}\overset{\zeta\rightarrow0}
{=}-1+\frac{2\zeta}{\eta}\left(  -\xi\pm\sqrt{\xi^{2}-\eta^{2}}\right)
=-1+2\zeta\frac{-\left(  \kappa^{-1}+\kappa\right)  \pm2}{\kappa^{-1}-\kappa
}\,,
\end{equation}
On the other hand, we find from Eq. (\ref{R-small-zeta})
\begin{equation}
\zeta\frac{\partial R}{\partial\zeta}\overset{\zeta\rightarrow0}
{=}-1+2\frac{\kappa^{1/2}-\kappa^{-1/2}}{\kappa^{1/2}+\kappa^{-1/2}}
\zeta+O\left(  \zeta^{3/2}\right)  \,.
\end{equation}
Comparing these two expressions, we conclude that we must have
\begin{equation}
\frac{\kappa^{1/2}-\kappa^{-1/2}}{\kappa^{1/2}+\kappa^{-1/2}}=\frac{-\left(
\kappa^{-1}+\kappa\right)  \pm2}{\kappa^{-1}-\kappa}\,.
\end{equation}
This identity holds if we choose the upper sign in the numerator of the RHS.
Now Eq. (\ref{zeta-d-zeta-R}) takes the form
\begin{equation}
\frac{\partial R}{\partial\zeta}=-\frac{1}{\zeta}+2\frac{\eta\zeta-\xi
+\sqrt{\xi^{2}-\eta^{2}-\eta C\zeta}}{\eta(1+\zeta^{2})-\left(  2\xi-C\right)
\zeta}\,.\label{dR-d-zeta-res}
\end{equation}
One can easily integrate this equation. The integration constant is fixed by
Eq. (\ref{R-small-zeta}). The calculation of the integral yields
\begin{equation}
R(\zeta)=-\ln\zeta+F(\zeta)-F(0)\,,\label{R-via-F}
\end{equation}
where
\begin{equation}
F(\zeta)=4\frac{\nu_{1}\ln\left[  u(\zeta)-\nu_{1}\right]  -\nu_{2}\ln\left[
u(\zeta)-\nu_{2}\right]  }{\nu_{1}-\nu_{2}}\,,
\end{equation}
\begin{equation}
\nu_{1,2}=\frac{-4Y_{3}\pm\sqrt{(K_{1}-K_{3})(K_{2}-K_{3})}}{|K_{1}-K_{2}|}\,,
\end{equation}
\begin{equation}
u(\zeta)=\frac{2\sqrt{\left(  K_{1}-K_{3}+2Y_{3}\right)  \left(  K_{2}
-K_{3}+2Y_{3}\right)  -2Y_{3}(K_{1}-K_{2})\zeta}}{|K_{1}-K_{2}|}\,.
\end{equation}
Now we insert Eqs. (\ref{R-via-F}) and (\ref{zeta-w-def}) into Eq.
(\ref{W-R-def})
\begin{equation}
W(g_{1},g_{2})=\ln g_{2}+\frac{1}{4}F\left(  \left(  \frac{g_{1}}{g_{2}
}\right)  ^{2}\right)  -\frac{1}{4}F(0)\,.\label{W-g-rotator-res}
\end{equation}

Thus we have computed function $W(g_{1},g_{2})$ which determines the
exponential part of the large-$N_{c}$ behavior (\ref{Phi-A-W}) for the system
with the Hamiltonian (\ref{H-J}). One should keep in mind that function
$F(\zeta)$ has branch points. At small $\zeta$ this function is regular.
According to Eq. (\ref{R-small-zeta})
\begin{equation}
F(\zeta)\overset{\zeta\rightarrow0}{=}F(0)+2\frac{\kappa^{1/2}-\kappa^{-1/2}
}{\kappa^{1/2}+\kappa^{-1/2}}\zeta+O\left(  \zeta^{2}\right)  \,.
\end{equation}
Therefore for small $|g_{1}/g_{2}|$ our result for $W(g_{1},g_{2})$ is
unambiguous. However, at larger $|g_{1}/g_{2}|$ one can meet cuts. In this
case the determination of the relevant branch of $W(g_{1},g_{2})$ will require
a special analysis.

\section{Models with the nontrivial vacuum}

\label{Nontrivial-vacuum-section}

\setcounter{equation}{0} 

\subsection{Color singlet states with $MN_{c}$ quarks}

\label{States-2Nc-section}

Our previous work was devoted to models of the baryon wave function based on
systems with the trivial vacuum. In such systems the baryon wave function
comes directly from the solution of the Schr\"{o}dinger equation for the
baryon (\ref{Schroedinger-baryon}). In other words, we worked with models
which have quarks but no antiquarks. Within this class of models we could
check the general properties of the large-$N_{c}$ limit like the exponential
large-$N_{c}$ behavior (\ref{Phi-A-W}) and the factorization of the
preexponential factors $A_{B}$ (\ref{A-factorization-1}).

Now we want to show that these general large-$N_{c}$ properties also hold in
more complicated models containing antiquark degrees of freedom in addition to
quarks. It is convenient to formulate these models in terms of the old Dirac
picture when both vacuum and baryons represented as a result of the occupation
of the bare vacuum with quarks. In these models the physical vacuum contains
$MN_{c}$ quarks whereas the baryon is made of $(M+1)N_{c}$ quarks where $M$ is
some integer number. This allows us to define the quark wave function of
baryons via the transition matrix element (\ref{bar-vac-transition}).

We want to start from the case $M=1$ in order to concentrate on the idea of
the method and to avoid unnecessary complications. The generalization for
arbitrary $M$ is discussed in Sec. \ref{Generalization-M-subsection} in terms
of the Schr\"{o}dinger approach. In Secs. \ref{Path-integral-section},
\ref{Equivalence-section} the case of arbitrary $M$ will be analyzed using the
path integral method. It should be stressed that the case $M=1$ from which we
want to start is rather nonphysical: the vacuum made of $N_{c}$ quarks is a
fermion for odd $N_{c}$ whereas the baryon made of $2N_{c}$ quarks is always a
boson. Nevertheless the $M=1$ case is a good starting point for understanding
the dynamics of models including quark-antiquark pairs.

\subsection{Color singlet states with $2N_{c}$ quarks}

The color singlet states with $2N_{c}$ quarks above the \emph{bare vacuum}
$|\Omega\rangle$ can be represented in the form
\begin{equation}
|B\rangle=\sum\limits_{i_{k}j_{k}}B_{j_{1}j_{2}\ldots j_{N_{c}}}^{i_{1}
i_{2}\ldots i_{N_{c}}}\prod\limits_{c=1}^{N_{c}}a_{i_{c}c}^{+}\prod
\limits_{c^{\prime}=1}^{N_{c}}a_{j_{c^{\prime}}c^{\prime}}^{+}|\Omega
\rangle\,.\label{Psi-F}
\end{equation}
Without any loss of generality we can assume that $B_{j_{1}j_{2}\ldots
j_{N_{c}}}^{i_{1}i_{2}\ldots i_{N_{c}}}$ is symmetric under permutations of
upper indices $i_{1}i_{2}\ldots i_{N_{c}}$ as well as under permutations of
lower indices $j_{1}j_{2}\ldots j_{N_{c}}$:
\begin{equation}
B_{j_{1}j_{2}\ldots j_{N_{c}}}^{i_{1}i_{2}\ldots i_{N_{c}}}=B_{\{j_{1}
j_{2}\ldots j_{N_{c}}\}}^{\{i_{1}i_{2}\ldots i_{N_{c}}\}}\,.
\end{equation}
The Fermi statistics of quarks
\begin{equation}
a_{i_{k}c}^{+}a_{j_{k}c^{\prime}}^{+}=-a_{j_{k}c^{\prime}}^{+}a_{i_{k}c}^{+}
\end{equation}
allows us to antisymmetrize $B_{j_{1}j_{2}\ldots j_{N_{c}}}^{i_{1}i_{2}\ldots
i_{N_{c}}}$ with respect to $i_{1}\leftrightarrow j_{1}$:
\begin{equation}
B_{j_{1}j_{2}\ldots j_{N_{c}}}^{i_{1}i_{2}\ldots i_{N_{c}}}\rightarrow
\frac{1}{2}\left(  B_{j_{1}j_{2}\ldots j_{N_{c}}}^{i_{1}i_{2}\ldots i_{N_{c}}
}-B_{i_{1}j_{2}\ldots j_{N_{c}}}^{j_{1}i_{2}\ldots i_{N_{c}}}\right)  \,.
\end{equation}
We can continue this process antisymmetrizing in all pairs $i_{k}
\leftrightarrow j_{k}$. After that the original symmetry in
$\{i_{1}i_{2}\ldots i_{N_{c}}\}$ and in $\{j_{1}j_{2}\ldots j_{N_{c}}\}$
is lost but
we still have the symmetry with respect to the permutations of any pair
$\{i_{m}j_{m}\}$ with another pair $\{i_{n}j_{n}\}$.

Note that our construction of tensors
$B_{j_{1}j_{2}\ldots j_{N_{c}}}^{i_{1}i_{2}\ldots i_{N_{c}}}$ based on

1) symmetrization in $\{i_{1}i_{2}\ldots i_{N_{c}}\}$,

2) symmetrization in $\{j_{1}j_{2}\ldots j_{N_{c}}\}$,

3) antisymmetrizations $[i_{1}j_{1}]$, 
$[i_{2}j_{2}]$,\ldots, $[i_{N_{c}}j_{N_{c}}]$

\noindent coincides with the construction of tensors associated with the
rectangular Young tableau containing two rows and $N_{c}$ columns.

Assuming that the tensor $B_{j_{1}j_{2}\ldots j_{N_{c}}}^{i_{1}i_{2}\ldots
i_{N_{c}}}$ obeys the above symmetry properties, let us construct the
``generating functional''
\begin{equation}
\tilde{\Phi}_{B}(\gamma)=\sum\limits_{i_{k}j_{k}}B_{j_{1}j_{2}\ldots j_{N_{c}
}}^{i_{1}i_{2}\ldots i_{N_{c}}}\gamma_{i_{1}j_{1}}\gamma_{i_{2}j_{2}}
\ldots\gamma_{i_{N_{c}}j_{N_{c}}}\,,\label{Xi-F}
\end{equation}
depending on the antisymmetric matrix ``source''
\begin{equation}
\gamma_{ij}=-\gamma_{ji}\,.\label{h-antisymmetric}
\end{equation}
This construction generalizes the case of states with $N_{c}$ quarks
(\ref{Phi-g-general-def}). We use the tilded notation $\tilde{\Phi}$ in order
to distinguish the functional $\tilde{\Phi}_{B}(\gamma)$ depending on the
antisymmetric tensors $\gamma_{ij}$ from the functional $\tilde{\Phi}_{B}(g) $
depending on ``vectors'' $g_{i}$. According to Eqs. (\ref{Psi-F}),
(\ref{Xi-F}) the functional $\tilde{\Phi}_{B}(\gamma)$ corresponds to the
transition matrix element between the baryon $|B\rangle$ and the \emph{bare}
vacuum $|\Omega\rangle$. On the other hand, the functional $\Phi_{B}(g)$ is
associated with the transition matrix element (\ref{Phi-transition-def})
between the baryon $|B\rangle$ and the \emph{physical} vacuum $|0\rangle$.

Now we have the correspondence between the states $|B\rangle$ (\ref{Psi-F})
and functions $\tilde{\Phi}_{B}(\gamma)$ (\ref{Xi-F}):
\begin{equation}
|B\rangle\rightarrow\tilde{\Phi}_{B}(\gamma)
\end{equation}
Within this correspondence operators $T_{mn}$ (\ref{T-via-a-0}) are mapped to
\begin{equation}
T_{mn}=\sum\limits_{c=1}^{N_{c}}a_{mc}^{+}a_{nc}\quad\rightarrow\quad
\sum\limits_{j}\gamma_{mj}\frac{\partial}{\partial\gamma_{nj}}
\,.\label{a-h-2-correspondence}
\end{equation}
According to Eq. (\ref{h-antisymmetric}) tensors $\gamma_{nj}$ are
antisymmetric. Therefore we must be careful about the normalization of the
derivative $\partial/\partial\gamma_{nj}$. Our choice is
\begin{equation}
\frac{\partial}{\partial\gamma_{pq}}\gamma_{mn}=\delta_{mp}\delta_{nq}
-\delta_{np}\delta_{mq}\,.\label{dh-def}
\end{equation}
Obviously we have
\begin{equation}
\sum\limits_{ij}\gamma_{ij}\frac{\partial}{\partial\gamma_{ij}}\tilde{\Phi
}_{B}(\gamma)=2N_{c}\tilde{\Phi}_{B}(\gamma)\,.\label{Phi-M2-homgen}
\end{equation}

Now we turn to the Schr\"{o}dinger equation. For simplicity we will assume
that
\begin{equation}
L_{mn}=0\label{L-0-sch-M2}
\end{equation}
in the Hamiltonian (\ref{H-general}). Using the $\gamma$ representation
(\ref{a-h-2-correspondence}) for this Hamiltonian, we can rewrite the
Schr\"{o}dinger equation
\begin{equation}
H\tilde{\Phi}_{B}(\gamma)=\mathcal{E}_{B}\tilde{\Phi}_{B}(\gamma)\,
\end{equation}
in the form
\begin{equation}
\frac{1}{2N_{c}}\sum\limits_{mnpq}V_{mnpq}\left(  \sum\limits_{j}\gamma
_{mj}\frac{\partial}{\partial\gamma_{nj}}\right)  \left(  \sum\limits_{k}
\gamma_{pk}\frac{\partial}{\partial\gamma_{qk}}\right)  \tilde{\Phi}
_{B}(\gamma)=\mathcal{E}_{B}\tilde{\Phi}_{B}(\gamma)\,.\label{dh-Schr-M2}
\end{equation}

\subsection{Large-$N_{c}$ limit}

At large $N_{c}$ we use the standard ansatz for the generating function
corresponding to the $2N_{c}$ quark state:
\begin{equation}
\tilde{\Phi}_{B}(\gamma)=N_{c}^{\nu_{B}}\tilde{A}_{B}(\gamma)\exp\left[
N_{c}W_{\mathrm{bar}}(\gamma)\right]  \,.\label{Phi-exp-M2}
\end{equation}
Inserting this ansatz into Eq. (\ref{Phi-M2-homgen}), we find
\begin{equation}
\sum\limits_{ij}\gamma_{ij}\frac{\partial}{\partial\gamma_{ij}}W_{\mathrm{bar}
}(\gamma)=2\,.\label{h-dh-k2}
\end{equation}
Combining the large-$N_{c}$ expression (\ref{Phi-exp-M2}) with the
Schr\"{o}dinger equation (\ref{dh-Schr-M2}) and using the $1/N_{c}$ expansion
for the energy
\begin{equation}
\mathcal{E}_{B}=N_{c}E_{0}+\Delta E_{B}+O(N_{c}^{-1})\,,
\end{equation}
we find in the leading order
\begin{equation}
\frac{1}{2}\sum\limits_{mnpq}V_{mnpq}\mathcal{T}_{mn}(\gamma)\mathcal{T}
_{pq}(\gamma)=E_{0}\,,\label{V-TT-E}
\end{equation}
where
\begin{equation}
\mathcal{T}_{mn}(\gamma)\equiv\sum\limits_{j}\gamma_{mj}\frac{\partial
W_{\mathrm{bar}}(\gamma)}{\partial\gamma_{nj}}\,.
\end{equation}
In the next-to-leading order we obtain
\begin{equation}
\sum\limits_{mnpq}\left[  V_{mnpq}\mathcal{T}_{mn}(\gamma)\sum\limits_{k}
\gamma_{pk}\frac{\partial\ln\tilde{A}_{B}(\gamma)}{\partial\gamma_{qk}
}+\frac{1}{2}V_{mnpq}\sum\limits_{j}\gamma_{mj}\frac{\partial\mathcal{T}
_{pq}(\gamma)}{\partial\gamma_{nj}}\right]  =\Delta E_{B}\,.\label{A-h2}
\end{equation}

\subsection{Next-to-leading order}

In the NLO we have equation (\ref{A-h2}). The structure of this equation
allows for the solutions of the form
\begin{equation}
\tilde{A}_{B}(\gamma)=\tilde{A}^{(0)}(\gamma)\prod\limits_{\alpha}\left[
\xi_{\alpha}(\gamma)\right]  ^{n_{\alpha}}\,,\label{A-factorization}
\end{equation}
\begin{equation}
\Delta E_{B}=\Delta E_{0}+\sum\limits_{\alpha}n_{\alpha}\Delta E_{\alpha}\,,
\end{equation}
where $\tilde{A}^{(0)}(\gamma)$ obeys the inhomogeneous equation (\ref{A-h2})
\begin{equation}
\sum\limits_{mnpq}\left[  V_{mnpq}\mathcal{T}_{mn}(\gamma)\sum\limits_{k}
\gamma_{pk}\frac{\partial\ln\tilde{A}^{(0)}(\gamma)}{\partial\gamma_{qk}
}+\frac{1}{2}V_{mnpq}\sum\limits_{j}\gamma_{mj}\frac{\partial\mathcal{T}
_{pq}(\gamma)}{\partial\gamma_{nj}}\right]  =\Delta E_{0}\,.
\end{equation}
and $\xi_{k}(\gamma)$ are solutions of the homogeneous equation
\begin{equation}
\sum\limits_{mnpq}V_{mnpq}\mathcal{T}_{mn}(\gamma)\left[  \sum\limits_{k}
\gamma_{pk}\frac{\partial\ln\xi_{\alpha}(\gamma)}{\partial\gamma_{qk}}\right]
=\Delta E_{\alpha}\,.
\end{equation}

\subsection{Overlap matrix elements}

\label{Overlap-section}

Now let us consider two baryon states $|B_{1}\rangle$ and $|B_{2}\rangle$. Let
us calculate the overlap matrix element for these two states
\begin{align}
\langle B_{1}|B_{2}\rangle &  =2^{N_{c}}\sum\limits_{i_{k}j_{k}}\left[
\left(  B_{1}\right)  _{j_{1}j_{2}\ldots j_{N_{c}}}^{i_{1}i_{2}\ldots
i_{N_{c}}}\right]  ^{\ast}\left(  B_{2}\right)  _{j_{1}j_{2}\ldots j_{N_{c}}
}^{i_{1}i_{2}\ldots i_{N_{c}}}\nonumber\\
&  =\frac{\left(  2N_{c}\right)  ^{N_{c}}}{N_{c}!}\frac{\int d\gamma
d\gamma^{\ast}\exp\left(  -N_{c}\sum\limits_{ij}\gamma_{ij}\gamma_{ij}^{\ast
}\right)  \left[  \tilde{\Phi}_{B_{1}}(\gamma)\right]  ^{\ast}\left[
\tilde{\Phi}_{B_{2}}(\gamma)\right]  }{\int d\gamma d\gamma^{\ast}\exp\left(
-N_{c}\sum\limits_{ij}\gamma_{ij}\gamma_{ij}^{\ast}\right)  }\nonumber\\
&  \rightarrow\frac{\left(  2N_{c}\right)  ^{N_{c}}}{N_{c}!}\frac{\int d\gamma
d\gamma^{\ast}\exp\left\{  -N_{c}\left\{  \sum\limits_{ij}\gamma_{ij}
\gamma_{ij}^{\ast}+\left[  W_{\mathrm{bar}}(\gamma)\right]  ^{\ast
}+W_{\mathrm{bar}}(\gamma)\right\}  \right\}  \,\left[  \tilde{A}_{B_{1}
}(\gamma)\right]  ^{\ast}\left[  \tilde{A}_{B_{2}}(\gamma)\right]  }{\int
d\gamma d\gamma^{\ast}\exp\left(  -N_{c}\sum\limits_{ij}\gamma_{ij}\gamma
_{ij}^{\ast}\right)  }\,.
\end{align}
At large $N_{c}$ the integral over $\gamma$ can be computed using the saddle
method. The saddle point equation is
\begin{equation}
2\gamma_{ij}^{\ast}=-\left.  \frac{\partial W_{\mathrm{bar}}(\gamma)}
{\partial\gamma_{ij}}\right|  _{\gamma=\gamma^{(0)}}\,.\label{h-SP-1}
\end{equation}
We use notation $\gamma=\gamma^{(0)}$ for the solution of this equation.
Combining Eqs. (\ref{h-SP-1}) and (\ref{h-dh-k2}), we find
\begin{equation}
\sum\limits_{ij}\gamma_{ij}^{(0)}\gamma_{ij}^{(0)\ast}=1\,.
\end{equation}

\subsection{From the equation for $W_{\mathrm{bar}}(\gamma)$ to the Hartree equation}

Eq. (\ref{V-TT-E}) for $W_{\mathrm{bar}}(\gamma)$ is valid for any $\gamma$.
If we take this equation at the saddle point $\gamma^{(0)}$ (\ref{h-SP-1}),
then we arrive at the equation
\begin{equation}
E_{0}=\sum\limits_{mnpq}2V_{mnpq}\left(  \sum\limits_{j}\gamma_{mj}
^{(0)}\gamma_{nj}^{(0)\ast}\right)  \left(  \sum\limits_{k}\gamma_{pk}
^{(0)}\gamma_{qk}^{(0)\ast}\right)  \,.\label{Pre-Hartree-M2}
\end{equation}
This is nothing else but the expression for the Hartree energy
(\ref{E-0-general}) corresponding to the case $M=2$, $L_{mn}=0$:
\begin{equation}
E_{0}=\frac{1}{2}\sum\limits_{r,s=1}^{M}\sum\limits_{n_{1}n_{2}n_{3}n_{4}
}V_{n_{1}n_{2}n_{3}n_{4}}\phi_{n_{1}}^{r\ast}\phi_{n_{2}}^{r}\phi_{n_{3}
}^{s\ast}\phi_{n_{4}}^{s}\,.\label{E0-Hartree-M2}
\end{equation}
Indeed, taking
\begin{equation}
\gamma_{mn}^{(0)}=\frac{1}{\sqrt{2}}\sum\limits_{r,s=1}^{2}\varepsilon
_{rs}\left(  \phi_{m}^{r}\phi_{n}^{s}\right)  ^{\ast}\,\quad(\varepsilon
_{12}=-\varepsilon_{21}=1),
\end{equation}
we find
\begin{equation}
\sum\limits_{j}\gamma_{mj}^{(0)}\gamma_{nj}^{(0)\ast}=\frac{1}{2}
\sum\limits_{s}\phi_{m}^{s\ast}\phi_{n}^{s}\,.
\end{equation}
Inserting this expression into Eq. (\ref{Pre-Hartree-M2}), we reproduce the
Hartree expression for the energy (\ref{E0-Hartree-M2}).

\subsection{Saddle point equation for the baryon wave function}

We have explained above how to solve the Schr\"{o}dinger equation for the
baryon state $|B\rangle$ (\ref{Psi-F}) described by the generating function
$\tilde{\Phi}_{B}(\gamma)$ (\ref{Xi-F}). Now let us turn to the vacuum. In our
model the \emph{physical} vacuum $|0\rangle$ state contains $N_{c}$ quarks put
into the \emph{bare} vacuum $|\Omega\rangle$. This means that the physical
vacuum of this model is described by the same equations as the baryon in
simpler models with the trivial vacuum which were studied earlier. Therefore
we can apply the old baryon representation (\ref{Phi-g-general-def}) to our
new vacuum:
\begin{align}
|0\rangle &  =\sum\limits_{i_{1}\ldots i_{N_{c}}}\psi_{\mathrm{vac}}
^{i_{1}\ldots i_{N_{c}}}\prod\limits_{c=1}^{N_{c}}a_{i_{c}c}^{+}|\Omega
\rangle\,,\\
\Phi_{\mathrm{vac}}(k) &  =\sum\limits_{i_{1}\ldots i_{N_{c}}}\psi
_{\mathrm{vac}}^{i_{1}\ldots i_{N_{c}}}k_{i_{1}}\ldots k_{i_{N_{c}}}\,.
\end{align}
Here we use notation $k$ for the ``source'' argument of the function
$\Phi_{\mathrm{vac}}$ instead of $g$ as it was in Eq. (\ref{Phi-g-general-def}).

Now we want to compute the matrix element
\begin{gather}
\langle0|\prod\limits_{c}\left(  \sum\limits_{j}g_{j}a_{jc}\right)
|B\rangle=2^{N_{c}}\sum\limits_{i_{m}j_{m}}\psi_{\mathrm{vac}}^{i_{1}\ldots
i_{N_{c}}}g_{j_{1}}\ldots g_{j_{N_{c}}}\left(  B_{\mathrm{bar}}\right)
_{j_{1}\ldots j_{N_{c}}}^{i_{1}\ldots i_{N_{c}}}\nonumber\\
=\frac{1}{\left(  N_{c}!\right)  ^{2}}\left.  \left(  \sum\limits_{ij}
g_{j}\frac{\partial}{\partial k_{i}^{\ast}}\frac{\partial}{\partial\gamma
_{ij}}\right)  ^{N_{c}}\left[  \Phi_{\mathrm{vac}}(k)\right]  ^{\ast}
\tilde{\Phi}_{B}(\gamma)\right|  _{k=\gamma=0}\nonumber\\
=\frac{N_{c}^{N_{c}}}{N_{c}!}\left.  \left[  \exp\left(  \sum\limits_{ij}
\frac{1}{N_{c}}g_{j}\frac{\partial}{\partial k_{i}^{\ast}}\frac{\partial
}{\partial\gamma_{ij}}\right)  \right]  \left[  \Phi_{\mathrm{vac}}(k)\right]
^{\ast}\tilde{\Phi}_{B}(\gamma)\right|  _{k=\gamma=0}\nonumber\\
=\frac{N_{c}^{N_{c}}}{N_{c}!}\left[  \int dQdQ^{\ast}\exp\left(  -\frac{N_{c}
}{2}\sum\limits_{ij}Q_{ij}Q_{ij}^{\ast}\right)  \right]  ^{-1}\int dQdQ^{\ast
}\exp\left(  -\frac{N_{c}}{2}\sum\limits_{ij}Q_{ij}Q_{ij}^{\ast}\right)
\nonumber\\
\times\left.  \left[  \exp\sum\limits_{ij}\left(  Q_{ij}^{\ast}g_{j}
\frac{\partial}{\partial k_{i}^{\ast}}+\frac{1}{2}Q_{ij}\frac{\partial
}{\partial\gamma_{ij}}\right)  \right]  \left[  \Phi_{\mathrm{vac}}(k)\right]
^{\ast}\tilde{\Phi}_{B}(\gamma)\right|  _{k=\gamma=0}\nonumber\\
=\frac{N_{c}^{N_{c}}}{N_{c}!}\frac{\int dQdQ^{\ast}\exp\left(  -\frac{N_{c}
}{2}\sum\limits_{ij}Q_{ij}Q_{ij}^{\ast}\right)  \left[  \Phi_{\mathrm{vac}
}(Qg^{\ast})\right]  ^{\ast}\tilde{\Phi}_{B}(Q)}{\int dQdQ^{\ast}\exp\left(
-\frac{N_{c}}{2}\sum\limits_{ij}Q_{ij}Q_{ij}^{\ast}\right)  }\,.
\end{gather}
Here
\begin{align}
\Phi_{\mathrm{vac}}(k) &  =N_{c}^{\nu_{\mathrm{vac}}}A_{\mathrm{vac}}
(k)\exp\left[  N_{c}W_{\mathrm{vac}}(k)\right]  \,,\\
\tilde{\Phi}_{B}(\gamma) &  =N_{c}^{\tilde{\nu}_{B}}\tilde{A}_{B}(\gamma
)\exp\left[  N_{c}W_{\mathrm{bar}}(\gamma)\right]  \,.
\end{align}
Note that the universal functional $W_{\mathrm{bar}}(\gamma)$ describes the
large $N_{c}$ behavior of $\tilde{\Phi}_{B}(\gamma)$ for all low-lying baryons
$B$. In the same way the functional $W_{\mathrm{vac}}(k)$ describes not only
the exponential behavior of the vacuum $\Phi_{\mathrm{vac}}(k)$ but also the
``meson'' functionals $\Phi_{\mathrm{mes}}(\gamma)$ corresponding to the
$O(N_{c}^{0})$ excitations above the vacuum.

Thus at large $N_{c}$ we have
\begin{gather}
\langle0|\prod\limits_{c=1}^{N_{c}}\left(  \sum\limits_{j}g_{j}a_{jc}\right)
|B\rangle=\frac{e^{N_{c}}}{\sqrt{2\pi N_{c}}}N_{c}^{\nu_{\mathrm{vac}}
+\tilde{\nu}_{B}}\nonumber\\
\times\frac{\int dQdQ^{\ast}\left[  A_{\mathrm{vac}}(Qg^{\ast})\right]
\tilde{A}_{B}(Q)\exp N_{c}\left\{  \left[  W_{\mathrm{vac}}(Qg^{\ast})\right]
^{\ast}+W_{\mathrm{bar}}(Q)-\frac{1}{2}\sum\limits_{ij}Q_{ij}Q_{ij}^{\ast
}\right\}  }{\int dQdQ^{\ast}\exp\left(  -\frac{N_{c}}{2}\sum\limits_{ij}
Q_{ij}Q_{ij}^{\ast}\right)  }\,.\label{M2-ME-SP}
\end{gather}
At large $N_{c}$ the integral over $Q,Q^{\ast}$ can be computed using the
saddle point method. The deformation of the integration contour in the saddle
point method can lead to the violation of the complex conjugation relation
between $Q_{ij}$ and $Q_{ij}^{\ast}$. Therefore we introduce an independent
notation for $Q_{ij}^{\ast}$
\begin{equation}
Q_{ij}^{\ast}\rightarrow u_{ij}\,.
\end{equation}
The saddle point equations are
\begin{align}
\frac{\partial}{\partial u_{ij}}\left\{  \left[  W_{\mathrm{vac}}(u^{\ast
}g^{\ast})\right]  ^{\ast}+W_{\mathrm{bar}}(Q)-\frac{1}{2}\sum\limits_{mn}
Q_{mn}u_{mn}\right\}   &  =0\,,\label{dr-SP-1}\\
\frac{\partial}{\partial Q_{ij}}\left\{  \left[  W_{\mathrm{vac}}(u^{\ast
}g^{\ast})\right]  ^{\ast}+W_{\mathrm{bar}}(Q)-\frac{1}{2}\sum\limits_{mn}
Q_{mn}u_{mn}\right\}   &  =0\,.\label{dr-SP-2}
\end{align}

Now we obtain from Eqs. (\ref{dr-SP-1}), (\ref{dr-SP-2})
\begin{align}
k_{i}  &  =u_{ij}^{\ast}g_{j}^{\ast}\,,\label{k-2-WF-SP-1}\\
g_{j}\left[  \frac{\partial W_{\mathrm{vac}}(k)}{\partial k_{i}}\right]
^{\ast}-g_{i}\left[  \frac{\partial W_{\mathrm{vac}}(k)}{\partial k_{j}
}\right]  ^{\ast}  &  =Q_{ij}\,,\label{k-2-WF-SP-2}\\
\frac{\partial W_{\mathrm{bar}}(Q)}{\partial Q_{ij}}  &  =u_{ij}
\,.\label{k-2-WF-SP-3}
\end{align}
These equations determine
\begin{equation}
Q_{ij}=Q_{ij}(g)\,,\quad u_{ij}=u_{ij}(g)\,.
\end{equation}
Combining Eqs. (\ref{k-2-WF-SP-3}) and (\ref{h-dh-k2}), we find
\begin{equation}
\sum\limits_{ij}Q_{ij}(g)u_{ij}(g)=2\,.\label{r-q-one}
\end{equation}

\subsection{Functionals $W(g)$ and $A_{B}(g)$ from the saddle point method}

Applying the saddle-point method to Eq. (\ref{M2-ME-SP}), we find
\begin{gather}
\langle0|\prod\limits_{c=1}^{N_{c}}\left(  \sum\limits_{j}g_{j}a_{jc}\right)
|B\rangle=N_{c}^{\nu_{B}}J(g)\left[  A_{\mathrm{vac}}(u^{\ast}g^{\ast
})\right]  ^{\ast}\tilde{A}_{B}(Q)\nonumber\\
\times\exp N_{c}\left\{  \left[  W_{\mathrm{vac}}(u^{\ast}g^{\ast})\right]
^{\ast}+W_{\mathrm{bar}}(Q_{ij})\right\}  _{Q=Q(g),u=u(g)}\,.
\end{gather}
Here $J(g)$ stands from the contribution of the Jacobian corresponding to the
fluctuations around the saddle point and $N_{c}^{\nu_{B}}$ accumulates all
powers of $N_{c}$ coming from various sources. We can rewrite the above
representation in the form
\begin{equation}
\langle0|\prod\limits_{c=1}^{N_{c}}\left(  \sum\limits_{j}g_{j}a_{jc}\right)
|B\rangle=N_{c}^{\nu_{B}}A_{B}(g)\exp\left[  N_{c}W(g)\right]  \,,
\end{equation}
where
\begin{equation}
W(g)=\left\{  \left[  W_{\mathrm{vac}}(u^{\ast}g^{\ast})\right]  ^{\ast
}+W_{\mathrm{bar}}(Q_{ij})\right\}  _{Q=Q(g),u=u(g)}
\,,\label{W-eff-W-bar-W-vac}
\end{equation}
\begin{equation}
A_{B}(g)=\left.  J(g)\left[  A_{\mathrm{vac}}(u^{\ast}g^{\ast})\right]
^{\ast}\tilde{A}_{B}(Q)\right|  _{Q=Q(g),u=u(g)}\,.
\end{equation}

Using the factorization of $\tilde{A}_{B}(Q_{ij})$ (\ref{A-factorization}), we
arrive at the factorized form of $A_{B}(g)$
\begin{equation}
A_{B}(g)=A^{(0)}(g)\prod\limits_{k}A_{k}(g)\label{A-factorization-model}
\end{equation}
with
\begin{align}
A_{k}(g)  & =\xi_{k}\left(  Q(g)\right)  \,,\\
A^{(0)}(g)  & =J(g)\tilde{A}^{(0)}\left(  Q(g)\right)  \left\{
A_{\mathrm{vac}}\left(  \left[  u(g)g\right]  ^{\ast}\right)  \right\}
^{\ast}\,.
\end{align}

Our result (\ref{A-factorization-model}) shows the mechanism of the general
factorization (\ref{A-factorization-1}) of functionals $A_{B}(g)$ in models
with the nontrivial vacuum.

\subsection{Representations for $W(g)$ in terms of trajectories}

\label{W-Sch-trajectories-section}

In our current model, the physical vacuum state contains $N_{c}$ quarks
occupying the bare vacuum. This means our physical \emph{vacuum} can be
described by equations derived in Sec. \ref{Classical-dynamics-section} for
\emph{baryons}. Thus the functional $W_{\mathrm{vac}}$ is given by Eqs.
(\ref{S-int-2}) and (\ref{W-S-id}). In the special case $L_{mn}=0$ with which
we are dealing now [See Eq. (\ref{L-0-sch-M2})], the result for
$W_{\mathrm{vac}}$ is especially simple and is given by Eq. (\ref{W-ln-I-0}):
\begin{equation}
W_{\mathrm{vac}}(k)=\ln I_{\mathrm{vac}}(k)\quad\left(  \mathrm{if}
\,L_{mn}=0\right)  \,,\label{W-vac-L-0}
\end{equation}
where $I_{\mathrm{vac}}(g)$ is determined by the trajectories obeying
asymptotic conditions (\ref{q-I}), (\ref{p-I}) and (\ref{q-0-g})
\begin{gather}
q_{n}^{\mathrm{vac}}(t)\overset{t\rightarrow-\infty}{=}\delta_{n0}
I_{\mathrm{vac}}(k)\exp\left(  \varepsilon_{\mathrm{vac}}^{1}t\right)
,\label{vac-boundary-1}\\
p_{n}^{\mathrm{vac}}(t)\overset{t\rightarrow-\infty}{=}\frac{\delta_{n0}
}{I_{\mathrm{vac}}(k)}\exp\left(  -\varepsilon_{\mathrm{vac}}^{1}t\right)
\,,\label{vac-boundary-2}\\
q_{n}^{\mathrm{vac}}(0)=k_{n}\,.\label{vac-boundary-3}
\end{gather}
These trajectories obey condition (\ref{pq-one}):
\begin{equation}
\sum\limits_{n}p_{n}^{\mathrm{vac}}(t)q_{n}^{\mathrm{vac}}
(t)=1\,.\label{pq-vac-1}
\end{equation}
According to Eqs. (\ref{pq-t-0}), (\ref{dS-dq0}) and (\ref{W-S-id}) we have
\begin{equation}
p_{n}^{\mathrm{vac}}(0)=\frac{\partial W_{\mathrm{vac}}(k)}{\partial k_{n}
}\,.\label{p-d-W-k}
\end{equation}

One can generalize the work done in Sec. \ref{Classical-dynamics-section} for
the case of states containing $N_{c}$ quarks above the bare vacuum and derive
a similar representation for $W_{\mathrm{bar}}(g)$:
\begin{equation}
W_{\mathrm{bar}}(Q)=\ln I_{\mathrm{bar}}(Q)\quad\left(  \mathrm{if}
\,L_{mn}=0\right)  \,.\label{W-bar-L-0}
\end{equation}
Here $Q_{ij}$ is the boundary $t=0$ value of some trajectory 
$Q_{ij}^{\mathrm{bar}}(t)$ in the space of antisymmetric tensors
\begin{equation}
Q_{ij}^{\mathrm{bar}}(0)=Q_{ij}\,.\label{bar-boundary-0}
\end{equation}
and the parameter $I_{\mathrm{bar}}(Q)$ is determined by the $t\rightarrow
-\infty$ asymptotic behavior of these trajectories
\begin{gather}
Q_{ij}^{\mathrm{bar}}(t)\overset{t\rightarrow-\infty}{=}\varepsilon
_{ij}I_{\mathrm{bar}}(Q)\exp\left[  \left(  \varepsilon_{\mathrm{bar}}
^{1}+\varepsilon_{\mathrm{bar}}^{2}\right)  t\right]
\,,\label{bar-boundary-1}\\
P_{ij}^{\mathrm{bar}}(t)\overset{t\rightarrow-\infty}{=}\frac{\varepsilon
_{ij}}{2I_{\mathrm{bar}}(Q)}\exp\left[  -\left(  \varepsilon_{\mathrm{bar}
}^{1}+\varepsilon_{\mathrm{bar}}^{2}\right)  t\right]
\,.\label{bar-boundary-2}
\end{gather}
Here $\varepsilon_{ij}$ is the antisymmetric tensor
\begin{equation}
\varepsilon_{ij}=\left\{
\begin{array}
[c]{ll}
1 & \mathrm{if}\,i=1,j=2,\\
-1 & \mathrm{if}\,i=2,j=1,\\
0 & \,\mathrm{otherwise}\,.
\end{array}
\right.
\end{equation}
We assume that the values $i=1,2$ correspond to the two occupied
single-particle levels of the baryon state in the Hartree picture. The
Hamilton equation describing the dynamics in the space of antisymmetric
tensors $Q_{ij}^{\mathrm{bar}}$, $P_{ij}^{\mathrm{bar}}$ will be discussed in
Sec. \ref{Equivalence-section}. By analogy with Eq. (\ref{p-d-W-k}) we have
\begin{equation}
P_{ij}^{\mathrm{bar}}(0)=\frac{1}{2}\frac{\partial W_{\mathrm{bar}}
(Q)}{\partial Q_{ij}}\,.\label{p-d-W-r}
\end{equation}
The analog of Eq. (\ref{pq-vac-1}) is
\begin{equation}
\sum\limits_{ij}P_{ij}^{\mathrm{bar}}(t)Q_{ij}^{\mathrm{bar}}
(t)=1\,.\label{PQ-2}
\end{equation}
The constant on the RHS of Eq. (\ref{p-d-W-r}) is fixed by relations
(\ref{h-dh-k2}).

The above representations (\ref{W-vac-L-0}) for $W_{\mathrm{vac}}(k)$ and
(\ref{W-bar-L-0}) for $W_{\mathrm{bar}}(Q)$ are independent of each other.
However, in the expression (\ref{W-eff-W-bar-W-vac}) for the functional $W(g)
$
\begin{equation}
W_{\mathrm{Sch}}(g)=\left\{  W_{\mathrm{vac}}\left[  k(g^{\ast})\right]
\right\}  ^{\ast}+W_{\mathrm{bar}}\left[  Q(g)\right]
\label{W-eff-W-bar-W-vac-2}
\end{equation}
the variables $Q$ and $k$ on the RHS become $g$ ($g^{\ast}$) dependent. We use
the label \emph{Sch} in $W_{\mathrm{Sch}}(g)$ in order to stress that we are
dealing with the Schr\"{o}dinger approach. Later we will show how the
functional $W(g)$ can be calculated in the path integral approach [using
notation $W_{\mathrm{pi}}(g)$]. Certainly the result of the calculation should
not depend on the method but in order to avoid confusion at the intermediate
steps we prefer to use different notations for the two different methods.

The dependence of $k$ and $Q$ on $g$ is described by equations
(\ref{k-2-WF-SP-1})--(\ref{k-2-WF-SP-3}):
\begin{gather}
\sum\limits_{j}g_{j}\frac{\partial W_{\mathrm{bar}}(Q)}{\partial Q_{ij}}
=k_{i}^{\ast}\,,\label{bar-vac-1}\\
g_{j}\left[  \frac{\partial W_{\mathrm{vac}}(k)}{\partial k_{i}}\right]
^{\ast}-g_{i}\left[  \frac{\partial W_{\mathrm{vac}}(k)}{\partial k_{j}
}\right]  ^{\ast}=Q_{ij}\,.\label{bar-vac-2}
\end{gather}
Thus the problem of the calculation of the functional $W_{\mathrm{Sch}}(g)$
reduces (in the case $L_{mn}=0$) to search for trajectories obeying boundary
conditions (\ref{vac-boundary-1})--(\ref{vac-boundary-3}),
(\ref{bar-boundary-0})--(\ref{bar-boundary-2}), (\ref{bar-vac-1}),
(\ref{bar-vac-2}). The asymptotic behavior of these trajectories at
$t\rightarrow\infty$ determines parameters $I_{\mathrm{vac}}(k)$ and
$I_{\mathrm{bar}}(Q) $. According to Eqs. (\ref{W-vac-L-0}), (\ref{W-bar-L-0})
and (\ref{W-eff-W-bar-W-vac-2}) we have
\begin{equation}
W_{\mathrm{Sch}}(g)=\left\{  \ln I_{\mathrm{vac}}\left[  k(g^{\ast})\right]
\right\}  ^{\ast}+\ln I_{\mathrm{bar}}\left[  Q(g)\right]  \quad\left(
\mathrm{if}\,L_{mn}=0\right)  \,.\label{W-sch-I-vac-bar}
\end{equation}

\subsection{Generalization for models with $M>1$}

\label{Generalization-M-subsection}

The previous analysis was devoted to ``baryon'' states containing $2N_{c}$
quarks above the bare vacuum. These states were described by functions
$\tilde{\Phi}_{B}(\gamma)$ depending on antisymmetric tensors $\gamma_{ij}$.
The generalization to the case of states with $MN_{c}$ quarks is
straightforward. In this case we must work with wave functions depending on
antisymmetric tensors $\gamma_{\lbrack i_{1}i_{2}\ldots i_{M}]}$ of rank $k$.
The analog of Eq. (\ref{a-h-2-correspondence}) is
\begin{equation}
T_{mn}=\sum\limits_{c=1}^{N_{c}}a_{mc}^{+}a_{nc}\rightarrow\frac{1}
{(M-1)!}\sum\limits_{i_{2}i_{3}\ldots i_{M}}\gamma_{mi_{2}i_{3}\ldots i_{M}
}\frac{\partial}{\partial\gamma_{ni_{2}i_{3}\ldots i_{M}}}
\,,\label{T-mn-M-general}
\end{equation}
where the derivative is normalized by the condition
\begin{equation}
\frac{\partial}{\partial\gamma_{i_{1}i_{2}i_{3}\ldots i_{M}}}\gamma
_{j_{1}j_{2}j_{3}\ldots j_{M}}=\det_{ab}\left\|  \delta_{i_{a}j_{b}}\right\|
\,.
\end{equation}

\section{Path integral approach}

\label{Path-integral-section}

\setcounter{equation}{0} 

\subsection{Advantage of the path integral approach}

Our previous analysis of the large-$N_{c}$ models was based on the operator
approach with the stationary Schr\"{o}dinger equation
(\ref{Schroedinger-baryon}) as a starting point. Now we want to study
large-$N_{c} $ models using the path integral method. In principle, both
Schr\"{o}dinger and path integral approaches must lead to the same results. In
Sec. \ref{Equivalence-section} we will demonstrate the equivalence of the
results derived using these two methods. However, in spite of this equivalence
we will see that the path integral method allows us to obtain in a
straightforward way rather interesting results which are not so obvious in the
Schr\"{o}dinger approach. The reason is that the path integral approach is
based on time-dependent trajectories. On the other hand, as we know from Sec.
\ref{Classical-dynamics-section}, in the Schr\"{o}dinger approach the
trajectories appear rather indirectly. In the Schr\"{o}dinger approach one
first derives the Hamilton--Jacobi equation for the ``action'' $W(g)$ and only
after that $W(g)$ can be interpreted in terms of trajectories. In the case of
models with the nontrivial vacuum the situation is even more involved, since
the trajectories appearing in the Schr\"{o}dinger approach belong to the phase
space of antisymmetric tensors. On the contrary, the classical dynamics coming
from the path integral approach can be formulated in simpler terms.

Since the path integral approach allows for a straightforward derivation of
the representation for the functional $W(g)$ in terms of simple classical
dynamics, we prefer to turn to the path integral formalism in this section.
The equivalence of the results obtained in this section with the results based
on the Schr\"{o}dinger equation will be established in Sec.
\ref{Equivalence-section}.

\subsection{Model}

We want to apply the path integral approach to models described by the
Hamiltonian (\ref{H-general}). For simplicity we will consider the case when
$L_{n_{1}n_{2}}=0$ in this Hamiltonian. For our aims it is convenient to
rewrite the Hamiltonian in the form
\begin{equation}
H=\frac{1}{2N_{c}}\sum\limits_{\alpha\beta}\left(  a^{+}\Gamma_{\alpha
}a\right)  V_{\alpha\beta}\left(  a^{+}\Gamma_{\beta}a\right)
\,,\label{H-PI-4-q}
\end{equation}
where
\begin{equation}
\left(  a^{+}\Gamma_{\alpha}a\right)  =\sum\limits_{c=1}^{N_{c}}
\sum\limits_{mn}a_{cm}^{+}\Gamma_{\alpha}^{mn}a_{cn}\,.
\end{equation}
The coefficients $V_{\alpha\beta}$ and $\Gamma_{\alpha}^{mn}$ are assumed to
obey the conditions
\begin{align}
V_{\alpha\beta}  &  =V_{\alpha\beta}^{\ast}\,,\quad V_{\alpha\beta}
=V_{\beta\alpha}\,,\label{V-h}\\
\left(  \Gamma_{\alpha}^{mn}\right)  ^{\ast}  &  =\Gamma_{\alpha}
^{nm}\,.\label{Gamma-h}
\end{align}

The expression (\ref{H-PI-4-q}) for the Hamiltonian corresponds to the choice
\begin{equation}
V_{n_{1}n_{2}n_{3}n_{4}}=\sum\limits_{\alpha}\Gamma_{\alpha}^{n_{1}n_{2}
}\Gamma_{\alpha}^{n_{3}n_{4}}
\end{equation}
in Eq. (\ref{H-general}) .

In the path integral approach this system is described by the action
\begin{equation}
S_{E}(a)=-\int dt\left[  \left(  a^{+}\partial_{t}a\right)  +\frac{1}{2N_{c}
}\sum\limits_{\alpha\beta}\left(  a^{+}\Gamma_{\alpha}a\right)  V_{\alpha
\beta}\left(  a^{+}\Gamma_{\beta}a\right)  \right]  \,.
\end{equation}
We use the same notation $a^{+},a$ for the Grassmann integration variables as
for the operators in Eq. (\ref{H-PI-4-q}). Strictly speaking, the transition
from the operator formalism to the path integral has to be accompanied by a
careful treatment of the problems of the operator ordering. However, in the
leading order of the $1/N_{c}$ expansion these subtleties are not important.

One can ``bosonize'' the theory introducing the path integral over the
auxiliary boson variable $\pi(t)$:
\begin{equation}
\exp S_{E}(a)=\frac{\int D\pi\exp\left\{  -\left[  a^{+}K(\pi)a\right]
+N_{c}S_{\mathrm{bos}}(\pi)\right\}  }{\int D\pi\exp\left[  N_{c}
S_{\mathrm{bos}}(\pi)\right]  }\,.\label{pi-bosonization}
\end{equation}
Here
\begin{equation}
S_{\mathrm{bos}}(\pi)=\int dtL_{\mathrm{bos}}(\pi)\,,\label{S-bos-pi}
\end{equation}
\begin{equation}
L_{\mathrm{bos}}(\pi)=\sum\limits_{\alpha\beta}\frac{1}{2}\pi_{\alpha}\left(
V^{-1}\right)  _{\alpha\beta}\pi_{\beta}\,,\label{L-bos-def}
\end{equation}
\begin{equation}
\left[  a^{+}K(\pi)a\right]  =\int dta^{+}K(\pi)a\,,
\end{equation}
\begin{equation}
K(\pi)=\partial_{t}+\sum\limits_{\alpha}\pi_{\alpha}\Gamma_{\alpha
}\,.\label{K-pi-def}
\end{equation}

\subsection{Path integral approach at finite $N_{c}$}

The generating functional $\Phi_{B}(g)$ for the baryon wave function
(\ref{Phi-transition-def}) can be represented in the form
\begin{equation}
\Phi_{B}(g)=\langle0|J(g,a,0)|B\rangle\,,
\end{equation}
where we use the short notation

\begin{align}
J(g,a,t_{1})  &  =\prod\limits_{c=1}^{N_{c}}\left[  \sum_{m}g_{m}\cdot
a_{mc}(t_{1})\right]  \,,\\
J(g^{\prime},a^{+},t_{2})  &  =\prod\limits_{c=1}^{N_{c}}\left[  \sum_{m}
g_{m}^{\prime}\cdot a_{mc}^{+}(t_{2})\right]  \,.
\end{align}
We work with the Euclidean time. Therefore operators $a(t)$ and $a^{+}(t)$ are
not Hermitean conjugate.

In order to study $\Phi_{B}(g)$ in the path integral approach we introduce the
correlation function
\begin{equation}
Z(g,g^{\prime},t_{1},t_{2})=\langle0|T\left\{  J(g,a,t_{1})J(g^{\prime}
,a^{+},t_{2})\right\}  |0\rangle\,.\label{Z-g-g-prime-def}
\end{equation}
The path integral representation for this correlation function is
\begin{equation}
Z(g,g^{\prime},t_{1},t_{2})=\frac{\int DaDa^{+}D\pi J(g,a,t_{1})J(g^{\prime
},a^{+},t_{2})\exp\left\{  -\left[  a^{+}K(\pi)a\right]  +N_{c}S_{\mathrm{bos}
}(\pi)\right\}  }{\int DaDa^{+}D\pi\exp\left\{  -\left[  a^{+}K(\pi)a\right]
+N_{c}S_{\mathrm{bos}}(\pi)\right\}  }\,.\label{PI-start}
\end{equation}

At large $t_{1}-t_{2}\rightarrow+\infty$ only the contribution of the lightest
baryon $|B\rangle$ survives in Eq. (\ref{Z-g-g-prime-def}):
\begin{equation}
Z(g,g^{\prime},t_{1},t_{2})\overset{t_{1}-t_{2}\rightarrow+\infty
}{\longrightarrow}\langle0|J(g,a,0)|B\rangle e^{-(t_{2}-t_{1})\left(
\mathcal{E}_{B}-\mathcal{E}_{\mathrm{vac}}\right)  }\langle B|J(g^{\prime
},a^{+},0)|0\rangle\,.\label{B-insertion}
\end{equation}

\subsection{Large $N_{c}$ limit}

Now we want to consider the limit of large $N_{c}$. In this case we have with
the exponential accuracy
\begin{equation}
\langle0|J(g,a,0)|B\rangle\sim e^{N_{c}W(g)}\,.\label{exp-NW}
\end{equation}
The difference $\mathcal{E}_{B}-\mathcal{E}_{\mathrm{vac}}$ appearing in Eq.
(\ref{B-insertion}) has the order $O(N_{c})$:
\begin{equation}
\mathcal{E}_{B}-\mathcal{E}_{\mathrm{vac}}=N_{c}\left(  E_{\mathrm{bar}
}-E_{\mathrm{vac}}\right)  +O(N_{c}^{0})\,.\label{DE-Nc-DE}
\end{equation}

If we combine the large-time limit with the large-$N_{c}$ limit
\begin{equation}
t_{1}-t_{2}\rightarrow\infty,\quad N_{c}\rightarrow\infty
\,,\label{Nc-t-double-limit}
\end{equation}
then we find from Eqs. (\ref{B-insertion}), (\ref{exp-NW}) and (\ref{DE-Nc-DE})
\begin{equation}
Z(g,g^{\prime},t_{1},t_{2})\sim\exp\left\{  N_{c}\left[  -\left(
E_{\mathrm{bar}}-E_{\mathrm{vac}}\right)  (t_{1}-t_{2})+W(g)+\left[
W(g^{\prime\ast})\right]  ^{\ast}\right]  \right\}
\,.\label{Z-expectation-wrong}
\end{equation}

\subsection{Calculation of the path integral}

First we can compute the Gaussian integral over quarks in Eq. (\ref{PI-start})
\begin{equation}
Z(g,g^{\prime},t_{1},t_{2})=\frac{\int D\pi\left\{  \left[  g\cdot\langle
t_{1}|K^{-1}(\pi)|t_{2}\rangle\cdot g^{\prime}\right]  \left[  \mathrm{Det}
K(\pi)\right]  \exp\left[  S_{\mathrm{bos}}(\pi)\right]  \right\}  ^{N_{c}}
}{\int D\pi\left\{  \left[  \mathrm{Det}K(\pi)\right]  \exp S_{\mathrm{bos}
}(\pi)\right\}  ^{N_{c}}}\,.\label{Z-path-int-def}
\end{equation}
At large $N_{c}$ the path integrals in the numerator and in the denominator
can be computed using the saddle point method. The numerator and the
denominator have different saddle points which will be denoted $\pi
^{\mathrm{cl}}$ and $\pi^{\mathrm{vac}}$, respectively. The result of the
saddle-point integration is
\begin{equation}
\frac{1}{N_{c}}\ln Z(g,g^{\prime},t_{1},t_{2})=S_{\mathrm{nonloc}}
(\pi^{\mathrm{cl}})+\Delta S_{\mathrm{bos}}\left(  \pi^{\mathrm{cl}}\right)
\,,\label{ln-Z-Nc-res-2}
\end{equation}
where
\begin{equation}
S_{\mathrm{nonloc}}(\pi)=\ln\left\{  \left[  g\cdot\langle t_{1}|K^{-1}
(\pi)|t_{2}\rangle\cdot g^{\prime}\right]  \mathrm{Det}\frac{K(\pi)}
{K(\pi^{\mathrm{vac}})}\right\}  \,,\label{S-nonloc-def}
\end{equation}
\begin{equation}
\Delta S_{\mathrm{bos}}\left(  \pi\right)  =S_{\mathrm{bos}}(\pi
)-S_{\mathrm{bos}}(\pi^{\mathrm{vac}})=\int_{-\infty}^{\infty}dt\left[
L_{\mathrm{bos}}(\pi)-L_{\mathrm{bos}}(\pi^{\mathrm{vac}})\right]
\,.\label{Delta-S-bos-def}
\end{equation}
The saddle point $\pi^{\mathrm{cl}}$ is given by the equation
\begin{equation}
\left.  \frac{\delta}{\delta\pi}\left[  S_{\mathrm{nonloc}}(\pi)+\Delta
S_{\mathrm{bos}}\left(  \pi\right)  \right]  \right|  _{\pi=\pi^{\mathrm{cl}}
}=0\,.\label{pi-SP}
\end{equation}
The solution $\pi^{\mathrm{cl}}$ is $t$ dependent. It also depends on
$g,g^{\prime},t_{1},t_{2}$:
\begin{equation}
\pi^{\mathrm{cl}}=\pi^{\mathrm{cl}}(t|g,g^{\prime},t_{1},t_{2}
)\,.\label{pi-cl-solution}
\end{equation}
On the contrary, the saddle point $\pi^{\mathrm{vac}}$ is a constant which can
be found from the saddle point equation
\begin{equation}
\left.  \frac{\delta}{\delta\pi}\ln\left\{  \left[  \mathrm{Det}K(\pi)\right]
\left[  \exp S_{\mathrm{bos}}(\pi)\right]  \right\}  \right|  _{\pi
=\pi^{\mathrm{vac}}}=0\,.\label{pi-vac-eq}
\end{equation}
At large times $t\gg t_{1}$ or $t\ll t_{2}$ we have
\begin{equation}
\lim_{t\rightarrow\pm\infty}\pi^{\mathrm{cl}}(t)=\pi^{\mathrm{vac}
}\,.\label{pi-asymp}
\end{equation}
Combining Eqs. (\ref{Z-expectation-wrong}) and (\ref{ln-Z-Nc-res-2}), we find
\begin{equation}
W(g)+\left[  W(g^{\prime\ast})\right]  ^{\ast}=\lim_{t_{1}-t_{2}
\rightarrow+\infty}\left[  S_{\mathrm{nonloc}}(\pi^{\mathrm{cl}})+\Delta
S_{\mathrm{bos}}\left(  \pi^{\mathrm{cl}}\right)  +\left(  t_{1}-t_{2}\right)
\left(  E_{\mathrm{bar}}-E_{\mathrm{vac}}\right)  \right]
\,.\label{W-vial-pi-cl}
\end{equation}

\subsection{Fermion formulation of the effective large-$N_{c}$ theory}

In principle, Eq. (\ref{W-vial-pi-cl}) combined with the saddle point equation
(\ref{pi-SP}) contains everything what is needed for the calculation of the
functional $W(g)$. However, these equations involve the nonlocal action
$S_{\mathrm{nonloc}}(\pi^{\mathrm{cl}})$ defined in Eq. (\ref{S-nonloc-def}).
In order to solve the saddle point equation (\ref{pi-SP}) we would like to
find a local formulation for our effective large-$N_{c}$ theory. To this aim
we write the path integral representation for $S_{\mathrm{nonloc}}(\pi)$ over
auxiliary Grassmann variables $b$, $b^{+}$:
\begin{align}
&  \exp S_{\mathrm{nonloc}}(\pi)=\left[  g\cdot\langle t_{1}|K^{-1}
(\pi^{\mathrm{cl}})|t_{2}\rangle\cdot g^{\prime}\right]  \mathrm{Det}
\frac{K(\pi)}{K(\pi^{\mathrm{vac}})}\nonumber\\
&  =\frac{\int DbDb^{+}\left[  gb(t_{1})\right]  \left[  g^{\prime}b^{+}
(t_{2})\right]  \exp\tilde{S}(b,b^{+},\pi)}{\int DbDb^{+}\exp\tilde{S}
(b,b^{+},\pi^{\mathrm{vac}})}\,.\label{S-nonloc-PI}
\end{align}
Note that the fermion fields $b$, $b^{+}$ appearing here have \emph{no color}
indices. The action $\tilde{S}$ is
\begin{equation}
\tilde{S}(b,b^{+},\pi)=\int dt\left[  b^{+}K(\pi)b\right]
\,.\label{S-tilde-K}
\end{equation}
The path integral (\ref{S-nonloc-PI}) can be also interpreted in terms of the
operator approach to the theory of quantum fermions in the background
classical field $\pi$
\begin{equation}
\frac{\int DbDb^{+}\left[  gb(t_{1})\right]  \left[  g^{\prime}b^{+}
(t_{2})\right]  \exp\tilde{S}(b,b^{+},\pi)}{\int DbDb^{+}\exp\tilde{S}
(b,b^{+},\pi^{\mathrm{vac}})}=\langle0|\left[  gb(t_{1})\right]  \left[
g^{\prime}b^{+}(t_{2})\right]  |0\rangle_{\pi}\,.\label{ME-PI}
\end{equation}
For brevity we omit the \emph{in} and \emph{out} labels of the two ``vacuum''
states in this matrix element. One should also keep in mind that the Euclidean
evolution violates the Hermitian conjugation of operator $b(t)$ and $b^{+}
(t)$. Moreover even the hermiticity of the Hamiltonian can be violated.
Indeed, the background field $\pi^{(1)}$ comes from the saddle point equation
(\ref{pi-SP}) and the application of the saddle point method can be
accompanied by the deformation of the integration contour.

Combining Eqs. (\ref{S-nonloc-PI}) and (\ref{ME-PI}), we find
\begin{equation}
S_{\mathrm{nonloc}}(\pi)=\ln\langle0|\left[  gb(t_{1})\right]  \left[
g^{\prime}b^{+}(t_{2})\right]  |0\rangle_{\pi}\,.\label{Nc-1-int}
\end{equation}

Inserting Eq. (\ref{Nc-1-int}) into Eq. (\ref{W-vial-pi-cl}), we obtain
\begin{equation}
W(g)+\left[  W(g^{\prime\ast})\right]  ^{\ast}=\lim_{t_{1}-t_{2}
\rightarrow+\infty}\left\{  \ln\langle0|\left[  gb(t_{1})\right]  \left[
g^{\prime}b^{+}(t_{2})\right]  |0\rangle_{\pi^{\mathrm{cl}}}+\Delta
S_{\mathrm{bos}}\left(  \pi^{\mathrm{cl}}\right)  -\left(  t_{1}-t_{2}\right)
\left(  E_{\mathrm{bar}}-E_{\mathrm{vac}}\right)  \right\}
\,.\label{ln-g-psi-W}
\end{equation}

Using Eqs. (\ref{S-bos-pi}), (\ref{L-bos-def}), and (\ref{Nc-1-int}), we find
from the saddle point equation (\ref{pi-SP})
\begin{equation}
\left\{  \frac{\delta}{\delta\pi_{\alpha}(t)}\langle0|\left[  gb(t_{1}
)\right]  \left[  g^{\prime}b^{+}(t_{2})\right]  |0\rangle_{\pi}\right\}
_{\pi=\pi^{\mathrm{cl}}}=-\sum\limits_{\beta}\left(  V^{-1}\right)
_{\alpha\beta}\left(  \pi^{\mathrm{cl}}\right)  _{\beta}(t)\langle0|\left[
gb(t_{1})\right]  \left[  g^{\prime}b^{+}(t_{2})\right]  |0\rangle
_{\pi^{\mathrm{cl}}}\,.\label{pi-variation-SP}
\end{equation}
The variational derivative on the LHS has a simple interpretation in terms of
matrix elements in the operator formulation of the same theory:
\begin{equation}
\frac{\delta}{\delta\pi_{\alpha}(t)}\langle0|\left[  gb(t_{1})\right]  \left[
g^{\prime}b^{+}(t_{2})\right]  |0\rangle_{\pi}=-\langle0|T\left\{  \left[
gb(t_{1})\right]  \left[  b^{+}(t)\Gamma_{\alpha}b(t)\right]  \left[
g^{\prime}b^{+}(t_{2})\right]  \right\}  |0\rangle_{\pi}
\,.\label{pi-variation-ME}
\end{equation}
Combining equations (\ref{pi-variation-SP}) and (\ref{pi-variation-ME}), we
find
\begin{align}
&  \sum\limits_{\beta}\left(  V^{-1}\right)  _{\alpha\beta}\left(
\pi^{\mathrm{cl}}\right)  _{\beta}(t)\langle0|\left[  gb(t_{1})\right]
\left[  g^{\prime}b^{+}(t_{2})\right]  |0\rangle_{\pi^{\mathrm{cl}}
}\nonumber\\
&  =\langle0|T\left\{  \left[  gb(t_{1})\right]  \left[  b^{+}(t)\Gamma
_{\alpha}b(t)\right]  \left[  g^{\prime}b^{+}(t_{2})\right]  \right\}
|0\rangle_{\pi^{\mathrm{cl}}}\,.\label{pi-SP-ME}
\end{align}
This is a new form of the saddle point equation which determines
$\pi^{\mathrm{cl}}$.

\subsection{Separation of the $t_{1}$ and $t_{2}$ contributions}

Our aim is to compute the functional $W(g)$. However, representation
(\ref{ln-g-psi-W}) is written for the sum $W(g)+\left[  W(g^{\prime\ast
})\right]  ^{\ast}$. Now we want to extract information about $W(g)$ from this
equation. At finite $t_{1}$, $t_{2}$ the RHS of Eq. (\ref{ln-g-psi-W}) is a
nontrivial functional of $g$ and $g^{\prime}$. Only in the limit $t_{1}
-t_{2}\rightarrow+\infty$ the dependence on $g$ and $g^{\prime}$ reduces to
the sum of two independent terms 
$W(g)+\left[  W(g^{\prime\ast})\right]^{\ast}$.
This additive dependence is a consequence of the properties of the
classical solution $\pi^{\mathrm{cl}}(t)$ at large $t_{1}$ and $t_{2}$.

Eq. (\ref{ln-g-psi-W}) was derived in the limit $t_{1}-t_{2}\rightarrow
+\infty$. But if we look at the history of the derivation of this equation,
then we find that the limit $t_{1}-t_{2}\rightarrow+\infty$ was used only on
the RHS of this equation. In particular, the saddle point equation
(\ref{pi-SP-ME}) does not know anything about the limit 
$t_{1}-t_{2}\rightarrow+\infty$.

According to Eq. (\ref{pi-asymp}) at large $t\rightarrow\pm\infty$ (when $t\gg
t_{1}$ or $t\ll t_{2}$) the field $\pi^{\mathrm{cl}}(t)$ approaches the
constant configuration $\pi^{\mathrm{vac}}$.
At large $t_{1}-t_{2}\rightarrow+\infty$ we can also consider the intermediate region $t_{2}\ll
t\ll t_{1}$. In this region the field $\pi^{\mathrm{cl}}(t)$ is also
asymptotically static but with another value $\pi^{\mathrm{bar}}$. Thus we
have three regions where $\pi^{\mathrm{cl}}(t)$ becomes asymptotically static:
\begin{equation}
\pi^{\mathrm{cl}}(t)=\left\{
\begin{array}
[c]{ll}
\pi^{\mathrm{vac}}, & \mathrm{if\,}t\ll t_{2},\\
\pi^{\mathrm{bar}}\, & \mathrm{if\,}t_{2}\ll t\ll t_{1},\\
\pi^{\mathrm{vac}}\, & \mathrm{if\,}t_{1}\ll t\,.
\end{array}
\right.
\end{equation}
The constant configuration $\pi^{\mathrm{bar}}$ corresponds to the ``baryon''
state $|B\rangle$ of the effective fermion theory (remember that now we deal
with the effective fermion theory where fermions have no color indices).
Therefore the matrix elements appearing in Eq. (\ref{pi-SP-ME}) become
\begin{equation}
\langle0|T\left\{  \left[  gb(t_{1})\right]  \left[  g^{\prime}b^{+}
(t_{2})\right]  \right\}  |0\rangle_{\pi^{\mathrm{cl}}}\overset{t_{1}\gg
t_{2}}{=}\langle0|\left[  gb(t_{1})\right]  |B\rangle_{\pi}\langle B|\left[
g^{\prime}b^{+}(t_{2})\right]  |0\rangle_{\pi^{\mathrm{cl}}}
\,\,,\label{g-psi-g-psi-factorization}
\end{equation}
\begin{align}
&  \langle0|T\left\{  \left[  gb(t_{1})\right]  \left[  b^{+}(t)\Gamma
_{\alpha}b(t)\right]  \left[  g^{\prime}b^{+}(t_{2})\right]  \right\}
|0\rangle_{\pi^{\mathrm{cl}}}\nonumber\\
&  \overset{t_{1}\sim t\gg t_{2}}{=}\langle0|T\left\{  \left[  gb(t_{1}
)\right]  \left[  b^{+}(t)\Gamma_{\alpha}b(t)\right]  \right\}  |B\rangle
_{\pi^{\mathrm{cl}}}\langle B|\left[  g^{\prime}b^{+}(t_{2})\right]
|0\rangle_{\pi^{\mathrm{cl}}}\,.
\end{align}
Inserting these factorized expressions into the saddle point equation
(\ref{pi-SP-ME}), we find in the region $t_{1}\sim t\gg t_{2}$
\begin{equation}
\sum\limits_{\beta}\left(  V^{-1}\right)  _{\alpha\beta}\pi_{\beta}
(t)\langle0| \left[  gb(t_{1})\right]  |B\rangle_{\pi^{\mathrm{cl}}}
\overset{t_{1}\sim t\gg t_{2}}{=}\langle0|T\left\{  \left[  gb(t_{1})\right]
\left[  b^{+}(t)\Gamma_{\alpha}b(t)\right]  \right\}  |B\rangle_{\pi
^{\mathrm{cl}}}\,.\label{SP-reduced-1}
\end{equation}
Let us take the limit $t_{2}\rightarrow-\infty$ keeping $t_{1}$ and $t$ fixed.
In this limit Eq. (\ref{SP-reduced-1}) ``forgets'' about the behavior of
$\pi(t)$ at $t\sim t_{2}$ and at $t\ll t_{2}$. Hence we can replace
$\pi^{\mathrm{cl}}$ by its $t_{1}$ component
\begin{equation}
\pi^{(1)}(t)=\left\{
\begin{array}
[c]{ll}
\pi^{\mathrm{bar}}\, & \mathrm{if\,}t\ll t_{1},\\
\pi^{\mathrm{cl}}(t) & \,\mathrm{if\,}t\sim t_{1},\\
\pi^{\mathrm{vac}}\, & \mathrm{if\,}t\gg t_{1}.
\end{array}
\right. \label{pi-1-def}
\end{equation}
As a result, Eq. (\ref{SP-reduced-1}) becomes
\begin{equation}
\sum\limits_{\beta}\left(  V^{-1}\right)  _{\alpha\beta}\pi_{\beta}
^{(1)}(t)\langle0|\left[  gb(t_{1})\right]  |B\rangle_{\pi^{(1)}}
=\langle0|T\left\{  \left[  gb(t_{1})\right]  \left[  b^{+}(t)\Gamma_{\alpha
}b(t)\right]  \right\}  |B\rangle_{\pi^{(1)}}\,.\label{pi-t-res-0}
\end{equation}
Hence
\begin{equation}
\pi_{\alpha}^{(1)}(t)=\sum\limits_{\beta}V_{\alpha\beta}\frac{\langle
0|T\left\{  \left[  gb(t_{1})\right]  \left[  b^{+}(t)\Gamma_{\beta
}b(t)\right]  \right\}  |B\rangle_{\pi^{(1)}}}{\langle0|\left[  gb(t_{1}
)\right]  |B\rangle_{\pi^{(1)}}}\,.\label{pi-t-res}
\end{equation}
Working with the background field $\pi^{(1)}$, we can consider the limits
$t\rightarrow\pm\infty$ at fixed $t_{1}$
\begin{equation}
\langle0|T\left\{  \left[  gb(t_{1})\right]  \left[  b^{+}(t)\Gamma_{\alpha
}b(t)\right]  \right\}  |B\rangle_{\pi^{(1)}}\overset{t\ll t_{1}}{=}
\langle0|\left[  gb(t_{1})\right]  |B\rangle_{\pi^{(1)}}\langle B|\left[
b^{+}(t)\Gamma_{\alpha}b(t)\right]  |B\rangle_{\pi^{\mathrm{bar}}}\,,
\end{equation}
\begin{equation}
\langle0|T\left\{  \left[  gb(t_{1})\right]  \left[  b^{+}(t)\Gamma_{\alpha
}b(t)\right]  \right\}  |B\rangle_{\pi^{(1)}}\overset{t\gg t_{1}}{=}
\langle0|\left[  b^{+}(t)\Gamma_{\alpha}b(t)\right]  |0\rangle_{\pi
^{\mathrm{vac}}}\langle0|\left[  gb(t_{1})\right]  |B\rangle_{\pi^{(1)}}\,.
\end{equation}
Inserting these asymptotic expressions into Eq. (\ref{pi-t-res}), we obtain
\begin{equation}
\pi_{\alpha}^{\mathrm{vac}}=\sum\limits_{\beta}V_{\alpha\beta}\langle0|\left[
b^{+}(0)\Gamma_{\beta}b(0)\right]  |0\rangle_{\pi^{\mathrm{vac}}
}\,,\label{pi-vac-psi}
\end{equation}
\begin{equation}
\pi_{\alpha}^{\mathrm{bar}}=\sum\limits_{\beta}V_{\alpha\beta}\langle
B|\left[  b^{+}(0)\Gamma_{\beta}b(0)\right]  |B\rangle_{\pi^{\mathrm{bar}}
}\,.\label{pi-bar-psi}
\end{equation}
In Sec. \ref{Hartree-PI-section} we will show that these are nothing else but
the static Hartree equations for the vacuum and for the baryon in the
large-$N_{c}$ theory.

Note that the time-ordered product has a discontinuity at $t=t_{1}$:
\begin{equation}
\left[  \lim_{t\rightarrow t_{1}-0}-\lim_{t\rightarrow t_{1}+0}\right]
\langle0|T\left\{  \left[  gb(t_{1})\right]  \left[  b^{+}(t)\Gamma_{\alpha
}b(t)\right]  \right\}  |B\rangle_{\pi^{(1)}}=\langle0|\left[  g\Gamma
_{\alpha}b(t_{1})\right]  |B\rangle_{\pi^{(1)}}\,.
\end{equation}
Combining this with Eq. (\ref{pi-t-res}), we conclude that the field
$\pi^{(1)}(t)$ has a discontinuity at $t\rightarrow t_{1}$:
\begin{equation}
\left[  \lim_{t\rightarrow t_{1}-0}-\lim_{t\rightarrow t_{1}+0}\right]
\pi_{\alpha}^{(1)}(t)=\sum\limits_{\beta}V_{\alpha\beta}\frac{\langle0|\left[
g\Gamma_{\beta}b(t_{1})\right]  |B\rangle_{\pi^{(1)}}}{\langle0|\left[
gb(t_{1})\right]  |B\rangle_{\pi^{(1)}}}\,.\label{pi-1-discontinuity}
\end{equation}

Using the property (\ref{g-psi-g-psi-factorization}), we can rewrite Eq.
(\ref{ln-g-psi-W}) in the form
\begin{equation}
W(g)+\left[  W(g^{\prime\ast})\right]  ^{\ast}=\ln\langle0|\left[
gb(t_{1})\right]  |B\rangle_{\pi^{(1)}}+\ln\langle B|\left[  g^{\prime}
b^{+}(t_{2})\right]  |0\rangle_{\pi^{(2)}}+\Delta S_{\mathrm{bos}}\left(
\pi^{\mathrm{cl}}\right)  -\left(  t_{1}-t_{2}\right)  \left(  E_{\mathrm{bar}
}-E_{\mathrm{vac}}\right)  \,.\label{W-t1-t2}
\end{equation}
Here $\pi^{(2)}$ is the analog of $\pi^{(1)}$ for the vicinity of the $t_{2}$
time epoch:
\begin{equation}
\pi^{(2)}(t)=\left\{
\begin{array}
[c]{ll}
\pi^{\mathrm{vac}}\, & \mathrm{if\,}t\ll t_{2},\\
\pi^{\mathrm{cl}}(t) & \mathrm{if\,}\,t\sim t_{2},\\
\pi^{\mathrm{bar}}\, & \mathrm{if\,}t\gg t_{2}.
\end{array}
\right. \label{pi-2-def}
\end{equation}

Using the definition (\ref{Delta-S-bos-def}) of $\Delta S_{\mathrm{bos}}$ and
the properties (\ref{pi-1-def}), (\ref{pi-2-def}) of $\pi^{(1)}$ and
$\pi^{(2)}$, we obtain
\begin{gather}
\Delta S_{\mathrm{bos}}\left(  \pi^{\mathrm{cl}}\right)  =\int_{-\infty
}^{t_{2}}dt\left[  L_{\mathrm{bos}}(\pi^{\mathrm{cl}})-L_{\mathrm{bos}}
(\pi^{\mathrm{vac}})\right]  +\int_{t_{2}}^{t_{1}}dt\left[  L_{\mathrm{bos}
}(\pi^{\mathrm{cl}})-L_{\mathrm{bos}}(\pi^{\mathrm{bar}})\right] \nonumber\\
+\int_{t_{1}}^{\infty}dt\left[  L_{\mathrm{bos}}(\pi^{\mathrm{cl}
})-L_{\mathrm{bos}}(\pi^{\mathrm{vac}})\right]  +\left[  L_{\mathrm{bos}}
(\pi^{\mathrm{bar}})-L_{\mathrm{bos}}(\pi^{\mathrm{vac}})\right]  (t_{1}
-t_{2})\nonumber\\
=\int_{-\infty}^{t_{2}}dt\left[  L_{\mathrm{bos}}(\pi^{(2)})-L_{\mathrm{bos}
}(\pi^{\mathrm{vac}})\right]  +\int_{t_{2}}^{+\infty}dt\left[  L_{\mathrm{bos}
}(\pi^{(2)})-L_{\mathrm{bos}}(\pi^{\mathrm{bar}})\right] \nonumber\\
+\int_{-\infty}^{t_{1}}dt\left[  L_{\mathrm{bos}}(\pi^{(1)})-L_{\mathrm{bos}
}(\pi^{\mathrm{bar}})\right]  +\int_{t_{1}}^{\infty}dt\left[  L_{\mathrm{bos}
}(\pi^{(1)})-L_{\mathrm{bos}}(\pi^{\mathrm{vac}})\right] \nonumber\\
+\left[  L_{\mathrm{bos}}(\pi^{\mathrm{bar}})-L_{\mathrm{bos}}(\pi
^{\mathrm{vac}})\right]  (t_{1}-t_{2})\,.
\end{gather}
Inserting this into Eq. (\ref{W-t1-t2}) and separating the $t_{1}$ and $t_{2}
$ contributions, we find
\begin{align}
&  W(g)=\ln\langle0|\left[  gb(t_{1})\right]  |B\rangle_{\pi^{(1)}}
+\int_{-\infty}^{t_{1}}dt\left[  L_{\mathrm{bos}}(\pi^{(1)})-L_{\mathrm{bos}
}(\pi^{\mathrm{bar}})\right]  +\int_{t_{1}}^{\infty}dt\left[  L_{\mathrm{bos}
}(\pi^{(1)})-L_{\mathrm{bos}}(\pi^{\mathrm{vac}})\right] \nonumber\\
&  +t_{1}\left\{  \left(  E_{\mathrm{bar}}-E_{\mathrm{vac}}\right)  +\left[
L_{\mathrm{bos}}(\pi^{\mathrm{bar}})-L_{\mathrm{bos}}(\pi^{\mathrm{vac}
})\right]  \right\}  \,.\label{W-repr-t1}
\end{align}

Thus we have reduced the problem of the calculation of the functional $W(g)$
to the analysis of the fermion system in the ``self-consistent'' field
$\pi^{(1)}(t)$. Indeed, the field $\pi^{(1)}(t)$ is determined by Eq.
(\ref{pi-t-res}) combined with the discontinuity condition
(\ref{pi-1-discontinuity}) at $t_{1}$ and with the boundary conditions
(\ref{pi-1-def})
\begin{equation}
\lim_{t\rightarrow-\infty}\pi^{(1)}(t)=\pi^{\mathrm{bar}}\,,\quad
\lim_{t\rightarrow+\infty}\pi^{(1)}(t)=\pi^{\mathrm{vac}}
\,.\label{pi-1-t-asympt}
\end{equation}
Inserting this field $\pi^{(1)}(t)$ into Eq. (\ref{W-repr-t1}), we can find
$W(g)$. Note that the $g$ dependence appears via the discontinuity
(\ref{pi-1-discontinuity}) of $\pi^{(1)}(t)$ at $t_{1}$.

\subsection{Hartree equations from the path integral formalism}

\label{Hartree-PI-section}

The saddle point field $\pi^{(1)}(t)$ is determined by equation
(\ref{pi-t-res}). This equation contains the matrix elements
\begin{equation}
\langle0|\left[  gb(t_{1})\right]  |B\rangle_{\pi^{(1)}}\,,\label{ME-1}
\end{equation}
\begin{equation}
\langle0|T\left\{  \left[  gb(t_{1})\right]  \left[  b^{+}(t)\Gamma_{\beta
}b(t)\right]  \right\}  |B\rangle_{\pi^{(1)}}\,.\label{ME-2}
\end{equation}
of the effective theory with the $b$ fermions that have no color.

The matrix elements (\ref{ME-1}) and (\ref{ME-2}) are written for the theory
in the background $t$-dependent field $\pi^{(1)}$. The vacuum state
$\langle0|$ appears in these matrix elements as an \emph{out} state
corresponding to $t\rightarrow+\infty$. At $t\rightarrow+\infty$ we have
$\pi^{(1)}(t)\rightarrow\pi^{\mathrm{vac}}$ according to Eq.
(\ref{pi-1-t-asympt}). Therefore the state $\langle0|$ is the ground state of
the theory with the Hamiltonian associated with the quadratic form
(\ref{K-pi-def}) taken in the static background field $\pi^{(1)}$:
\begin{equation}
\langle0|\sum\limits_{\alpha}\pi_{\alpha}^{\mathrm{vac}}\left(  b^{+}
\Gamma_{\alpha}b\right)  =\langle0|E_{\mathrm{vac}}\,.
\end{equation}
This means that the \emph{physical} vacuum state $\langle0|$ could be thought
of as made of $M$ quarks above the \emph{bare} vacuum $\langle\Omega|$
\begin{equation}
\langle0|=\langle\Omega|\prod\limits_{s=1}^{M}\left[  \left(  \phi
_{\mathrm{vac}}^{s}\right)  ^{+}b\right]  =\langle\Omega|\prod\limits_{s=1}
^{M}\left[  \sum\limits_{m}\left(  \phi_{\mathrm{vac}}^{s}\right)  _{m}^{\ast
}b_{m}\right]  \,,\label{vac-via-b-Omega}
\end{equation}
where the ``single-particle'' wave functions $\phi_{\mathrm{vac}}^{S}$ are
solutions of the equation
\begin{equation}
\left(  \phi_{\mathrm{vac}}^{s}\right)  ^{+}\sum\limits_{\alpha}\pi_{\alpha
}^{\mathrm{vac}}\Gamma_{\alpha}=\left(  \phi_{\mathrm{vac}}^{s}\right)
^{+}\varepsilon_{\mathrm{vac}}^{s}\,.\label{pi-Hartree-vac-1}
\end{equation}
Using expression (\ref{vac-via-b-Omega}), we can compute the matrix element on
the RHS of Eq. (\ref{pi-vac-psi}):
\begin{equation}
\langle0|\left[  b^{+}(0)\Gamma^{\beta}b(0)\right]  |0\rangle_{\pi
^{\mathrm{vac}}}=\sum\limits_{s=1}^{M}\left(  \phi_{\mathrm{vac}}^{s}\right)
^{+}\Gamma^{\beta}\phi_{\mathrm{vac}}^{s}\,.
\end{equation}
Inserting this result into Eq. (\ref{pi-vac-psi}), we find
\begin{equation}
\pi_{\alpha}^{\mathrm{vac}}=\sum\limits_{\beta}V_{\alpha\beta}\sum
\limits_{s=1}^{M}\left(  \phi_{\mathrm{vac}}^{s}\right)  ^{+}\Gamma^{\beta
}\phi_{\mathrm{vac}}^{s}\,.\label{pi-Hartree-vac-2}
\end{equation}
Now we see that Eqs. (\ref{pi-Hartree-vac-1}) and (\ref{pi-Hartree-vac-2}) are
nothing else but the Hartree equations (\ref{h-phi-Hartree-general}),
(\ref{h-phi-Hartree-general-2}) for the Hamiltonian (\ref{H-PI-4-q}). In the
framework of these Hartree equations we have
\begin{equation}
E_{\mathrm{vac}}=\frac{1}{2}\sum\limits_{s=1}^{M}\varepsilon_{\mathrm{vac}
}^{s}\,.\label{E-vac-sum-eps}
\end{equation}

Instead of the second quantized representation for the physical vacuum
$|0\rangle$ we can use the language the of $M$-particle wave function written
in terms of the Slater determinant
\begin{equation}
|0\rangle\rightarrow\frac{1}{\sqrt{M!}}\det_{1\leq s,i\leq M}\left\|  \left(
\phi_{\mathrm{vac}}^{s}\right)  _{n_{i}}\right\|  \,.
\end{equation}

Similar equations can be written for the baryon state $|B\rangle$ which is
associated in the matrix elements (\ref{ME-1}), (\ref{ME-2}) with the limit
$t\rightarrow+\infty$:
\begin{equation}
|B\rangle=\prod\limits_{s=1}^{M+1}\left(  \phi_{\mathrm{bar}}^{s}b^{+}\right)
|\Omega\rangle=\left\{  \prod\limits_{s=1}^{M+1}\left[  \sum\limits_{k}\left(
\phi_{\mathrm{bar}}^{s}\right)  _{k}b_{k}^{+}\right]  \right\}  |\Omega
\rangle\,,\label{B-via-b-Omega}
\end{equation}
\begin{equation}
\sum\limits_{\alpha}\left(  \pi_{\alpha}^{\mathrm{bar}}\Gamma_{\alpha}\right)
\phi_{\mathrm{bar}}^{s}=\varepsilon_{\mathrm{bar}}^{s}\phi_{\mathrm{bar}}
^{s}\,,\label{bar-Hartree}
\end{equation}
\begin{equation}
\pi_{\alpha}^{\mathrm{bar}}=\sum\limits_{\beta}V_{\alpha\beta}\sum
\limits_{s=1}^{M+1}\left(  \phi_{\mathrm{bar}}^{s}\right)  ^{+}\Gamma^{\beta
}\phi_{\mathrm{bar}}^{s}\,,\label{bar-Hartree-2}
\end{equation}
\begin{equation}
E_{\mathrm{bar}}=\frac{1}{2}\sum\limits_{s=1}^{M+1}\varepsilon_{\mathrm{bar}
}^{s}\,.\label{E-sum-eps-vac}
\end{equation}
The state $|B\rangle$ corresponds to the Slater determinant
\begin{equation}
|B\rangle\rightarrow\frac{1}{\sqrt{(M+1)!}}\det_{1\leq s,i\leq M+1}\left\|
\left(  \phi_{\mathrm{bar}}^{s}\right)  _{n_{i}}\right\|  \,.
\end{equation}
We work with the pure 4-fermionic Hamiltonian (\ref{H-PI-4-q}) which has no
quadratic piece. Therefore we have additional simplifications. Combining Eqs.
(\ref{L-bos-def}), (\ref{pi-Hartree-vac-1}), (\ref{pi-Hartree-vac-2}),
(\ref{E-vac-sum-eps}), (\ref{E-sum-eps-vac}), we find
\begin{equation}
E_{\mathrm{vac}}=L_{\mathrm{bos}}(\pi^{\mathrm{vac}})=\sum\limits_{s=1}
^{M}\varepsilon_{\mathrm{vac}}^{s}=\frac{1}{2}\sum\limits_{\alpha\beta
}V_{\alpha\beta}\pi_{\alpha}^{\mathrm{vac}}\left(  V^{-1}\right)
_{\alpha\beta}\pi_{\beta}^{\mathrm{vac}}
\end{equation}
and similarly for the baryon
\begin{equation}
E_{\mathrm{bar}}=L_{\mathrm{bos}}(\pi^{\mathrm{bar}})=\sum\limits_{s=1}
^{M+1}\varepsilon_{\mathrm{bar}}^{s}=\frac{1}{2}\sum\limits_{\alpha\beta
}V_{\alpha\beta}\pi_{\alpha}^{\mathrm{bar}}\left(  V^{-1}\right)
_{\alpha\beta}\pi_{\beta}^{\mathrm{bar}}\,.
\end{equation}

Now we can simplify expression (\ref{W-repr-t1})
\begin{align}
&  W(g)=\ln\langle0|\left[  gb(t_{1})\right]  |B\rangle_{\pi^{(1)}}
+t_{1}\left[  \sum\limits_{s=1}^{M+1}\varepsilon_{\mathrm{bar}}^{s}
-\sum\limits_{s=1}^{M}\varepsilon_{\mathrm{vac}}^{s}\right] \nonumber\\
&  +\int_{-\infty}^{t_{1}}dt\left[  L_{\mathrm{bos}}(\pi^{(1)}
)-L_{\mathrm{bos}}(\pi^{\mathrm{bar}})\right]  +\int_{t_{1}}^{\infty}dt\left[
L_{\mathrm{bos}}(\pi^{(1)})-L_{\mathrm{bos}}(\pi^{\mathrm{vac}})\right]
\,.\label{W-repr-t2}
\end{align}

\subsection{Time dependence}

In the previous section we have shown how the static Hartree equations for the
baryon and for the vacuum appear in the context of the calculation of the
generating functional $W(g)$ for the baryon wave function. These static
equations correspond to the asymptotic limit (\ref{pi-1-t-asympt}) of the
field $\pi^{(1)}(t)$ at $t\rightarrow\pm\infty$.

Now we turn to the calculation of the time-dependent matrix elements
(\ref{ME-1}), (\ref{ME-2}). These matrix elements are written in the
Heisenberg representation with fixed states $\langle0|$ and $|B\rangle$ but
time-dependent operators $b(t)$ and $b^{+}(t)$. However, we can rewrite these
matrix elements in terms of the Schr\"{o}dinger representation. The
Schr\"{o}dinger representation is based on the time-dependent wave functions
corresponding to the static Heisenberg wave functions $\langle0|$
(\ref{vac-via-b-Omega}) and $|B\rangle$ (\ref{B-via-b-Omega}):
\begin{equation}
\langle0|\rightarrow\langle0,t|=\langle\Omega|\prod\limits_{s=1}^{M} \left[
\tilde{\phi}_{\mathrm{vac}}^{s}(t)b\right]  \,,\label{vac-0-t}
\end{equation}
\begin{equation}
|B\rangle\rightarrow|B,t\rangle=\prod\limits_{s=1}^{M+1}\left[  \phi
_{\mathrm{bar}}^{s}(t)b^{+}\right]  |\Omega\rangle\,.\label{B-t}
\end{equation}
We use notation $\tilde{\phi}_{\mathrm{vac}}^{s}(t)$ instead of $\left[
\phi_{\mathrm{vac}}^{s}(t)\right]  ^{+}$ because the Euclidean evolution is
not unitary.

The time dependence of the wave functions $\tilde{\phi}_{\mathrm{vac}}^{s}(t)
$ and $\phi_{\mathrm{bar}}^{s}(t)$ is controlled by the quadratic form
$K(\pi^{(1)})$ (\ref{K-pi-def}) which defines the $b,b^{+}$ theory
(\ref{S-nonloc-PI}), (\ref{S-tilde-K}). Obviously the role of the Hamiltonian
in the quadratic form $K(\pi^{(1)})$ is played by
\begin{equation}
h(t)=\sum\limits_{\alpha}\pi_{\alpha}^{(1)}(t)\Gamma_{\alpha}\,.
\end{equation}
The corresponding evolution operator is
\begin{equation}
U(\tau_{1},\tau_{2})\overset{\tau_{1}>\tau_{2}}{=}T\exp\left[  -\int
\limits_{\tau_{2}}^{\tau_{1}}dt\sum\limits_{\alpha}\pi_{\alpha}^{(1)}
(t)\Gamma_{\alpha}dt\right]  \,,\label{U-Texp}
\end{equation}
\begin{equation}
U(\tau_{1},\tau_{2})\overset{\tau_{1}<\tau_{2}}{=}\left[  U(\tau_{2},\tau
_{1})\right]  ^{-1}\,.\label{U-U-inv}
\end{equation}
This operator $U(\tau_{1},\tau_{2})$ controls the evolution of wave functions
$\phi_{\mathrm{vac}}^{s}(t)$ and $\phi_{\mathrm{bar}}^{s}(t)$:
\begin{equation}
\tilde{\phi}_{\mathrm{vac}}^{s}(t)=\lim_{\tau\rightarrow+\infty}\left(
\phi_{\mathrm{vac}}^{s}\right)  ^{+}U(\tau,t)\exp\left(  \tau\varepsilon
_{\mathrm{vac}}^{s}\right)  \,,\label{vac-t}
\end{equation}
\begin{equation}
\phi_{\mathrm{bar}}^{s}(t)=\lim_{\tau\rightarrow-\infty}U(t,\tau
)\phi_{\mathrm{bar}}^{s}\left(  -\tau\varepsilon_{\mathrm{bar}}^{s}\right)
\,.\label{Phi-n-bar-def}
\end{equation}
On the RHS we have the time independent wave functions 
$\phi_{\mathrm{vac}}^{s}$ and $\phi_{\mathrm{bar}}^{s}$ which are determined by the Hartree
equations (\ref{pi-Hartree-vac-1}), (\ref{pi-Hartree-vac-2}) for
$\phi_{\mathrm{vac}}^{s}$ and (\ref{bar-Hartree}), (\ref{bar-Hartree-2}) for
$\phi_{\mathrm{bar}}^{s}$.

For the matrix element (\ref{ME-1}), the reduction to the Schr\"{o}dinger
representation can be done as follows
\begin{equation}
\langle0|\left[  gb(t_{1})\right]  |B\rangle_{\pi^{(1)}}=\langle
0,t_{1}|\left(  gb\right)  |B,t_{1}\rangle_{\pi^{(1)}}=\langle\Omega|\left\{
\prod\limits_{r=1}^{M}\left[  \tilde{\phi}_{\mathrm{vac}}^{r}(t_{1})b\right]
\right\}  \left(  gb\right)  \left\{  \prod\limits_{s=1}^{M+1}\left[
\phi_{\mathrm{bar}}^{s}(t_{1})b^{+}\right]  \right\}  |\Omega\rangle
\,.\label{vac-g-bar-ME}
\end{equation}
Let us define
\begin{equation}
g(t)=gU(t_{1},t)\,,\label{g-t-def}
\end{equation}
using the evolution operator (\ref{U-Texp}). Next, let us introduce a special
notation for the set of $M+1$ functions including $\phi_{\mathrm{vac}}^{s}(t)$
and $g(t)$
\begin{equation}
\tilde{\phi}_{\mathrm{vac},g}^{r}(t)=\left\{
\begin{array}
[c]{cc}
\tilde{\phi}_{\mathrm{vac}}^{r}(t) & \mathrm{if}\,1\leq r\leq M\,,\\
g(t) & \mathrm{if}\,r=M+1\,.
\end{array}
\right. \label{Phi-vac-g}
\end{equation}
Then we find from Eq. (\ref{vac-g-bar-ME})
\begin{equation}
\langle0|\left[  gb(t_{1})\right]  |B\rangle_{\pi^{(1)}}=\langle
\Omega|\left\{  \prod\limits_{r=1}^{M+1}\left[  \tilde{\phi}_{\mathrm{vac}
,g}^{r}(t_{1})b\right]  \right\}  \left\{  \prod\limits_{s=1}^{M+1}\left[
\phi_{\mathrm{bar}}^{s}(t_{1})b^{+}\right]  \right\}  |\Omega\rangle
=\det_{1\leq r,s\leq M+1}D_{rs}(t_{1})\,,\label{vac-g-psi-B}
\end{equation}
where we have introduced the notation
\begin{equation}
D_{rs}(t)=\tilde{\phi}_{\mathrm{vac},g}^{r}(t)\cdot\phi_{\mathrm{bar}}
^{s}(t)\,.\label{D-via-Phi-def}
\end{equation}
Inserting Eqs. (\ref{vac-t}), (\ref{Phi-n-bar-def}) and using Eq.
(\ref{U-U-inv}), we find
\begin{equation}
D_{rs}(t)=\lim_{\tau_{1}\rightarrow+\infty}\lim_{\tau_{2}\rightarrow-\infty
}\exp\left(  \tau_{1}\varepsilon_{\mathrm{vac}}^{r}-\tau_{2}\varepsilon
_{\mathrm{bar}}^{s}\right)  \left(  \phi_{\mathrm{vac}}^{s}\right)  ^{+}
U(\tau_{1},\tau_{2})\phi_{\mathrm{bar}}^{s}\,.\label{D-vai-Phi-lim}
\end{equation}
We see that $D_{rs}(t)$ is $t$ independent
\begin{equation}
\frac{d}{dt}D_{rs}(t)=0\,,\label{D-t-independence}
\end{equation}
since the $t$ evolution (\ref{vac-t}), (\ref{g-t-def}) of $\tilde{\phi
}_{\mathrm{vac},g}^{r}(t)$ compensates the $t$ evolution (\ref{Phi-n-bar-def})
of $\phi_{\mathrm{bar}}^{s}(t)$.

By analogy with Eq. (\ref{vac-g-bar-ME}) we can write
\begin{equation}
\langle0|T\left\{  \left[  gb(t_{1})\right]  \left[  b^{+}(t)\Gamma_{\beta
}b(t)\right]  \right\}  |B\rangle_{\pi^{(1)}}=\left\{
\begin{array}
[c]{cc}
\langle0,t|\left[  g(t)b\right]  \left(  b^{+}\Gamma_{\beta}b\right)
|B,t\rangle_{\pi^{(1)}}\,\, & \mathrm{if}\,t<t_{1},\\
\langle0,t|\left(  b^{+}\Gamma_{\beta}b\right)  \left[  g(t)b\right]
|B,t\rangle_{\pi^{(1)}} & \mathrm{if}\,t>t_{1}.
\end{array}
\right. \label{p-multi}
\end{equation}
In the case $t<t_{1}$ this leads to
\begin{align}
&  \langle0|T\left\{  \left[  gb(t_{1})\right]  \left[  b^{+}(t)\Gamma_{\beta
}b(t)\right]  \right\}  |B\rangle_{\pi^{(1)}}\overset{t<t_{1}}{=}\langle
\Omega|\left\{  \prod\limits_{r=1}^{M+1}\left[  \tilde{\phi}_{\mathrm{vac}
,g}^{r}(t_{1})b\right]  \right\}  \left(  b^{+}\Gamma_{\beta}b\right)
\left\{  \prod\limits_{s=1}^{M+1}\left[  \phi_{\mathrm{bar}}^{s}(t_{1}
)b^{+}\right]  \right\}  |\Omega\rangle\nonumber\\
&  =\left(  \det_{1\leq r^{\prime},s^{\prime}\leq M+1}D_{r^{\prime}s^{\prime}
}\right)  \sum\limits_{r,s=1}^{M+1}\left(  D^{-1}\right)  _{sr} \left[
\tilde{\phi}_{\mathrm{vac},g}^{r}(t)\Gamma_{\beta}\phi_{\mathrm{bar}}
^{s}(t)\right]  \,.\label{ME-1-A-Phi}
\end{align}
Inserting Eqs. (\ref{vac-g-psi-B}) and (\ref{ME-1-A-Phi}) into Eq.
(\ref{pi-t-res}), we obtain
\begin{equation}
\pi_{\alpha}^{(1)}(t)\overset{t<t_{1}}{=}\sum\limits_{\beta}V_{\alpha\beta
}\sum\limits_{r,s=1}^{M+1}\left(  D^{-1}\right)  _{sr}\left[  \tilde{\phi
}_{\mathrm{vac},g}^{r}(t)\Gamma_{\beta}\phi_{\mathrm{bar}}^{s}(t)\right]
\,.\label{pi-via-Phi-0}
\end{equation}
According to Eq. (\ref{D-via-Phi-def}) $D_{rs}$ is a local (in $t$) function
of $\phi_{\mathrm{vac}}(t)$, $\phi_{\mathrm{bar}}(t)$ and $g(t)$. Therefore
Eq. (\ref{pi-via-Phi-0}) provides us a local expression for $\pi_{\alpha}(t)$
via $\phi_{\mathrm{vac}}(t)$, $\phi_{\mathrm{bar}}(t)$ and $g(t)$:
\begin{equation}
\pi_{\alpha}^{(1)}(t)\overset{t<t_{1}}{=}\Pi_{\alpha}^{(-)}\left(  \tilde
{\phi}_{\mathrm{vac}}(t),\phi_{\mathrm{bar}}(t),g(t)\right)
\,,\label{pi-plus}
\end{equation}
where
\begin{equation}
\Pi_{\alpha}^{(-)}\left(  \tilde{\phi}_{\mathrm{vac}},\phi_{\mathrm{bar}
},g\right)  =\sum\limits_{\beta}V_{\alpha\beta}\sum\limits_{r,s=1}
^{M+1}\left\{  \left[  \tilde{\phi}_{\mathrm{vac},g}\cdot\phi_{\mathrm{bar}
}\right]  ^{-1}\right\}  _{sr}\left[  \tilde{\phi}_{\mathrm{vac,}g}^{r}
\Gamma_{\beta}\phi_{\mathrm{bar}}^{s}\right]  \,.
\end{equation}
The $\tilde{\phi}_{\mathrm{vac},g}$ is defined similarly to Eq.
(\ref{Phi-vac-g}).

It is easy to see that the above argument can be repeated for the case
$t>t_{1}$:
\begin{equation}
\pi_{\alpha}^{(1)}(t)\overset{t>t_{1}}{=}\Pi_{\alpha}^{(+)}\left(  \tilde
{\phi}_{\mathrm{vac}}(t),\phi_{\mathrm{bar}}(t),g(t)\right) \label{pi-minus}
\end{equation}

Inserting Eq. (\ref{vac-g-psi-B}) into Eq. (\ref{W-repr-t2}), we find
\begin{align}
&  W(g)=\ln\det_{1\leq r,s\leq M+1}D_{rs}+t_{1}\left[  \sum\limits_{s=1}
^{M+1}\varepsilon_{\mathrm{bar}}^{s}-\sum\limits_{s=1}^{M}\varepsilon
_{\mathrm{vac}}^{s}\right] \nonumber\\
&  +\int_{-\infty}^{t_{1}}dt\left[  L_{\mathrm{bos}}(\pi^{(1)}
)-L_{\mathrm{bos}}(\pi^{\mathrm{bar}})\right]  +\int_{t_{1}}^{\infty}dt\left[
L_{\mathrm{bos}}(\pi^{(1)})-L_{\mathrm{bos}}(\pi^{\mathrm{vac}})\right]
\,.\label{W-repr-t3}
\end{align}
According to Eq. (\ref{D-t-independence}) $D_{rs}$ is independent of $t$.
However, $D_{rs}$ depends on $t_{1}$. This $t_{1}$ dependence is implicit.
Indeed, if we shift $t_{1}$ in the original path integral representation,
\begin{equation}
t_{1}\rightarrow t_{1}+\Delta T\,,\label{t1-shift}
\end{equation}
then this will lead to the corresponding shift of the saddle point $\pi^{(1)}
$
\begin{equation}
\pi^{(1)}(t)\rightarrow\pi^{(1)}(t-\Delta T)
\end{equation}
and to the shift in the evolution operator
\begin{equation}
U(\tau_{1},\tau_{2})\rightarrow U(\tau_{1}-\Delta T,\tau_{2}-\Delta T)\,.
\end{equation}
Therefore the matrix $D_{rs}$ (\ref{D-vai-Phi-lim}) will change as follows:
\begin{equation}
D_{rs}\rightarrow D_{rs}+\Delta T\left(  \varepsilon_{\mathrm{vac}}
^{r}-\varepsilon_{\mathrm{bar}}^{s}\right)  \,.\label{D-rs-shift}
\end{equation}
We see that the $\Delta T$ shifts of $t_{1}$ (\ref{t1-shift}) and $D_{rs}$
(\ref{D-rs-shift}) compensate each other in Eq. (\ref{W-repr-t3}) so that the
functional $W(g)$ is independent of $t_{1}$ as it should be. Using this
$t_{1}$ independence, we can simplify our equations by choosing
\begin{equation}
t_{1}=0\,.\label{t1-0}
\end{equation}
Then Eq. (\ref{W-repr-t3}) becomes
\begin{equation}
W(g)=\ln\det_{1\leq r,s\leq M+1}D_{rs}+\int_{-\infty}^{0}dt\left[
L_{\mathrm{bos}}(\pi^{(1)})-L_{\mathrm{bos}}(\pi^{\mathrm{bar}})\right]
+\int_{0}^{\infty}dt\left[  L_{\mathrm{bos}}(\pi^{(1)})-L_{\mathrm{bos}}
(\pi^{\mathrm{vac}})\right]  \,.\label{W-repr-t4}
\end{equation}
In the rest of the calculation we will assume the choice (\ref{t1-0}).

Using the expression (\ref{L-bos-def}) for $L_{\mathrm{bos}}$ and the result
(\ref{pi-via-Phi-0}) for $\pi^{(1)}$ at $t<0$, we find
\begin{align}
L_{\mathrm{bos}}(\pi^{(1)}) &  =\sum\limits_{\alpha\beta}\frac{1}{2}
\pi_{\alpha}^{(1)}\left[  \left(  V^{-1}\right)  _{\alpha\beta}\pi_{\beta
}^{(1)}\right]  =\sum\limits_{\alpha}\frac{1}{2}\pi_{\alpha}^{(1)}\left\{
\sum\limits_{r,s=1}^{M+1}\left(  D^{-1}\right)  _{sr}\left[  \tilde{\phi
}_{\mathrm{vac},g}^{r}(t)\Gamma_{\alpha}\phi_{\mathrm{bar}}^{s}(t)\right]
\right\} \nonumber\\
&  =\frac{1}{2}\left\{  \sum\limits_{r,s=1}^{M+1}\left(  D^{-1}\right)
_{sr}\tilde{\phi}_{\mathrm{vac},g}^{r}(t)\left(  \sum\limits_{\alpha}
\pi_{\alpha}^{(1)}\Gamma_{\alpha}\right)  \phi_{\mathrm{bar}}^{s}(t)\right\}
\,.
\end{align}
Here
\begin{equation}
\tilde{\phi}_{\mathrm{vac},g}^{r}(t)\left(  \sum\limits_{\alpha}\pi_{\alpha
}^{(1)}\Gamma_{\alpha}\right)  \phi_{\mathrm{bar}}^{s}(t)=\varepsilon
_{\mathrm{bar}}^{s}D_{rs}\,.
\end{equation}
Thus
\begin{equation}
L_{\mathrm{bos}}(\pi^{(1)}(t))=\frac{1}{2}\sum\limits_{r,s=1}^{M+1}
\varepsilon_{\mathrm{bar}}^{s}\,.\label{L-t-zero}
\end{equation}
We see that the RHS is $t$ independent. Taking the limit $t\rightarrow-\infty$
and using Eq. (\ref{pi-1-t-asympt}), we obtain
\begin{equation}
L_{\mathrm{bos}}(\pi^{\mathrm{bar}})=\frac{1}{2}\sum\limits_{r,s=1}
^{M+1}\varepsilon_{\mathrm{bar}}^{s}\,.\label{L-bar-zero}
\end{equation}
Combining Eqs. (\ref{L-t-zero}), (\ref{L-bar-zero}), we find
\begin{equation}
\int_{-\infty}^{0}dt\left[  L_{\mathrm{bos}}(\pi^{(1)})-L_{\mathrm{bos}}
(\pi^{\mathrm{bar}})\right]  =0\,.
\end{equation}
In a similar way one can show that
\begin{equation}
\int_{0}^{\infty}dt\left[  L_{\mathrm{bos}}(\pi^{(1)})-L_{\mathrm{bos}}
(\pi^{\mathrm{vac}})\right]  =0\,.
\end{equation}
As a result, Eq. (\ref{W-repr-t3}) can be simplified:
\begin{equation}
W_{\mathrm{p.i.}}(g)=\ln\det_{1\leq r,s\leq M+1}D_{rs}
\,.\label{W-general-res-0}
\end{equation}
We use the label \emph{p.i.} in order to emphasize that this result is
obtained in the path integral approach.

\subsection{Closed system of differential equations}

The $U$ evolution (\ref{U-Texp}) of functions 
$\tilde{\phi}_{\mathrm{vac}}^{s}(t)$, 
$\phi_{\mathrm{bar}}^{s}(t)$, $g(t)$ described by Eqs.
(\ref{vac-t}), (\ref{Phi-n-bar-def}), and (\ref{g-t-def}) can be represented
in terms of the differential equations
\begin{align}
\frac{d}{dt}\tilde{\phi}_{\mathrm{vac}}^{s}(t) &  =\tilde{\phi}_{\mathrm{vac}
}^{s}(t)\sum\limits_{\alpha}\pi_{\alpha}^{(1)}(t)\Gamma_{\alpha}
\,,\label{de-Phi-vac}\\
\frac{d}{dt}\phi_{\mathrm{bar}}^{s}(t) &  =-\sum\limits_{\alpha}\pi_{\alpha
}^{(1)}(t)\Gamma_{\alpha}\phi_{\mathrm{bar}}^{s}(t)\,,\label{de-Phi-bar}\\
\frac{d}{dt}g(t) &  =g(t)\sum\limits_{\alpha}\pi_{\alpha}^{(1)}(t)\Gamma
_{\alpha}\,.\label{dg-dt-eq}
\end{align}
Using expressions (\ref{pi-plus}) and (\ref{pi-minus}) for 
$\pi_{\alpha}^{(1)}$, we can rewrite these differential equations in the form
\begin{align}
\frac{d}{dt}\tilde{\phi}_{\mathrm{vac}}^{s}(t) &  =\tilde{\phi}_{\mathrm{vac}
}^{s}(t)\sum\limits_{\alpha}\Pi_{\alpha}^{(\pm)}\left(  \tilde{\phi
}_{\mathrm{vac}}(t),\phi_{\mathrm{bar}}(t),g(t)\right)  \Gamma_{\alpha
}\,,\label{de-1}\\
\frac{d}{dt}\phi_{\mathrm{bar}}^{s}(t) &  =-\sum\limits_{\alpha}\Pi_{\alpha
}^{(\pm)}\left(  \tilde{\phi}_{\mathrm{vac}}(t),\phi_{\mathrm{bar}
}(t),g(t)\right)  \Gamma_{\alpha}\phi_{\mathrm{bar}}^{s}(t)\,,\\
\frac{d}{dt}g(t) &  =g(t)\sum\limits_{\alpha}\Pi_{\alpha}^{(\pm)}\left(
\tilde{\phi}_{\mathrm{vac}}(t),\phi_{\mathrm{bar}}(t),g(t)\right)
\Gamma_{\alpha}\,,\label{de-3}
\end{align}
where $\Pi_{\alpha}^{(+)}$ should be used for $t>0$ and $\Pi_{\alpha}^{(-)}$
for $t<0$ (we work with $t_{1}=0$). This is a closed coupled system of
differential equations for functions $\phi_{\mathrm{vac}}^{s}(t),\phi
_{\mathrm{bar}}^{s}(t),g(t)$. In addition we have the boundary conditions
which follow from Eqs. (\ref{vac-t}), (\ref{Phi-n-bar-def}), and
(\ref{g-t-def})
\begin{equation}
\tilde{\phi}_{\mathrm{vac}}^{s}(t)\overset{t\rightarrow\infty}{\longrightarrow
}\exp\left(  t\varepsilon_{\mathrm{vac}}^{s}\right)  \left(  \phi
_{\mathrm{vac}}^{s}\right)  ^{+}\,,\label{Phi-t-vac-boundary}
\end{equation}
\begin{equation}
\phi_{\mathrm{bar}}^{s}(t)\overset{t\rightarrow-\infty}{\longrightarrow}
\exp\left(  -t\varepsilon_{\mathrm{bar}}^{s}\right)  \phi_{\mathrm{bar}}
^{s}\,,\label{Phi-t-bar-boundary}
\end{equation}
\begin{equation}
g(0)=g\,.\label{g-t-boundary}
\end{equation}
Note that the $g$ dependence appears via the boundary condition
(\ref{g-t-boundary}). Solving the system of differential equations
(\ref{de-1})--(\ref{de-3}) with the boundary conditions
(\ref{Phi-t-vac-boundary})--(\ref{g-t-boundary}), inserting the solution into
Eq. (\ref{W-general-res}) and using Eq. (\ref{D-via-Phi-def}), we can compute
the functional $W_{\mathrm{p.i.}}(g)$ (\ref{W-general-res}):
\begin{equation}
W_{\mathrm{p.i.}}(g)=\ln\det_{1\leq r,s\leq M+1}\left[  \tilde{\phi
}_{\mathrm{vac},g}^{r}(0)\cdot\phi_{\mathrm{bar}}^{s}(0)\right]
\,\,.\label{W-general-res}
\end{equation}

This construction solves the problem of the calculation of the functional
$W_{\mathrm{p.i.}}(g)$. However, one can wonder how this representation for
$W_{\mathrm{p.i.}}(g)$ derived in the framework of the path integral formalism
is related to our results obtained in the Schr\"{o}dinger approach. This
question is studied in the next section.

\section{Equivalence of the Schr\"{o}dinger and path integral approaches}

\label{Equivalence-section}

\setcounter{equation}{0} 

\subsection{Problem of equivalence}

Our analysis of models with the nontrivial vacuum has resulted in two
expressions for the generating functional $W(g)$. One expression
(\ref{W-sch-I-vac-bar}) was derived from the Schr\"{o}dinger equation using
the operator formalism. The other result (\ref{W-general-res}) for $W(g)$ was
obtained in the path integral formalism. Certainly both approaches are
equivalent and should lead to the same results. However, in order to see this
equivalence explicitly some extra work is needed. In Sec.
\ref{PI-trivial-vacuum-section} we demonstrate the equivalence for the simple
case of models with the trivial vacuum and after that turn to the much more
interesting models with the nontrivial vacuum.

\subsection{Models with the trivial vacuum}

\label{PI-trivial-vacuum-section}

Let us start from the simplest case when the physical vacuum coincides with
the bare one. Then the set of occupied vacuum levels is empty
\begin{equation}
\left\{  \tilde{\phi}_{\mathrm{vac}}^{s}(t)\right\}  =\emptyset\,
\end{equation}
and
\begin{equation}
E_{\mathrm{vac}}=0\,.
\end{equation}
The baryon is described by the Hartree equations
(\ref{bar-Hartree})--(\ref{E-sum-eps-vac}) corresponding to one occupied
level:
\begin{equation}
\sum\limits_{\alpha}\left(  \pi_{\alpha}^{\mathrm{bar}}\Gamma_{\alpha}\right)
\phi_{\mathrm{bar}}=\varepsilon_{\mathrm{bar}}\phi_{\mathrm{bar}
}\,,\label{bar-Hartree-0}
\end{equation}
\begin{equation}
\pi_{\alpha}^{\mathrm{bar}}=\sum\limits_{b}V_{\alpha\beta}\left(
\phi_{\mathrm{bar}}^{+}\Gamma_{\beta}\phi_{\mathrm{bar}}\right)
\,,\label{pi-bar-Phi-0}
\end{equation}
\begin{equation}
E_{\mathrm{bar}}=\frac{1}{2}\varepsilon_{\mathrm{bar}}\,.
\end{equation}
Since the physical vacuum coincides with the bare one, we have
\begin{equation}
\langle0,t|=\langle0|\,,
\end{equation}
\begin{equation}
\langle0|b=0\,.
\end{equation}
Therefore the matrix element (\ref{p-multi}) vanishes at $t>t_{1}$. According
to Eq. (\ref{t1-0}) we work with $t_{1}=0$. Thus equation (\ref{p-multi})
results in
\begin{equation}
\langle0|T\left\{  \left[  gb(0)\right]  \left[  b^{+}(t)\Gamma_{\beta
}b(t)\right]  \right\}  |B\rangle_{\pi^{(1)}}=\theta(-t)\langle0|\left[
g(t)b\right]  \left(  b^{+}\Gamma_{\beta}b\right)  |B,t\rangle_{\pi^{(1)}
}\,=\theta(-t)\left[  g(t)\Gamma_{\beta}\phi_{\mathrm{bar}}(t)\right]
\,.\label{b-ME-triv-1}
\end{equation}

On the other hand, Eq. (\ref{vac-g-psi-B}) gives
\begin{equation}
\langle0|\left[  gb(0)\right]  |B\rangle_{\pi^{(1)}}=g(t)\phi_{\mathrm{bar}
}(t)=\mathrm{const}\,,\label{g-Phi-const}
\end{equation}
which is $t$ independent. Now we insert Eqs. (\ref{b-ME-triv-1}) and
(\ref{g-Phi-const}) into Eq. (\ref{pi-t-res}):
\begin{equation}
\pi_{\alpha}^{(1)}(t)=\theta(-t)\sum\limits_{\beta}V_{\alpha\beta
}\frac{\left[  g(t)\Gamma_{\beta}\phi_{\mathrm{bar}}(t)\right]  }{\left[
g(t)\phi_{\mathrm{bar}}(t)\right]  }\,.\label{pi-1-M0}
\end{equation}
According to Eqs. (\ref{de-Phi-bar}) and (\ref{dg-dt-eq}) we have the
differential equations
\begin{align}
\frac{d}{dt}\phi_{\mathrm{bar}}(t) &  =-\sum\limits_{\alpha}\pi_{\alpha}
^{(1)}(t)\Gamma_{\alpha}\phi_{\mathrm{bar}}(t)\,,\label{de-Phi-bar-M0}\\
\frac{d}{dt}g(t) &  =g(t)\sum\limits_{\alpha}\pi_{\alpha}^{(1)}(t)\Gamma
_{\alpha}\label{de-g-M0}
\end{align}
with the boundary conditions (\ref{Phi-t-bar-boundary}), (\ref{g-t-boundary})
\begin{equation}
\phi_{\mathrm{bar}}(t)\overset{t\rightarrow-\infty}{\longrightarrow}
\exp\left(  -t\varepsilon_{\mathrm{bar}}\right)  \phi_{\mathrm{bar}
}\,,\label{Phi-bar-as-M0}
\end{equation}
\begin{equation}
g(0)=g\,.\label{g0-M0}
\end{equation}
According to Eqs. (\ref{pi-1-t-asympt}), (\ref{bar-Hartree-0}), and
(\ref{de-g-M0}) the asymptotic behavior of $g(t)$ at $t\rightarrow-\infty$ is
\begin{equation}
g(t)\overset{t\rightarrow-\infty}{=}\mathrm{const\,}\phi_{\mathrm{bar}}
^{+}\exp\left(  t\varepsilon_{\mathrm{bar}}\right)  \,\label{g-as-M0}
\end{equation}
and the path integral result (\ref{W-general-res}) for $W(g)$ becomes
\begin{equation}
W_{\mathrm{p.i.}}(g)=\ln\left[  g(t)\phi_{\mathrm{bar}}(t)\right]
\,.\label{W-g-Phi-bar}
\end{equation}
Thus the path integral calculation of the functional $W_{\mathrm{p.i.}}(g)$ is
reduced to solving differential equations (\ref{de-Phi-bar-M0}),
(\ref{de-g-M0}) with $\pi^{(1)}$ given by Eq. (\ref{pi-1-M0}) and with
boundary conditions (\ref{Phi-bar-as-M0}), (\ref{g0-M0}), (\ref{g-as-M0}).

According to Eq. (\ref{g-Phi-const}) the RHS of Eq. (\ref{W-g-Phi-bar}) is $t
$ independent. Taking the limit $t\rightarrow-\infty$ and inserting the
asymptotic behavior (\ref{g-as-M0}), we can determine the constant on the RHS
of (\ref{g-as-M0}):
\begin{equation}
g(t)\overset{t\rightarrow-\infty}{=}e^{W_{\mathrm{p.i.}}(g)}\mathrm{\,}
\phi_{\mathrm{bar}}^{+}\exp\left(  t\varepsilon_{\mathrm{bar}}\right)  \,.
\end{equation}

Let us show that this path integral result $W_{\mathrm{p.i.}}(g)$ is
equivalent to the Schr\"{o}dinger approach discussed in Sec.
\ref{Classical-dynamics-section}. Note that our path integral treatment dealt
with the case $L_{mn}=0$. In the Schr\"{o}dinger approach this case was
discussed in Sec. \ref{L-0-case-section}.

In order to match the equations of Sec. \ref{L-0-case-section} with the
equations of the path integral approach one has to use the following
``dictionary'' connecting the two formalisms:
\begin{align}
q(t) &  =g(t)\,,\label{q-PI}\\
p(t) &  =e^{-W_{\mathrm{p.i.}}(g)}\phi_{\mathrm{bar}}(t)\,,\label{p-PI}\\
\varepsilon_{0} &  =\varepsilon_{\mathrm{bar}}\,,\\
I(g) &  =\left[  g(t)\phi_{\mathrm{bar}}(t)\right]  \,.
\end{align}
On the LHS we have the quantities appearing in Sec. \ref{L-0-case-section}
within the Schr\"{o}dinger approach whereas the RHS is represented by the
objects which were used in the path integral method.

Now it is easy to see that the boundary conditions (\ref{q-I}), (\ref{p-I}),
and (\ref{q-0-g}) correspond to Eqs. (\ref{q-PI}), (\ref{p-PI}), and
(\ref{g0-M0}), respectively. The Hamilton equations (\ref{pq-Hamilton-eq}) are
mirrors of Eqs. (\ref{de-Phi-bar-M0})--(\ref{de-g-M0}).

\subsection{Variables $Q^{(L)}$, $P^{(L)}$}

In the previous section we have demonstrated the equivalence of the
Schr\"{o}dinger and path integral approaches in the case of systems with the
trivial vacuum. Now we want to consider the general case when the vacuum is nontrivial.

Our main lesson from the analysis of the systems with the trivial vacuum is
that the equivalence between the two approaches can be established on the
basis of the Hamiltonian description of trajectories used for the calculation
of the functional $W(g)$.

In the case of the nontrivial vacuum we already have the representation for
$W(g)$ in terms of trajectories obeying the differential equations
(\ref{de-Phi-vac})--(\ref{dg-dt-eq}). Now we want to rewrite these
differential equations in the Hamiltonian form.

To this aim we must introduce the canonical variables. Note that the original
differential equations (\ref{de-Phi-vac})--(\ref{dg-dt-eq}) are written in
terms of single-particle wave functions $\tilde{\phi}_{\mathrm{vac}}^{s}(t)$,
$\phi_{\mathrm{bar}}^{s}(t)$. As we will see below, the Hamiltonian formalism
is based on $M$- and $\left(  M+1\right)  $-particle wave functions including
the Slater determinants associated with the states (\ref{vac-0-t}) and
(\ref{B-t})
\begin{equation}
P_{m_{1}\ldots m_{M+1}}^{(M+1)}(t)=\underset{m_{1}\ldots m_{M+1}
}{\mathrm{Antisym}}\prod\limits_{s=1}^{M+1}\left[  \phi_{\mathrm{bar}}
^{s}(t)\right]  _{m_{s}}=\frac{1}{(M+1)!}\det_{1\leq r,s\leq M+1}\left[
\phi_{\mathrm{bar}}^{r}(t)\right]  _{m_{s}}\,,\label{P-M1-def}
\end{equation}
\begin{equation}
Q_{m_{1}\ldots m_{M}}^{(M)}(t)=\underset{m_{1}\ldots m_{M}}{\mathrm{Antisym}
}\prod\limits_{s=1}^{M}\left[  \tilde{\phi}_{\mathrm{vac}}^{s}(t)\right]
_{m_{s}}=\frac{1}{M!}\det_{1\leq r,s\leq M}\left[  \tilde{\phi}_{\mathrm{vac}
}^{r}(t)\right]  _{m_{s}}\,.\label{X-M-def}
\end{equation}
We assume that the antisymmetrization operator Antisym is normalized by the
condition
\begin{equation}
\left(  \mathrm{Antisym}\right)  ^{2}=\mathrm{Antisym}\,.
\end{equation}
Assuming the choice (\ref{t1-0}) $t_{1}=0$, we deal with the $M$-particle
states at $t>0$ and with the $(M+1)$-particle states at $t<0$. Therefore the
$M+1$ particle ``wave function'' $P_{m_{1}\ldots m_{M+1}}^{(M+1)}$ is relevant
for the description of the region $t<0$ whereas the variable $Q_{m_{1}\ldots
m_{M}}^{(M)}$ will be used at $t>0$.

Now we want to introduce the canonically conjugate variables for
$Q_{m_{1}\ldots m_{M}}^{(M)}$ and for $P_{m_{1}\ldots m_{M+1}}^{(M+1)}$:
\begin{equation}
P_{m_{1}\ldots m_{M}}^{(M)}(t)=(M+1)\sum\limits_{m_{M+1}}g_{m_{M+1}
}(t)\underset{m_{1}\ldots m_{M+1}}{\mathrm{Antisym}}\prod\limits_{s=1}
^{M+1}\left[  \phi_{\mathrm{bar}}^{s}(t)\right]  _{m_{s}}\,,\label{P-M-def}
\end{equation}
\begin{equation}
Q_{m_{1}\ldots m_{M+1}}^{(M+1)}(t)=\underset{m_{1}\ldots m_{M+1}
}{\mathrm{Antisym}}\left\{  g_{m_{M+1}}(t)\prod\limits_{s=1}^{M}\left[
\tilde{\phi}_{\mathrm{vac}}^{s}(t)\right]  _{m_{s}}\right\}  =\frac{1}
{(M+1)!}\det\left[  \tilde{\phi}_{\mathrm{vac},g}^{r}(t)\right]  _{m_{s}
}\,.\label{X-M1-def}
\end{equation}
On the RHS of the Eq. (\ref{X-M1-def}) we use notation $\tilde{\phi
}_{\mathrm{vac},g}^{r}(t)$ (\ref{Phi-vac-g}).

\subsection{Differential equations for $Q^{(L)}$, $P^{(L)}$}

Now let us rewrite the differential equations (\ref{de-Phi-vac}
)--(\ref{dg-dt-eq}) in terms of variables $Q^{(M)}$, $P^{(M)}$ for $t>0$ and
in terms of $Q^{(M+1)}$, $P^{(M+1)}$ for $t<0$:
\begin{align}
\frac{d}{dt}Q_{m_{1}\ldots m_{L}}^{(L)}(t)  &  =\sum\limits_{\alpha}
\pi_{\alpha}^{(1)}(t)\left[  \Gamma_{\alpha}\cdot Q^{(L)}\right]
_{m_{1}\ldots m_{L}}\,,\label{dx-dt-1}\\
\frac{d}{dt}P_{m_{1}\ldots m_{L}}^{(L)}(t)  &  =-\sum\limits_{\alpha}
\pi_{\alpha}^{(1)}(t)\left[  P^{(L)}\cdot\Gamma_{\alpha}\right]  _{m_{1}\ldots
m_{L}}\,,\label{dp-dt-1}
\end{align}
where
\begin{equation}
L=\left\{
\begin{array}
[c]{ll}
M & \mathrm{if}\,t>0\,,\\
M+1 & \mathrm{if}\,t<0\,,
\end{array}
\right. \label{L-via-M}
\end{equation}
and
\begin{align}
\left[  \Gamma_{\alpha}\cdot P^{(L)}\right]  _{m_{1}\ldots m_{L}}  &
=\sum\limits_{n}\left[  \left(  \Gamma_{\alpha}\right)  _{m_{1}n}
P_{nm_{2}\ldots m_{L}}^{(L)}(t)+\left(  \Gamma_{\alpha}\right)  _{m_{2}
n}P_{m_{1}n\ldots m_{L}}^{(L)}(t)+\ldots+\left(  \Gamma_{\alpha}\right)
_{m_{L}n}P_{m_{1}\ldots m_{L-1}n}^{(L)}(t)\right]  \,,\\
\left[  Q^{(L)}\cdot\Gamma_{\alpha}\right]  _{m_{1}\ldots m_{L}}  &
=\sum\limits_{n}\left[  Q_{nm_{2}\ldots m_{L}}^{(L)}(t)\left(  \Gamma_{\alpha
}\right)  _{nm_{1}}+Q_{m_{1}n\ldots m_{L}}^{(L)}(t)\left(  \Gamma_{\alpha
}\right)  _{nm_{2}}+\ldots+Q_{m_{1}\ldots m_{L-1}n}^{(L)}(t)\left(
\Gamma_{\alpha}\right)  _{nm_{L}}\right]  \,.
\end{align}
The field $\pi_{\alpha}^{(1)}(t)$ is given by
\begin{equation}
\pi_{\alpha}^{(1)}(t)=\sum\limits_{\beta}V_{\alpha\beta}\frac{\left[
Q^{(L)}(t)\cdot\Gamma_{\beta}\cdot P^{(L)}(t)\right]  }{\left[  Q^{(L)}
(t)\cdot P^{(L)}(t)\right]  }\,,\label{pi-1-via-X-P}
\end{equation}
where $L$ is defined by Eq. (\ref{L-via-M}) and
\begin{align}
\left[  Q^{(L)}\cdot P^{(L)}\right]   &  =\sum\limits_{m_{1}\ldots m_{L}
}Q_{m_{1}\ldots m_{L}}^{(L)}P_{m_{1}\ldots m_{L}}^{(L)}
\,,\label{X-P-contraction}\\
\left[  Q^{(L)}\cdot\Gamma_{\alpha}\cdot P^{(L)}\right]   &  =L\sum
\limits_{knm_{2}\ldots m_{L}}Q_{km_{2}\ldots m_{L}}^{(L)}\left(
\Gamma_{\alpha}\right)  _{kn}P_{nm_{2}\ldots m_{L}}^{(L)}\,.
\end{align}

Obviously
\begin{equation}
\frac{d}{dt}\left[  Q^{(L)}(t)\cdot P^{(L)}(t)\right]  =0\label{XP-L-const}
\end{equation}
and
\begin{equation}
\left(  M+1\right)  \sum\limits_{m_{1}\ldots m_{M+1}}\left[  Q^{(M+1)}(0)\cdot
P^{(M+1)}(0)\right]  =\left[  Q^{(M)}(0)\cdot P^{(M)}(0)\right]
\,.\label{X-P-contr-equal}
\end{equation}

\subsection{Hamiltonian interpretation of differential equations}

Let us introduce the Poisson bracket
\begin{equation}
\left\{  Q_{m_{1}\ldots m_{L}},P_{n_{1}\ldots n_{L}}\right\}  =\frac{1}
{L!}\det_{jk}\left\|  \delta_{m_{j}n_{k}}\right\|  \,.
\end{equation}
Then
\begin{align}
\left\{  Q_{m_{1}\ldots m_{L}}^{(L)},\left[  Q^{(L)}\cdot\Gamma_{\beta}\cdot
P^{(L)}\right]  \right\}   &  =\left[  Q^{(L)}\cdot\Gamma_{\beta}\right]
_{m_{1}\ldots m_{L}}\,,\\
\left\{  P_{m_{1}\ldots m_{L}}^{(L)},\left[  Q^{(L)}\cdot\Gamma_{\beta}\cdot
P^{(L)}\right]  \right\}   &  =-\left[  \Gamma_{\beta}\cdot Q^{(L)}\right]
_{m_{1}\ldots m_{L}}\,.
\end{align}
Now Eqs. (\ref{dx-dt-1}) and (\ref{dp-dt-1}) take the form
\begin{align}
\frac{d}{dt}Q_{m_{1}m_{2}\ldots m_{L}}^{(L)}  &  =\sum\limits_{\alpha}
\pi_{\alpha}(t)\left\{  Q_{m_{1}m_{2}\ldots m_{L}}^{(L)},\left[  Q\cdot
\Gamma_{\alpha}\cdot P\right]  \right\}  \,,\\
\frac{d}{dt}P_{m_{1}m_{2}\ldots m_{L}}^{(L)}  &  =\sum\limits_{\alpha}
\pi_{\alpha}(t)\left\{  P_{m_{1}m_{2}\ldots m_{L}}^{(L)},\left[  Q\cdot
\Gamma_{\alpha}\cdot P\right]  \right\}  \,.
\end{align}

Inserting $\pi_{a}^{(1)}$ from Eq. (\ref{pi-1-via-X-P}), we can rewrite these
equations in the form
\begin{align}
\frac{d}{dt}Q_{m_{1}m_{2}\ldots m_{L}}^{(L)} &  =\frac{\left\{  Q_{m_{1}
m_{2}\ldots m_{L}}^{(L)},H\right\}  }{\left[  Q^{(L)}(t)\cdot P^{(L)}
(t)\right]  }\,,\label{Q-L-H-eq}\\
\frac{d}{dt}P_{m_{1}m_{2}\ldots m_{L}}^{(L)} &  =\frac{\left\{  P_{m_{1}
m_{2}\ldots m_{L}}^{(L)},H\right\}  }{\left[  Q^{(L)}(t)\cdot P^{(L)}
(t)\right]  }\,\,,\label{P-L-H-eq}
\end{align}
where the Hamiltonian $H$ is
\begin{equation}
H=\frac{1}{2}\sum\limits_{\alpha\beta}V_{\alpha\beta}\left[  Q\cdot
\Gamma_{\alpha}\cdot P\right]  \left[  Q\cdot\Gamma_{\beta}\cdot P\right]
\,.\label{H-VPQ-antisym}
\end{equation}
According to Eq. (\ref{XP-L-const}) the denominator $\left[  Q^{(L)}(t)\cdot
P^{(L)}(t)\right]  $ is $t$ independent along any solution so that this
denominator can be treated as a constant.

\subsection{Boundary conditions}

Using Eq. (\ref{X-M-def}), we can rewrite the boundary condition
(\ref{Phi-t-vac-boundary}) in the form
\begin{equation}
Q_{m_{1}\ldots m_{M}}^{(M)}(t)\overset{t\rightarrow\infty}{\longrightarrow
}\frac{1}{M!}\exp\left(  t\sum\limits_{p=1}^{M}\varepsilon_{\mathrm{vac}}
^{p}\right)  \det_{1\leq r,s\leq M}\left(  \phi_{\mathrm{vac}}^{r}\right)
_{m_{s}}^{\ast}\,.\label{X-M-t-plus-infty}
\end{equation}

Similarly we find from Eqs. (\ref{Phi-t-bar-boundary}) and (\ref{P-M1-def})
\begin{equation}
P_{m_{1}\ldots m_{M+1}}^{(M+1)}(t)\overset{t\rightarrow-\infty}
{\longrightarrow}\frac{1}{(M+1)!}\exp\left(  -t\sum\limits_{p=1}
^{M+1}\varepsilon_{\mathrm{bar}}^{p}\right)  \det_{1\leq r,s\leq M+1}\left(
\phi_{\mathrm{bar}}^{r}\right)  _{m_{s}}\,.\label{P-N1-past}
\end{equation}
Comparing Eqs. (\ref{X-M-def}) and (\ref{X-M1-def}), we find at $t=0$
\begin{equation}
Q_{m_{1}\ldots m_{M+1}}^{(M+1)}(0)=\underset{m_{1}\ldots m_{M+1}
}{\mathrm{Antisym}}\left[  Q_{m_{1}\ldots m_{M}}^{(M)}(0)g_{m_{M+1}}\right]
\,.\label{Q-g-boundary-PI}
\end{equation}
Similarly, the comparison of Eqs. (\ref{P-M1-def}) and (\ref{P-M-def}) results in
\begin{equation}
P_{n_{1}\ldots n_{M}}^{(M)}(0)=(M+1)\sum\limits_{m_{M+1}}g_{m_{M+1}}
P_{m_{1}\ldots m_{M+1}}^{(M+1)}(0)\,.\label{P-g-boundary-PI}
\end{equation}

\subsection{Expression for $W(g)$}

According to Eqs. (\ref{D-via-Phi-def}), (\ref{P-M1-def}), and (\ref{X-M1-def})
we have
\begin{equation}
\det_{1\leq r,s\leq M+1}D_{rs}=(M+1)!\left[  Q^{(M+1)}(t)\cdot P^{(M+1)}
(t)\right]  \,.
\end{equation}
Combining this with Eq. (\ref{W-general-res-0}), we find
\begin{equation}
W_{\mathrm{p.i.}}(g)=\ln\left\{  (M+1)!\left[  Q^{(M+1)}(t)\cdot
P^{(M+1)}(t)\right]  \right\}  \,.\label{W-PI-res-1}
\end{equation}
Later we will check the equivalence of this result with the expression
(\ref{W-sch-I-vac-bar}) obtained in the Schr\"{o}dinger approach.

Note that the RHS of Eq. (\ref{W-PI-res-1}) is $t$ independent according to
Eq. (\ref{XP-L-const}). Using Eqs. (\ref{XP-L-const}) and
(\ref{X-P-contr-equal}), we can also write
\begin{equation}
W_{\mathrm{p.i.}}(g)=\ln\left\{  M!\left[  Q^{(M)}(t)\cdot P^{(M)}(t)\right]
\right\}  \,.\label{W-via-XP-PI}
\end{equation}

\subsection{Asymptotic representation}

The asymptotic behavior of $P_{m_{1}\ldots m_{M}}^{(M)}(t)$ at $t\rightarrow
+\infty$ is given by the expression
\begin{equation}
P_{m_{1}\ldots m_{M}}^{(M)}(t)\overset{t\rightarrow\infty}{\longrightarrow
}I(g)\exp\left(  -t\sum\limits_{p=1}^{M}\varepsilon_{\mathrm{vac}}^{p}\right)
\det_{1\leq r,s\leq M}\left(  \phi_{\mathrm{vac}}^{r}\right)  _{m_{s}
}\label{P-M-t-plus-infty}
\end{equation}
with some constant $I(g)$. According to Eq. (\ref{XP-L-const}) the combination
$\left[  Q^{(M)}(t)\cdot P^{(M)}(t)\right]  $ is $t$ independent so that it
can be computed via its limit at $t\rightarrow+\infty$. Using Eqs.
(\ref{X-M-t-plus-infty}) and (\ref{P-M-t-plus-infty}), we obtain
\begin{equation}
\left[  Q^{(M)}(t)\cdot P^{(M)}(t)\right]  =\lim_{t\rightarrow+\infty}\left[
Q^{(M)}(t)\cdot P^{(M)}(t)\right]  =I(g)\,.\label{XP-I}
\end{equation}
Combining this with Eq. (\ref{X-P-contr-equal}), we find
\begin{equation}
\left[  Q^{(M+1)}(t)\cdot P^{(M+1)}(t)\right]  =\frac{I(g)}{M+1}
\,.\label{XP-2}
\end{equation}
Taking the limit $t\rightarrow-\infty$ in this equation, we can fix the
constant in the asymptotic expression
\begin{equation}
Q_{m_{1}\ldots m_{M+1}}^{(M+1)}(t)\overset{t\rightarrow-\infty}
{\longrightarrow}\frac{I(g)}{M+1}\exp\left(  t\sum\limits_{p=1}^{M+1}
\varepsilon_{\mathrm{bar}}^{p}\right)  \det_{1\leq r,s\leq M+1}\left(
\phi_{\mathrm{bar}}^{r}\right)  _{m_{s}}^{\ast}\,.
\end{equation}

Inserting expression (\ref{XP-I}) into Eq. (\ref{W-via-XP-PI}), we obtain
\begin{equation}
W_{\mathrm{p.i.}}(g)=\ln I(g)\,.\label{W-ln-I}
\end{equation}

\subsection{Comparison of path integral and Schr\"{o}dinger results}

\label{PI-nontrivial-vacuum-section}

We have computed the functional $W(g)$ using the operator (the
Schr\"{o}dinger) approach and the path integral method. The two results are
given by Eqs. (\ref{W-sch-I-vac-bar}) and (\ref{W-ln-I}), respectively. Our
path integral result (\ref{W-ln-I}) is more general since it deals with models
where the physical ground state contains $N_{c}M$ quarks with arbitrary $M$
whereas the Schr\"{o}dinger result (\ref{W-sch-I-vac-bar}) corresponds to
$M=1$.

Comparing equations (\ref{W-sch-I-vac-bar}) and (\ref{W-ln-I}), we see that
they give the same result for $W(g)$ if
\begin{equation}
I(g)=\left\{  I_{\mathrm{vac}}\left(  k(g^{\ast})\right)  \right\}  ^{\ast
}I_{\mathrm{bar}}\left(  Q(g)\right)  \,.\label{I-Iv-Ib-identity}
\end{equation}
The functional $I(g)$ arises in the path integral approach whereas the
functionals $I_{\mathrm{vac}}\left(  k(g^{\ast})\right)  $ and
$I_{\mathrm{bar}}\left(  Q(g)\right)  $come from the Schr\"{o}dinger approach.
The aim of this section is to give an independent derivation of the identity
(\ref{I-Iv-Ib-identity}) which directly establishes the equivalence of the
path integral and Schr\"{o}dinger methods.

The functionals appearing in Eq. (\ref{I-Iv-Ib-identity}) can be represented
in terms of trajectories. In the path integral approach the trajectory
representation for $I(g)$ arises via the saddle point method. The
Schr\"{o}dinger representation for $W(g)$ was formulated in Sec.
\ref{W-Sch-trajectories-section} in terms of trajectories $q_{i}
^{\mathrm{vac}}(t)$, $p_{i}^{\mathrm{vac}}(t)$ for the vacuum sector and
trajectories $Q_{ij}^{\mathrm{bar}}(t)$, $P_{ij}^{\mathrm{bar}}(t)$ for the
baryon sector. Let us show that the connection between the trajectories of the
Schr\"{o}dinger approach and the trajectories $P^{(L)}(t),Q^{(L)}(t)$ (with
$L=1,2$) of the path integral approach is given by equations
\begin{align}
Q_{i}^{(1)}(t) &  =I_{\mathrm{vac}}^{\ast}\left[  p_{i}^{\mathrm{vac}
}(-t)\right]  ^{\ast}\,,\label{Q1-via-p-vac}\\
P_{i}^{(1)}(t) &  =I_{\mathrm{bar}}\left[  q_{i}^{\mathrm{vac}}(-t)\right]
^{\ast}\,,\label{P1-via-q-vac}
\end{align}
\begin{align}
Q_{ij}^{(2)}(t) &  =\frac{1}{2}I_{\mathrm{vac}}^{\ast}Q_{ij}^{\mathrm{bar}
}(t)\,,\label{Q-Q-bar}\\
P_{ij}^{(2)}(t) &  =I_{\mathrm{bar}}P_{ij}^{\mathrm{bar}}(t)\,.\label{P-P-bar}
\end{align}
Remember that the trajectories $q_{i}^{\mathrm{vac}}(t)$,
$p_{i}^{\mathrm{vac}}(t)$ were introduced in 
Sec. \ref{W-Sch-trajectories-section}
for $t<0$ whereas the trajectories $P^{(1)}(t)$, $Q^{(1)}(t)$ are defined at
$t>0$. Note that the equations for the trajectories are invariant under the
combination of the time reversal transformation $t\rightarrow-t$ with the
complex conjugation and with the exchange $P^{(1)}\leftrightarrow Q^{(1)}$.
Therefore functions $P^{(1)}(t)$, $Q^{(1)}(t)$ defined by Eqs.
(\ref{Q1-via-p-vac}) and (\ref{P1-via-q-vac}) automatically obey the equations
of motion.

Next, the equation for trajectories are invariant under the rescaling
transformations
\begin{equation}
Q^{(L)}\rightarrow a_{L}Q^{(L)}\,,\quad P^{(L)}\rightarrow b_{L}P^{(L)}\,.
\end{equation}
Therefore various $t$ independent factors appearing in Eqs.
(\ref{Q1-via-p-vac})--(\ref{P-P-bar}) do not violate the equations of motion.

Let us summarize. Eqs. (\ref{Q1-via-p-vac})--(\ref{P-P-bar}) express the
trajectories $Q^{(L)}(t)$, $P^{(L)}(t)$ of the path integral approach via the
trajectories $q^{\mathrm{vac}}(t)$, $p^{\mathrm{vac}}(-t)$, 
$Q^{\mathrm{bar}}(t)$, $Q^{\mathrm{bar}}(t)$ of the Schr\"{o}dinger approach. If the
Schr\"{o}dinger trajectories obey the equations of motion associated with the
Schr\"{o}dinger approach then the trajectories $Q^{(L)}(t)$, $P^{(L)}(t)$
defined by Eqs. (\ref{Q1-via-p-vac})--(\ref{P-P-bar}) satisfy the classical
equations motion (\ref{Q-L-H-eq}), (\ref{P-L-H-eq}) corresponding to the path
integral approach.

The next step is to check that the boundary conditions
(\ref{X-M-t-plus-infty})--(\ref{P-g-boundary-PI}) of the path integral approach also follow from the
equations of the Schr\"{o}dinger approach.

According to Eq. (\ref{vac-boundary-3}) we have
\begin{equation}
P^{(1)}(0)=I_{\mathrm{bar}}k^{\ast}\,.\label{P1-k}
\end{equation}
Using Eqs. (\ref{p-d-W-k}) and (\ref{Q1-via-p-vac}), we find
\begin{equation}
\frac{\partial W_{\mathrm{vac}}(k_{n})}{\partial k_{n}}=\frac{1}
{I_{\mathrm{vac}}}\left[  Q_{n}^{(1)}(0)\right]  ^{\ast}\,.
\end{equation}
Let us insert this equation into Eq. (\ref{bar-vac-2}):
\begin{equation}
\frac{1}{I_{\mathrm{vac}}^{\ast}}\left[  g_{j}Q_{i}^{(1)}(0)-g_{i}Q_{j}
^{(1)}(0)\right]  =Q_{ij}\,.
\end{equation}
According to Eqs. (\ref{bar-boundary-0}) and (\ref{P-P-bar}) we have
\begin{equation}
I_{\mathrm{vac}}^{\ast}Q_{ij}=2Q_{ij}^{(2)}(0)\,.
\end{equation}
Thus
\begin{equation}
g_{j}Q_{i}^{(1)}(0)-g_{i}Q_{j}^{(1)}(0)=2Q_{ij}^{(2)}(0)\,.
\end{equation}
This is nothing else but the boundary condition (\ref{Q-g-boundary-PI}) of the
path integral approach.

Now let us check the boundary condition (\ref{P-g-boundary-PI}). We start from
equation (\ref{bar-vac-1}) and insert $k$ from Eq. (\ref{P1-k})
\begin{equation}
\sum\limits_{j}\frac{\partial W_{\mathrm{bar}}(Q)}{\partial Q_{ij}}
g_{j}=\frac{1}{I_{\mathrm{bar}}}P_{i}^{(1)}(0)\,.\label{dW-g-P1}
\end{equation}
Next we combine Eq. (\ref{p-d-W-r}) with Eq. (\ref{P-P-bar})
\begin{equation}
\frac{\partial W_{\mathrm{bar}}(Q)}{\partial Q_{ij}}=\frac{2}{I_{\mathrm{bar}
}}P_{ij}^{(2)}(0)\,.\label{dW-P2}
\end{equation}
Inserting Eq. (\ref{dW-P2}) into Eq. (\ref{dW-g-P1}), we find
\begin{equation}
2\sum\limits_{j}P_{ij}^{(2)}(0)g_{j}=P_{i}^{(1)}(0)\,.
\end{equation}
This shows that the boundary condition (\ref{P-g-boundary-PI}) of the path
integral approach also holds.

Combining Eqs. (\ref{vac-boundary-1})--(\ref{vac-boundary-2}) with Eqs.
(\ref{P1-via-q-vac}), (\ref{Q1-via-p-vac}), we find
\begin{gather}
P_{n}^{(1)}(t)\overset{t\rightarrow\infty}{=}\delta_{n0}I_{\mathrm{bar}
}I_{\mathrm{vac}}\exp\left(  -\varepsilon_{\mathrm{vac}}^{1}t\right)
\,,\label{P1-as}\\
Q_{n}^{(1)}(t)\overset{t\rightarrow+\infty}{=}\delta_{n0}\exp\left(
\varepsilon_{\mathrm{vac}}^{1}t\right)  \,.\label{Q1-as}
\end{gather}
The last equation agrees with Eq. (\ref{X-M-t-plus-infty}).

Next, according to Eqs. (\ref{bar-boundary-1}), (\ref{bar-boundary-2}),
(\ref{Q-Q-bar}), (\ref{P-P-bar}) we have
\begin{gather}
Q_{ij}^{(2)}(t)\overset{t\rightarrow-\infty}{=}\frac{1}{2}\varepsilon
_{ij}I_{\mathrm{vac}}^{\ast}I_{\mathrm{bar}}\exp\left[  \left(  \varepsilon
_{\mathrm{bar}}^{1}+\varepsilon_{\mathrm{bar}}^{2}\right)  t\right]  \,,\\
P_{ij}^{(2)}(t)\overset{t\rightarrow-\infty}{=}\frac{1}{2}\varepsilon_{ij}
\exp\left[  -\left(  \varepsilon_{\mathrm{bar}}^{1}+\varepsilon_{\mathrm{bar}
}^{2}\right)  t\right]  \,.
\end{gather}
The last equation agrees with Eq. (\ref{P-N1-past}).

Thus we have checked that Eqs. (\ref{Q1-via-p-vac})--(\ref{P-P-bar})
connecting the trajectories of the path integral approach with the
trajectories of the Schr\"{o}dinger approach are compatible both with the
equations of motion and with the boundary conditions.

Now we can turn to the proof of the equality of functionals
$W_{\mathrm{Sch}}(g)$ and $W_{\mathrm{p.i.}}(g)$ associated with 
the Schr\"{o}dinger and path
integral methods. As was explained above, this proof reduces to the derivation
of the identity (\ref{I-Iv-Ib-identity}). Using Eqs. (\ref{XP-L-const}),
(\ref{P1-as}), (\ref{Q1-as}) we can compute $I(g)$ (\ref{XP-I}), (\ref{XP-I})
\begin{equation}
I(g)=\left[  Q^{(1)}(t)\cdot P^{(1)}(t)\right]  =2\left[  Q^{(2)}(t)\cdot
P^{(2)}(t)\right]  =I_{\mathrm{vac}}^{\ast}I_{\mathrm{bar}}\,.
\end{equation}
This proves relation (\ref{I-Iv-Ib-identity}). Thus
\begin{equation}
W_{\mathrm{Sch}}(g)=W_{\mathrm{p.i.}}(g)\,.
\end{equation}
The equivalence of the Schr\"{o}dinger and path integral representations for
$W(g)$ is checked.

\subsection{Analyticity and time contour}

The functional $\Phi_{B}(g)$ is holomorphic in $g$ by construction
(\ref{Phi-transition-def}). This analyticity is inherited in the large-$N_{c}$
limit (\ref{Phi-mod-exponential}) by the functional $W(g)$, although this
limit can be accompanied by the appearance of singularities in $W(g)$. The
representation for $W(g)$ in terms of classical trajectories is compatible
with the analyticity of $W(g)$. Indeed, the Hamiltonian (\ref{H-VPQ-antisym})
used in this representations is a holomorphic functions of coordinates and
momenta. However, the configuration of trajectories used for the construction
of $W(g)$ can lead to the singularities in $W(g)$. Examples of these
singularities can be found in the analytical expressions for $W(g)$ in the
rotator model analyzed in Sec. \ref{Asymmetric-rotator-section} [see Eq.
(\ref{W-g-rotator-res})] and in the naive quark model studied in Ref.
\cite{Pobylitsa-04}.

The analyticity of $W(g)$ in $g$ is closely related to the analyticity of the
corresponding trajectories in $t$. In our path integral construction we used
Euclidean time. However, the choice of the Euclidean time is not necessary.
For example, in the Schr\"{o}dinger approach the functional $W(g)$ can be
represented in a form which requires no time and no trajectories. The
Schr\"{o}dinger representation is formulated in terms of the action which is
defined without explicit usage of the time variable. But if one wants to
represent this action in terms of trajectories, then one has to use some
version of the time variable. Naively the choice of the time contour
(Euclidean, Minkowski or more complicated) is not important. However, even in
simple systems we can meet singularities restricting the freedom of the choice
of the time contour. Our general representation for the functional $W(g)$ in
terms of trajectories ignored this problem which requires an additional
analysis in any particular model.

\section{Conclusions}

\setcounter{equation}{0} 

\subsection{Main results}

The main aim of this work was to check the consistency of the assumptions used
for the construction of the $1/N_{c}$ expansion for the baryon wave function
in QCD. We have concentrated on those nonperturbative features which cannot be
directly proved in QCD. On the other hand, these properties can be studied in
detail in models preserving the basic structure of the $1/N_{c}$ expansion in
QCD. Our check is quite successful: in the analysis of models we could
explicitly demonstrate all basic features:

-- exponential large-$N_{c}$ behavior of the generating functional for the
baryon wave function,

-- universality of this behavior,

-- factorization of the preexponential terms.

In the absence of the direct proof of these properties in QCD, this model
analysis combined with the independent arguments of Refs.
\cite{Pobylitsa-04,Pobylitsa-05} (based on the evolution equation, spin-flavor
symmetry and soft-pion theorem) gives us a rather solid self-consistent
picture of the baryon wave function in large-$N_{c}$ QCD.

The main criterion for our choice of models was the possibility to perform the
direct calculation of the $1/N_{c}$ expansion. In the simplest models we could
obtain analytical results. In more general models the problem has been reduced
to the solution of a coupled system of differential equations corresponding to
the large-$N_{c}$ effective classical dynamics.

\subsection{Phenomenological applications}

Although the construction of realistic models was not the aim of this work,
our analysis still has a direct relation to phenomenology.

1) The practical applications of the $1/N_{c}$ expansion which can be of
interest for phenomenology are based on the large-$N_{c}$ contracted
spin-flavor $SU(2N_{f})$ symmetry. In the problem of the baryon wave function
we have a rather nonstandard manifestation of this symmetry
\cite{Pobylitsa-05}. If we \emph{assume} this realization of the spin-flavor
symmetry, then we can derive interesting results like the connection between
the distribution amplitudes of nucleon and $\Delta$ resonance. The aim of this
paper was to \emph{justify this assumption} by checking the structure of the
$1/N_{c}$ expansion with a direct calculation.

2) The complete analysis of the effects related to the spin-flavor symmetry
requires a thorough study of the next-to-next-to leading order (NNLO) of the
$1/N_{c}$ expansion. In the current analysis we did not go so far. Most of the
work was concentrated on the leading order (LO). We could also check the
factorization properties appearing in the next-to-leading order (NLO).
However, the subtle NNLO contributions crucial for the understanding of the
spin-flavor symmetry effects remained beyond the scope of this paper.
Nevertheless the successful LO and NLO check of the picture suggested in Refs.
\cite{Pobylitsa-04,Pobylitsa-05} gives us additional evidence for the validity
of this picture.

3) Although the results of this paper are presented in a simple
quantum-mechanical form, they can be directly rewritten in field-theoretical
terms relevant for the phenomenological models with four-fermionic interaction
of the Nambu--Jona-Lasinio type.

\subsection{Other aspects}

The methods used in our analysis of large-$N_{c}$ models are also interesting
from the point of view of the relation between the large-$N$ systems and
classical dynamics. Our study of models has also revealed a connection between
the equations for the generating functional $W(g)$ and such traditional
equations of the many-body physics as the Hartree equation (including its
time-dependent version) and RPA equation. Although the methods developed here
cannot be applied directly to large-$N_{c}$ QCD, they can be used in some
special QCD problems like the diagonalization of the matrix of anomalous
dimensions of the leading twist baryon operators \cite{Pobylitsa-04}.

\textbf{Acknowledgments.} I appreciate discussions with Ya.I. Azimov, V.
Braun, V. Bunakov, T. Cohen, K. Goeke, V.~Isakov, G. Kirilin, N. Kivel, L.N.
Lipatov, A. Manashov, N. Manton, N. Nikolaev, M.V. Polyakov, A. Shuvaev, and
N. Stefanis. I am especially grateful to V.Yu. Petrov who attracted my
attention to the problem of the baryon wave function at large $N_{c}$. This
work was supported by DFG\ and BMBF.

\end{document}